\documentclass[twocolumn]{aastex631}
\usepackage{amssymb}
\usepackage{graphicx}
\usepackage{grffile}
\usepackage{epsfig}
\usepackage{epstopdf}
\usepackage[T1]{fontenc}
\usepackage{apjfonts}
\usepackage{natbib}
\usepackage{url}
\usepackage{CJK}
\usepackage{amsmath} 
\usepackage{xcolor}
\newcommand{\msun}{\ensuremath{\rm M_\odot}}

\newcommand{\Ha}{\ensuremath{\rm H\alpha}}
\newcommand{\Hb}{\ensuremath{\rm H\beta}}

\newcommand{\lya}{\ensuremath{\rm Ly\alpha}}
\newcommand{\wlya}{$W_{\rm Ly\alpha}$}

\newcommand{\NII}{[\ion{N}{2}]}
\newcommand{\kms}{km\,s\ensuremath{^{-1}}}
\newcommand{\ztwo}{\ensuremath{z\sim2}}

\newcommand{\OII}{[\ion{O}{2}]}
\newcommand{\OIII}{[\ion{O}{3}]}

\newcommand{\HI}{\ion{H}{1}}

\newcommand{\expnt}[2]{\ensuremath{#1 \times 10^{#2}}}   
\newcommand{\fluxunits}{\ensuremath{\mathrm{erg}\,\mathrm{s}^{-1}\,\mathrm{cm}^{-2}}}

\newcommand{\sbunits}{\ensuremath{\mathrm{erg}\,\mathrm{s}^{-1}\,\mathrm{cm}^{-2}\,\mathrm{arcsec}^{-2}}}

\shorttitle{The Circumgalactic Medium of Extreme Emission Line Galaxies at $z\sim2$}
\shortauthors{Erb et al.}

\submitjournal{AAS Journals}
\turnoffeditone
\turnoffedittwo

\begin{document}
\begin{CJK*}{UTF8}{gbsn}

\title{\large{The Circumgalactic Medium of Extreme Emission Line Galaxies at $\pmb{z\sim2}$: Resolved Spectroscopy and Radiative Transfer Modeling of Spatially Extended Ly$\pmb{\alpha}$\ Emission in the KBSS-KCWI Survey\footnote{Based on data obtained at the W.M. Keck Observatory, which is operated as a scientific partnership among the California Institute of Technology, the University of California, and NASA, and was made possible by the generous financial support of the W.M. Keck Foundation.}}}

\correspondingauthor{Dawn K. Erb}
\email{erbd@uwm.edu}

\author[0000-0001-9714-2758]{Dawn K. Erb}
\affil{The Leonard E.\ Parker Center for Gravitation, Cosmology and Astrophysics, Department of Physics,\\ University of Wisconsin-Milwaukee, 3135 N Maryland Avenue, Milwaukee, WI 53211, USA}

\author[0000-0001-5113-7558]{Zhihui Li (李智慧)}
\affil{Cahill Center for Astronomy and Astrophysics, California Institute of Technology, 1200 E California Blvd, MC 249-17, Pasadena, CA 91125, USA}

\author[0000-0002-4834-7260]{Charles C. Steidel}
\affil{Cahill Center for Astronomy and Astrophysics, California Institute of Technology, 1200 E California Blvd, MC 249-17, Pasadena, CA 91125, USA}

\author[0000-0003-4520-5395]{Yuguang Chen (陈昱光)}
\affil{Cahill Center for Astronomy and Astrophysics, California Institute of Technology, 1200 E California Blvd, MC 249-17, Pasadena, CA 91125, USA}
\affil{Department of Physics and Astronomy, University of California Davis, 1 Shields Avenue, Davis, CA 95616, USA}

\author[0000-0003-2491-060X]{Max Gronke}
\affil{Max-Planck Institute for Astrophysics, Karl-Schwarzschild-Str. 1, D-85741 Garching, Germany}

\author[0000-0001-6369-1636]{Allison L. Strom}
\affil{Department of Physics and Astronomy and Center for Interdisciplinary Exploration and Research in Astrophysics (CIERA), Northwestern University, 2145 Sheridan Road, Evanston, IL 60208, USA}

\author[0000-0002-6967-7322]{Ryan F. Trainor}
\affil{Department of Physics and Astronomy, Franklin \& Marshall College, 637 College Avenue, Lancaster, PA 17603, USA}

\author[0000-0002-8459-5413]{Gwen C. Rudie}
\affil{The Observatories of the Carnegie Institution for Science, 813 Santa Barbara Street, Pasadena, CA 91101, USA}

\begin{abstract}
The resonantly scattered \lya\ line illuminates the extended halos of neutral hydrogen in the circumgalactic medium of galaxies. We present integral field Keck Cosmic Web Imager observations of double-peaked, spatially extended \lya\ emission in 12 relatively low-mass ($M_{\star} \sim10^9$ \msun) $z\sim2$ galaxies characterized by extreme nebular emission lines. Using individual spaxels and small bins as well as radially binned profiles of larger regions, we find that for most objects in the sample the \lya\ blue-to-red peak ratio increases, the peak separation decreases, and the fraction of flux emerging at line center increases with radius. We use new radiative transfer simulations to model each galaxy with a clumpy, multiphase outflow with radially varying outflow velocity, and self-consistently apply the same velocity model to the low ionization interstellar absorption lines. These models reproduce the trends of peak ratio, peak separation and trough depth with radius, and broadly reconcile outflow velocities inferred from \lya\ and absorption lines. The galaxies in our sample are well-described by a model in which neutral, outflowing clumps are embedded in a hotter, more highly ionized inter-clump medium (ICM), whose residual neutral content produces absorption at the systemic redshift. The peak ratio, peak separation and trough flux fraction are primarily governed by the line-of-sight component of the outflow velocity, the \HI\ column density, and the residual neutral density in the ICM respectively. Azimuthal asymmetries in the line profile further suggest non-radial gas motions at large radii and variations in the \HI\ column density in the outer halos. 
\end{abstract}


\section{Introduction}
\label{sec:intro}
A star-forming galaxy in the early universe is the nexus of a complex interchange of gas between stars and a nested series of gaseous reservoirs:\ the interstellar medium (ISM), consisting of the gas among the stars, the circumgalactic medium (CGM), and finally the intergalactic medium (IGM). As the transition region between the stars and the IGM, most of the key processes of galaxy evolution are modulated through the CGM (see \citealt{Tumlinson2017} for a recent review). Outflows powered by star formation drive gas out of the galaxy and into the CGM, where it may be reaccreted onto the galaxy or continue on to leave the galaxy entirely (e.g.\ \citealt{Veilleux2020}). New fuel for star formation is accreted through the CGM, likely via dense, cold streams of gas \citep{Keres2005,Dekel2006}. Ionizing photons that make their way out of the galaxy must also traverse the neutral hydrogen in the CGM \citep{Rudie2013,Steidel2018}.

The strongest emission line arising from gas in the CGM is due to the \lya\ transition of hydrogen, and deep observations have now revealed that the diffuse, distant universe is aglow with \lya\ emission \citep{Wisotzki2018,Ouchi2020}. This emission arises primarily from faint halos extending to tens of kpc around galaxies, but is also seen in the form of larger nebulae (``blobs'', e.g. \citealt{Fynbo1999,Steidel2000}) and filaments (e.g.\ \citealt{Cantalupo2014,Umehata2019,Daddi2021}). The initial detections of spatially extended \lya\ emission surrounding typical star-forming galaxies at high redshifts came from stacked, narrowband images \citep{Steidel11,Momose2014,Momose2016,Xue2017}, but more recently the Multi-Unit Spectroscopic Explorer (MUSE, \citealt{muse}) at the ESO-VLT and the Keck Cosmic Web Imager (KCWI, \citealt{kcwi2010,kcwi2012}) have enabled the study of individual halos around galaxies from $2\lesssim z \lesssim 6$ \citep{Wisotzki2016,Leclercq2017,Wisotzki2018,Erb18,Leclercq2020,Chen2021}.

A number of different mechanisms have been proposed to account for this extended \lya\ emission. Perhaps the most straightforward of these is the resonant scattering of \lya\ photons produced in galaxies by neutral hydrogen in the CGM \citep{Zheng2011,Kusakabe2019,Byrohl2021}, with the \lya\ profile then reflecting the kinematics and geometry of the CGM gas. Other possible sources of the \lya\ halo emission include \textit{in situ} photoionization (fluorescence), either by ionizing radiation escaping from the galaxy or by an external radiation field \citep{Kollmeier2010,Cantalupo2012,Mas-Ribas2016}; cooling radiation from infalling gas \citep{Haiman2000,Dijkstra2009,Faucher-Giguere2010,Lake2015}; or \lya\ emission from faint satellite galaxies \citep{Mas-Ribas2017}. Multiple mechanisms may contribute in a given halo, with their relative importance varying with radius \citep{Mitchell2021}. Recent theoretical results suggest that the \lya\ properties of gaseous halos are primarily influenced by galactic outflows within $\sim50$ kpc, while cold accretion flows dominate at larger radii \citep{Chung2019}; this result is in agreement with observations that find a transition between outflow and inflow-dominated kinematics at similar radius \citep{Chen2020}.

While most studies of \lya\ halos to date have focused on the spatial distribution of the emission via imaging (either from narrowband filters or reconstructed from integral field unit [IFU] data cubes), a number of recent IFU studies have analyzed spectral variations in the extended emission, using small samples of gravitationally lensed \citep{Patricio2016,Claeyssens2019,Solimano2022} and unlensed \citep{Erb18,Leclercq2020} galaxies at $z>2$. The inclusion of spectroscopic information has the potential to be a powerful discriminant among the proposed emission mechanisms, although the resonant nature of \lya\ emission has made the extraction of physical quantities from observed spectra difficult, even in the case of single, spatially integrated line profiles. Due to multiple scatterings, the emergent profile depends on the kinematics, geometry and density of neutral hydrogen and on the dust content (see \citealt{Dijkstra2014} for a review). The strongest peak of the observed \lya\ profile is almost always redshifted relative to the systemic redshift of the galaxy due to backscattering from a receding galactic outflow, and when the opacity to \lya\ photons in the outflow is relatively low (usually seen in lower mass, highly ionized galaxies) a secondary, blueshifted peak may be visible as well. In the local universe the separation between the two peaks has been observed to correlate with the escape of ionizing Lyman continuum radiation, with objects with narrower peak separations having higher escape fractions \citep{Verhamme2017,Izotov2018}.

Spatially resolved spectroscopic studies of individual \lya\ halos have so far mostly been based on MUSE data, and have therefore necessarily focused on galaxies at $z>3$. These MUSE studies have analyzed the spectral properties of \lya\ emitters with a single peak, finding that the velocity shift of the line is generally smaller for higher surface brightness regions and that there is a correlation between the width and velocity shift of the line, with broader emission often tending to come from the outer halo \citep{Claeyssens2019,Leclercq2020,Solimano2022}. At $z=2.3$, \citet{Erb18} studied a single low-mass galaxy with KCWI, measuring variations in the peak ratio and separation of the double-peaked \lya\ profile across the extended halo and finding that higher blue-to-red peak ratios and narrower separations tended to be found at larger radii. These spectroscopic studies have been broadly interpreted in the context of the resonant scattering of \lya\ photons in a galactic outflow, but definitive models for the observed trends have yet to be constructed.

A number of radiative transfer (RT) codes have successfully reproduced the \lya\ profiles of large numbers of spatially integrated spectra, generally by modeling the outflow as a spherical, expanding shell (e.g.\ \citealt{Verhamme2008,Verhamme2015,Hashimoto2015,Yang2017,Gronke2017}). These models have provided constraints on the properties of the scattering medium, while also indicating that even within the simplified regime of the shell model the interpretation of the \lya\ profile is complex. In general, the separation between the two peaks of the line increases with increasing \HI\ column density, while the relative strength of the blue peak decreases with increasing velocity of the shell (e.g.\ \citealt{Verhamme2015}). However, the physical parameters inferred from  shell models do not always match constraints on the gas obtained from interstellar absorption and nebular emission lines (e.g. \citealt{Kulas2012,Leitherer13,Orlitova2018}), with the models predicting lower outflow velocities and higher intrinsic line widths. More generally, the outflowing gas in real galaxies is multiphase and spans a wide range in velocity, in contrast to the single value assumed by the shell models (e.g.\ \citealt{Steidel10}).

Alternatively, \lya\ RT has been studied in a more realistic multiphase, clumpy medium, where cool, \HI\ clumps are embedded in a hot, highly ionized inter-clump medium (ICM) (e.g. \citealt{Neufeld91, Hansen06, Dijkstra12, Laursen13, Duval14, Gronke2016a}). In this multiphase, clumpy model, the kinematics, covering factor, and column density of the clumps, along with the residual \HI\ number density in the ICM, act together to shape the morphology of the \lya\ profile. Such a clumpy model converges to the monolithic shell model in the limit of being ``very clumpy'' (i.e.\ having $\sim$ 1000 clumps on average per line of sight), but its unique flexibility offers the possibility of obtaining more physically reasonable parameters of the gaseous medium that are consistent with other observations \citep{Li22b}.

These models were first applied to fitting the KCWI-observed \lya\ profiles of several regions in the $z=3.1$ \lya\ blob SSA22-LAB1 \citep{Steidel2000} by \citet{Li21}, who managed to reproduce the diverse morphologies of the observed profiles with reasonable physical parameters of the gaseous medium. Notably, they found that many of the observed \lya\ profiles have significant residual fluxes at the line center, which correspond to relatively few clumps per line-of-sight and low residual \HI\ density in the ICM. In addition, the very broad \lya\ wings can be reproduced by large random velocity dispersions of the clumps, but are hard to explain in the context of shell models without requiring unphysically large widths of the {\it{intrinsic}} profiles of the \lya\ emission. 

Follow-up work by \citet{Li22a} modeled the \lya\ profiles of another $z=3.1$ \lya\ blob, SSA22-LAB2, with both the multiphase, clumpy models and shell models. They identified a significant correlation between the shell expansion velocity and the clump outflow velocity, and found that the multiphase, clumpy model may alleviate the inconsistencies between the shell model parameters and the observational data. Moreover, for the first time, they attempted to use radially-binned models to fit the spatially resolved \lya\ profiles. They found that the \lya\ profiles at different impact parameters can be reproduced self-consistently assuming a common central source, and that the variation of the clump outflow velocity with respect to impact parameter can be explained by a line-of-sight projection effect of a radial outflow. In this paper, we build on the methodology of \citet{Li22a} and continue to model spatially resolved \lya\ spectra with the multiphase, clumpy model. 

We analyze the spectral properties of spatially extended \lya\ emission for a sample of 12 relatively low-mass, low-metallicity galaxies at $z\sim2$, using integral field spectroscopy from KCWI. Our focus on low-mass galaxies with extreme nebular line emission is motivated by the likely importance of faint galaxies to reionization (e.g.\ \citealt{Kuhlen12,Robertson2015}) and by the observed and expected connections between the \lya\ profile and Lyman continuum escape \citep{Dijkstra2016,Verhamme2017,Izotov2018,Steidel2018}. Escaping ionizing radiation must travel through the CGM, and spatially resolved models of extended \lya\ emission offer the possibility of obtaining constraints on the physical conditions in the multiphase CGM gas. The double-peaked nature of \lya\ emission from highly ionized sources also provides additional constraints on the models; all 12 of our targets have double-peaked profiles, which we quantify in both individual spaxels and binned regions before modeling the results with state-of-the-art radiative transfer codes. 

We describe our sample selection, observations, and data reduction in Section \ref{sec:data}, and measure the global properties of the \lya\ emission in Section \ref{sec:global}. In Section \ref{sec:spaxelspectra} we quantify the \lya\ profiles across the extended halos, measuring the line morphology in individual spaxels and small spatial bins. We bin the data with larger regions in Section \ref{sec:averagespectra}, to measure both average properties and maximum and minimum gradients in peak ratio and separation. In Section \ref{sec:modeling} we apply new models to the both the spatially resolved \lya\ emission and the rest-frame UV interstellar absorption lines, and we summarize our results and discuss their implications in Section \ref{sec:discuss}. We assume the \citet{planck2018cosmo} values of the cosmological parameters, $H_0=67.7$ \kms\ Mpc$^{-1}$, $\Omega_{\rm m}=0.31$, and $\Omega_{\Lambda}=0.69$; with these values, 1 arcsec subtends a distance of 8.4 proper kpc at $z=2.3$, the median redshift of our sample. 

\begin{deluxetable*}{l c c c c c c c c c r r c}[t]
\tablecaption{Targets Observed}
\label{tab:targets}
\tabletypesize{\footnotesize}
\tablehead{    
\colhead{ID} & 
\colhead{RA} & 
\colhead{Dec} & 
\colhead{$\mathcal{R}$} & 
\colhead{$M_{\rm UV}$} &  
\colhead{$\beta$} & 
\colhead{$z_{\rm neb}$} & 
\colhead{$\log(M_{\star}/M_{\odot})$}  &
\colhead{SFR}  &
\colhead{sSFR}  &
\colhead{$t_{\rm exp}$} \\ 
\colhead{} & 
\colhead{(J2000)} & 
\colhead{(J2000)} & 
\colhead{(AB mag)} & 
\colhead{(AB mag)} & 
\colhead{} & 
\colhead{} & 
\colhead{}  & 
\colhead{(\msun\ yr$^{-1}$)}  & 
\colhead{(Gyr$^{-1}$)}  & 
\colhead{(hr)} \\
\colhead{(1)} & 
\colhead{(2)} & 
\colhead{(3)} & 
\colhead{(4)} & 
\colhead{(5)}  & 
\colhead{(6)} &
\colhead{(7)}  & 
\colhead{(8)}  & 
\colhead{(9)} &
\colhead{(10)} &
\colhead{(11)}
}
\startdata
Q0142-BX165 & 01:45:16.867 & $-$09:46:03.47 & 23.51 & $-$21.62 & $-$1.90 & 2.3577 & \phantom{1}9.13 & 24.4 & 18.1 & 5.0 \\
Q0142-BX186 & 01:45:17.484 & $-$09:45:07.99 & 25.32 & $-$19.81 & $-$1.24 & 2.3569 & \phantom{1}8.59 & 12.2 & 31.3 & 5.0 \\
Q0207-BX87 & 02:09:44.234 & $-$00:04:13.51 & 23.84 & $-$21.15 & $-$1.72 & 2.1924 & \phantom{1}9.48 & \phantom{1}8.3 & \phantom{1}2.8 & 4.7 \\
Q0207-BX144 & 02:09:49.209 & $-$00:05:31.67 & 23.75 & $-$21.22 & $-$2.03 & 2.1682 & \phantom{1}9.22 & 18.9 & 11.4 & 4.5 \\
Q0449-BX110 & 04:52:17.201 & $-$16:39:40.64 & 23.94 & $-$21.17 & $-$1.72 & 2.3355 & \phantom{1}9.29 & 18.2 & \phantom{1}9.3 & 5.0 \\
Q0449-BX115 & 04:52:17.861 & $-$16:39:45.36 & 24.88 & $-$20.23 & $-$2.28 & 2.3348 & \phantom{1}8.90 & \phantom{1}2.1 & \phantom{1}2.6 & 5.0 \\
Q0821-MD36 & 08:21:11.410 & $+$31:08:29.44 & 24.48 & $-$20.82 & $-$1.62 & 2.5830 & \phantom{1}9.12 & 24.1 & 18.3 & 5.1 \\
Q1549-BX102 & 15:51:55.982 & $+$19:12:44.20 & 24.32 & $-$20.67 & $-$1.64 & 2.1934 & \phantom{1}9.64 & \phantom{1}6.0 & \phantom{1}1.4 & 5.0 \\
Q1700-BX729 & 17:01:27.773 & $+$64:12:29.48 & 24.02 & $-$21.14 & $-$1.87 & 2.3993 & 10.10 & 24.5 & \phantom{1}1.9 & 4.3 \\
Q2206-BX151 & 22:08:48.674 & $-$19:42:25.42 & 23.91 & $-$21.09 & $-$2.10 & 2.1974 & \phantom{1}9.97 & \phantom{1}5.5 & \phantom{1}0.6 & 4.9 \\
Q2343-BX418 & 23:46:18.571 & $+$12:47:47.36 & 23.99 & $-$21.10 & $-$2.05 & 2.3054 & \phantom{1}8.68 & 14.4 & 30.0 & 4.8 \\
Q2343-BX660 & 23:46:29.433 & $+$12:49:45.55 & 24.17 & $-$20.81 & $-$1.87 & 2.1742 & \phantom{1}8.73 & 13.9 & 25.8 & 5.0 \\
\enddata
\tablenotetext{}{\textbf{Columns:}\ (1) Galaxy ID; (2) Right ascension in hours, minutes and seconds; (3) Declination in degrees, minutes and seconds; (4) Observed $\mathcal{R}$-band AB magnitude; (5) Absolute UV magnitude at $\sim2100$ \AA; (6) Rest-frame UV slope $\beta$ measured from $G-\mathcal{R}$ color; (7) Systemic redshift from rest-frame optical nebular emission lines; (8) Stellar mass from SED fit; (9) Star formation rate from \Ha\ luminosity (see Section \ref{sec:sample}); (10) Specific star formation rate SFR/$M_{\star}$; (11) Total KCWI integration time.}
\end{deluxetable*}

\begin{figure*}[htbp]
\centerline{\epsfig{angle=00,width=\hsize,file=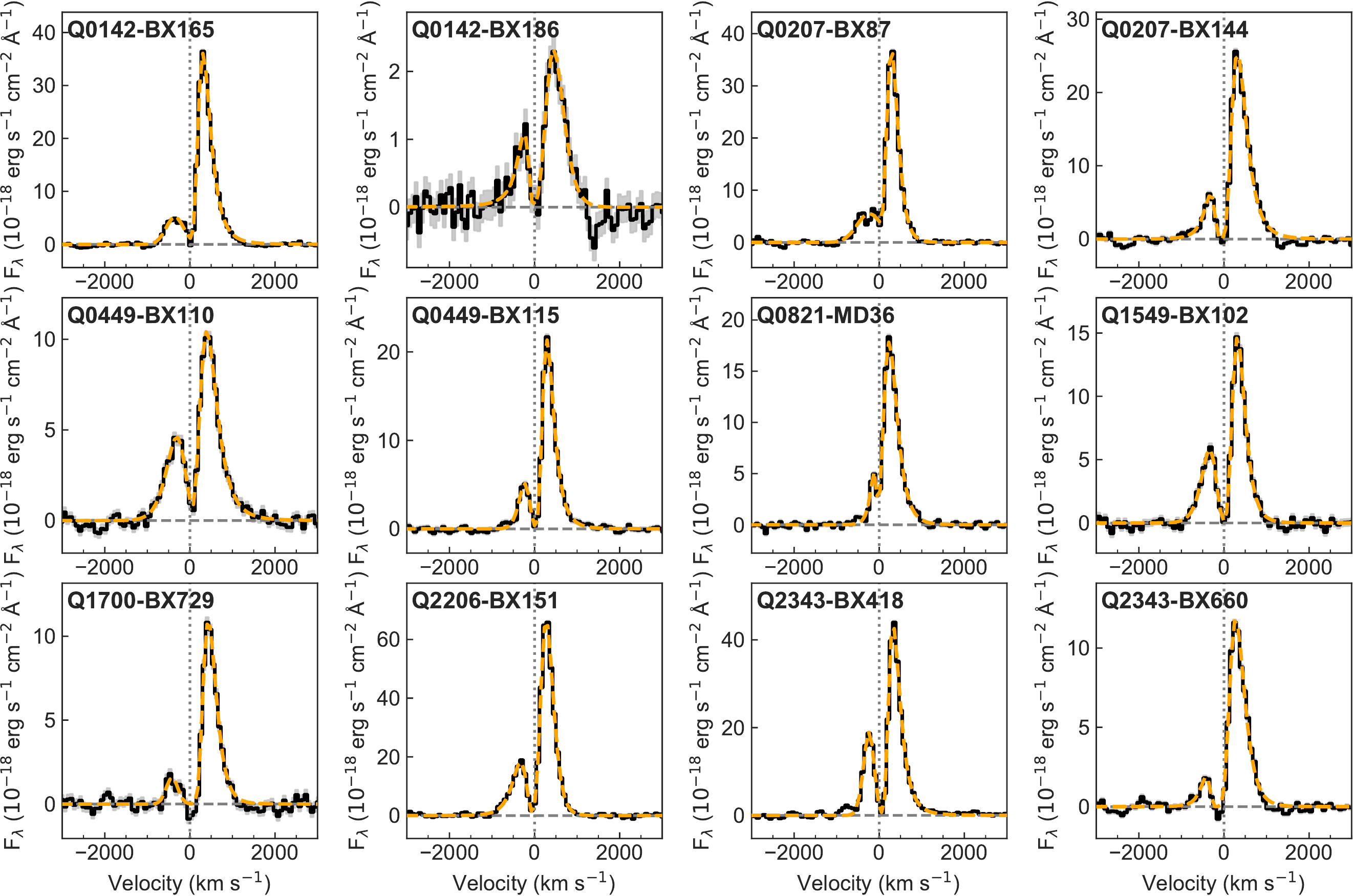}}
\caption{Continuum-subtracted \lya\ profiles from spatially integrated spectra, with double asymmetric Gaussian fits (discussed in Section \ref{sec:spaxelspectra}) shown in orange.}
\label{fig:lyaprofiles}
\end{figure*}

\begin{figure*}[htbp]
\centerline{\epsfig{angle=00,width=\hsize,file=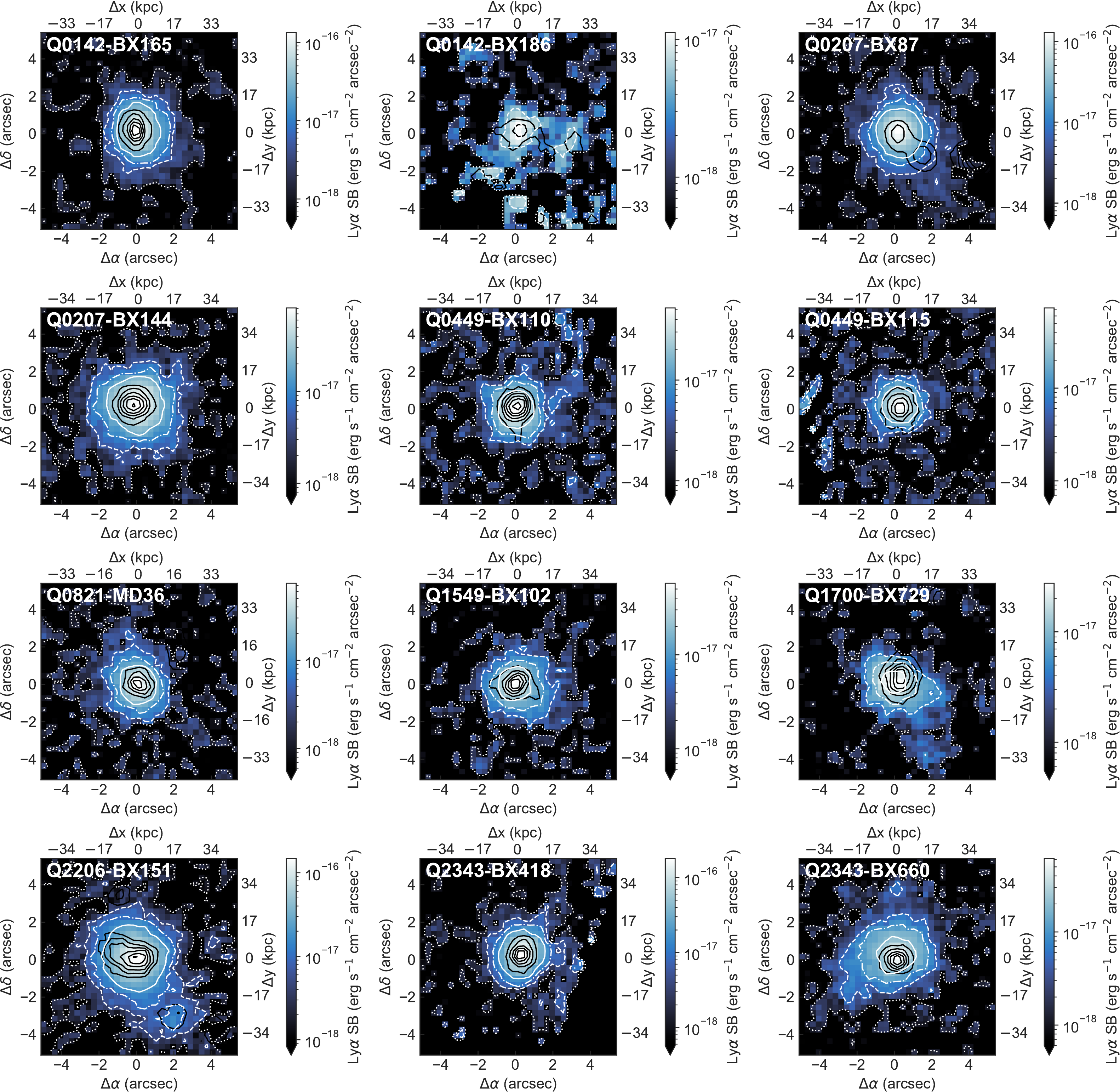}}
\caption{Continuum-subtracted \lya\ images. White contours show the \lya\ surface brightness, with the same levels in each panel:\  \expnt{1}{-18} (dotted),  \expnt{5}{-18}  (dashed), \expnt{1}{-17}  (dash-dot), and \expnt{2}{-17} (solid) \sbunits, and black contours indicate the adjacent UV continuum measured in a rest-frame 75 \AA\ window redward of the \lya\ emission line.}
\label{fig:lyaimages}
\end{figure*}

\section{Sample, Observations and Data Reduction}
\label{sec:data}

\begin{deluxetable*}{l c c c c c c c c c r r c}[h]
\tablecaption{Nebular Emission Line Measurements}
\label{tab:nebular}
\tabletypesize{\footnotesize}
\tablehead{    
\colhead{ID} & 
\colhead{$F_{\rm H\alpha}$} & 
\colhead{$F_{\rm H\beta}$} & 
\colhead{\Ha/\Hb} & 
\colhead{[\ion{N}{2}]/\Ha} & 
\colhead{[\ion{O}{3}]/\Hb} & 
\colhead{O32} & 
\colhead{O2}  &
\colhead{$n_e$} \\ 
\colhead{} & 
\colhead{($10^{-16}$ \fluxunits)} & 
\colhead{($10^{-16}$ \fluxunits)} & 
\colhead{} & 
\colhead{} & 
\colhead{} & 
\colhead{} & 
\colhead{}  & 
\colhead{(cm$^{-3}$)} \\
\colhead{(1)} & 
\colhead{(2)} & 
\colhead{(3)} & 
\colhead{(4)}  & 
\colhead{(5)} &
\colhead{(6)} &
\colhead{(7)} &
\colhead{(8)} &
\colhead{(9)}
}
\startdata
Q0142-BX165 & 1.51 $\pm$ 0.102 & 0.42 $\pm$ 0.003 & 3.56 $\pm$ 0.24 & 0.025 $\pm$ 0.006 & 6.30 $\pm$ 0.06 & 3.29 $\pm$ 0.76 & 1.32 $\pm$ 0.12 & 94 $\pm$ 96 \\
Q0142-BX186 & 0.60 $\pm$ 0.031 & 0.15 $\pm$ 0.005 & 4.04 $\pm$ 0.25 & $<$0.054 & 7.48 $\pm$ 0.29 & 3.75 $\pm$ 0.82 & 1.31 $\pm$ 0.18 & 100 $\pm$ 144 \\
Q0207-BX87 & 0.96 $\pm$ 0.081 & 0.37 $\pm$ 0.007 & 2.62 $\pm$ 0.23 & 0.041 $\pm$ 0.009 & 6.29 $\pm$ 0.16 & 6.45 $\pm$ 1.93 & 0.95 $\pm$ 0.12 & 572 $\pm$ 260 \\
Q0207-BX144 & 1.53 $\pm$ 0.032 & 0.44 $\pm$ 0.024 & 3.44 $\pm$ 0.20 & $<$0.031 & 6.15 $\pm$ 0.37 & 5.90 $\pm$ 1.16 & 1.24 $\pm$ 0.09 & 157 $\pm$ 84 \\
Q0449-BX110 & 1.10 $\pm$ 0.014 & 0.30 $\pm$ 0.013 & 3.65 $\pm$ 0.17 & $<$0.036 & 7.21 $\pm$ 0.33 & 3.86 $\pm$ 0.61 & 0.99 $\pm$ 0.05 & 488 $\pm$ 94 \\
Q0449-BX115 & ... & 0.13 $\pm$ 0.021 & ... & ... & 5.45 $\pm$ 0.92 & ... & ... & ... \\
Q0821-MD36 & 1.14 $\pm$ 0.048 & 0.31 $\pm$ 0.027 & 3.66 $\pm$ 0.35 & $<$0.063 & 6.94 $\pm$ 0.64 & 4.42 $\pm$ 1.41 & 0.89 $\pm$ 0.04 & 708 $\pm$ 112 \\
Q1549-BX102 & 0.69 $\pm$ 0.132 & 0.29 $\pm$ 0.011 & 2.42 $\pm$ 0.47 & $<$0.090 & 5.24 $\pm$ 0.22 & 3.87 $\pm$ 2.71 & 1.01 $\pm$ 0.13 & 448 $\pm$ 223 \\
Q1700-BX729 & 1.57 $\pm$ 0.019 & 0.46 $\pm$ 0.026 & 3.41 $\pm$ 0.20 & 0.059 $\pm$ 0.007 & 5.78 $\pm$ 0.34 & ... & ... & ... \\
Q2206-BX151\tablenotemark{a} & 0.63 $\pm$ 0.030 & ... & ... & ... & ... & ... & ... & ... \\
Q2343-BX418 & 1.21 $\pm$ 0.027 & 0.39 $\pm$ 0.016 & 3.09 $\pm$ 0.14 & 0.043 $\pm$ 0.008 & 7.32 $\pm$ 0.30 & 7.26 $\pm$ 1.12 & 0.93 $\pm$ 0.07 & 609 $\pm$ 156 \\
Q2343-BX660 & 1.63 $\pm$ 0.060 & 0.79 $\pm$ 0.032 & 2.06 $\pm$ 0.11 & 0.017 $\pm$ 0.005 & 6.93 $\pm$ 0.28 & 14.69 $\pm$ 2.77 & 1.15 $\pm$ 0.11 & 258 $\pm$ 128 \\
\enddata
\tablenotetext{}{\textbf{Columns:}\ (1) Galaxy ID; (2) Observed \Ha\ flux from Keck-MOSFIRE; (3) Observed \Hb\ flux; (4) \Ha/\Hb\ flux ratio; (5) [\ion{N}{2}]$\lambda$6585/\Ha\ flux ratio, with 3$\sigma$ upper limits for [\ion{N}{2}] non-detections; (6) [\ion{O}{3}]$\lambda$5008/\Hb\ flux ratio; (7) O32~$\equiv$~[\ion{O}{3}]$\lambda\lambda$4960,5008/[\ion{O}{2}]$\lambda\lambda$3727,3729, corrected for extinction; (8) Observed [\ion{O}{2}] flux ratio, O2~$\equiv$~[\ion{O}{2}]$\lambda$3729/[\ion{O}{2}]$\lambda$3727; (9) Electron density from [\ion{O}{2}] ratio. \edit1{When marked with ..., observations are unavailable.}}

\tablenotetext{a}{\Ha\ flux measurement from Keck-NIRSPEC, \citet{Kulas2012}.}
\end{deluxetable*}

\subsection{Sample selection and properties}
\label{sec:sample}
Target selection for this study was motivated by the simultaneous goals of characterizing the CGM in relatively low-mass, extreme emission line galaxies and improving our ability to extract physical information from double-peaked \lya\ emission. Because low-mass, low-metallicity, and highly ionized galaxies tend to exhibit strong, double-peaked \lya\ emission (e.g.\ \citealt{Henry2015,Erb16,Trainor2016,Verhamme2017,Matthee2021}), these objectives largely lead to the same targets.

Our targets are drawn from the Keck Baryonic Structure Survey (KBSS; \citealt{Rudie2012,Steidel2014,Strom2017}) of star-forming galaxies at $z\sim2$. Selection is primarily based on nebular emission line measurements from KBSS-MOSFIRE \citep{Steidel2014,Strom2017}:\ five of our 12 targets are drawn from the sample of \citet{Erb16}, who studied the \lya\ properties of \ztwo\ galaxies with extreme nebular emission line ratios placing them in the upper left corner of the \NII/\Ha\ vs.\ \OIII/\Hb\ ``BPT'' diagnostic diagram \citep{bpt}, with $\log(\mbox{\NII}/\Ha) \leq -1.1$ and $\log (\mbox{\OIII}/\Hb) \geq 0.75$. Galaxies in this region of the diagram lie at the low metallicity, high ionization end of the star-forming sequence, and the \ztwo\ galaxies in our sample have typical metallicities $12+\log(\rm O/H)\approx8.0$ (see \citealt{Erb16} for discussion). 

Four additional targets (Q0142-BX186, Q0449-BX110, Q0821-MD36, and Q1700-BX729) meet the nebular line ratio criteria used for the \citet{Erb16} paper but were identified later, and the remaining three objects, Q0449-BX115, Q1549-BX102 and Q2206-BX151, were selected from among the strongest \lya-emitters (LAEs) in the \ztwo\ KBSS sample. Q1549-BX102 lies just outside the emission line selection region, and Q0449-BX115 and Q2206-BX151 cannot be placed on the diagram due to insufficient data (Q0449-BX115 is detected in \OIII\ but not \Ha, and Q2206-BX151 has only \Ha\ observations). The median redshift of the sample is $z_{\rm med} = 2.32$, and all of the targets fall above the canonical LAE threshold, with rest-frame equivalent width \wlya~$>20$ \AA\ measured from long-slit spectroscopy.

The sample galaxies also have very high equivalent width \OIII\ $\lambda 5008$ emission, with $W_{[\rm O\,\sc{III}]} = 870$ \AA\ measured from a composite $H$-band spectrum (we do not measure individual equivalent widths because the continuum is noisy in many of the individual spectra); this value is comparable to that of $z\sim1$--2 reionization-era analogs selected for extreme \OIII\ emission \citep{Tang2019}.

Global properties of the galaxies are given in Table \ref{tab:targets}, and nebular emission line measurements in Table \ref{tab:nebular}. The sample is largely blue and bright, with 75\% of the objects brighter than $M^*_{\rm UV}=-20.70$ at $z\sim2.3$ \citep{Reddy2009} and median UV slope $\beta_{\rm med}=-1.87$. The median stellar mass of the sample is \expnt{1.5}{9} \msun, from modeling the spectral energy distributions with the BPASSv2.2 stellar population synthesis models \citep{BPASSv2.2} and assuming the SMC extinction law \citep{Gordon2003} and an initial mass function with slope $-2.35$ over the range 0.5--100 \msun\ and $-1.35$ between 0.1--0.5 \msun. This median stellar mass implies a halo mass of $\sim$\expnt{3}{11} \msun\  \citep{Girelli2020}, roughly three times lower than the typical halo mass of the KBSS parent sample \citep{Adelberger2005,Conroy2008,Trainor2012}; the corresponding virial radius is $\sim60$ kpc. The SED modeling also indicates that the galaxies are young and relatively unreddened, with median age 100 Myr and median $E(B-V)_{\rm cont}=0.05$. 

We use the \Ha/\Hb\ ratio and the SMC extinction law to correct the nebular emission lines for internal reddening, finding median $E(B-V)_{\rm neb}=0.14$. Two galaxies in the sample (Q0207-BX87 and Q2343-BX660) have \Ha/\Hb\ less than the theoretical value, here assumed to be 2.79 corresponding to Case B recombination at an electron temperature of 15,000 K. These galaxies are assigned $E(B-V)_{\rm neb}=0$. For the two objects that do not have measurements of the Balmer decrement we instead use reddening measurements from the SED fitting, $E(B-V)_{\rm cont}=0.0$ for both Q0449-BX115 and Q2206-BX151. 

Star formation rates are computed from the dust-corrected \Ha\ luminosity using the calibration of \citet{Theios2019}, who calculate the conversion between SFR and \Ha\ luminosity for the BPASSv2.2 stellar models used for the SED fitting described above. The resulting SFRs range from 2 to 25 \msun\ yr$^{-1}$, with a median of 14 \msun\ yr$^{-1}$. The galaxies with the two lowest SFRs in the sample, Q0449-BX115 and Q2206-BX151, are also the two for which we have determined the reddening using results from the SED fitting; $E(B-V)_{\rm cont}$ is typically smaller than $E(B-V)_{\rm neb}$, so for these two objects we have potentially underestimated the extinction correction and therefore also the SFR (in fact $E(B-V)_{\rm cont}=0$ for both, so no extinction corrections were applied). In addition, Q2206-BX151 is the only object that has not been observed with MOSFIRE. The \Ha\ flux measurement from Keck-NIRSPEC is reported by \citet{Kulas2012}, and has significant systematic uncertainties due to slit losses and the difficulties of accurate flux calibration (\citealt{Erb2006} estimated a typical factor of $\sim2$ slit loss correction for NIRSPEC observations of \Ha\ emission at $z\sim2$). 

From the stellar masses and SFRs we calculate the specific star formation rate, sSFR~$\equiv$~SFR/$M_{\star}$, finding a sample median of 10.4 Gyr$^{-1}$, more than a factor of four larger than the KBSS-MOSFIRE sample median of 2.4 Gyr$^{-1}$ \citep{Strom2017}. In other words, most of the galaxies in this sample lie significantly above the $z\sim2$ SFR-stellar mass relation (\citealt{Reddy2012,Whitaker2014}, but note that samples remain incomplete at the low masses characteristic of our targets).

The [\ion{O}{3}]/[\ion{O}{2}] ratios\footnote{O32~$\equiv$~[\ion{O}{3}]$\lambda\lambda$4960,5008/[\ion{O}{2}]$\lambda\lambda$3727,3729, corrected for extinction.} provide an estimate of the degree of excitation in the \ion{H}{2} regions. As expected given the sample selection criterion of high [\ion{O}{3}]/\Hb\ ratios, our targets fall at the upper end of the O32 distribution for the KBSS sample \citep{Strom2017}. High [\ion{O}{3}]/\Hb\ and O32 are both associated with high equivalent width \lya\ emission (e.g.\ \citealt{Nakajima2016,Trainor2019}). 

We calculate the electron density from the [\ion{O}{2}]$\lambda$3729/[\ion{O}{2}]$\lambda$3727 ratio, which has a median value of 1.01, slightly lower than the ratios of 1.13--1.16 found for composite spectra of KBSS galaxies at $z\sim2$ \citep{Steidel2014,Strom2017}. This lower ratio indicates that many of the galaxies in our sample have higher than average electron densities. Using the \OII\ electron density calibration of \citet{Sanders2016}, we find median $n_e \sim 450$ cm$^{-3}$. The nebular line ratios and electron temperatures are listed in Table \ref{tab:nebular}.
 
In summary, the galaxies studied here are relatively low-mass, highly star-forming, and luminous, with high specific star formation rates and nebular line ratios that place them at the upper end of their parent sample in ionization and electron density. They are not typical of star-forming galaxies at $z\sim2$, but may more closely resemble galaxies observed in the reionization era (e.g.\ \citealt{Stark2017}).

\subsection{Observations}
\label{sec:obs}
The 12 targets were observed with KCWI over the course of a number of observing runs between September 2018 and August 2020. We used the Medium IFU with the BL grating, which provides a field of view of 16\farcs5~$\times$~ 20\farcs4 and spectral resolution $R\approx1800$. As detailed in Table \ref{tab:targets}, total integration times were approximately five hours per target, divided into individual 1200 s exposures between which we rotated the field by $10$--$90^{\circ}$.

\subsection{Data reduction}
\label{sec:redux}
The KCWI data were reduced using procedures described in detail by \citet{Chen2021}, but we give an overview of the method here. Each KCWI exposure was reduced using the official data reduction pipeline (DRP) written in IDL.\footnote{\href{https://github.com/Keck-DataReductionPipelines/KcwiDRP}{https://github.com/Keck-DataReductionPipelines/KcwiDRP}} The DRP conducts overscan and bias subtraction, cosmic-ray removal, flat-fielding, sky subtraction, differential atmospheric refraction correction, and flux calibration, and assembles 2D spectra of the slices into a 3D data cube. A median-filtered cube was constructed for each data cube using a running boxcar filter of size 0\farcs69 $\times$ 4\farcs6 $\times$ 100 \AA. This median-filtered cube was subtracted from the original cube to remove low-frequency scattered light in both the spatial and spectral dimensions. 

The world coordinate system (WCS) of the post-DRP data cubes was corrected by cross-correlating the pseudo-white-light images of the data cubes with each other. Data cubes of multiple exposures for the same target were rotated to the north-up direction and resampled onto a common 3D grid of 0\farcs3 $\times$ 0\farcs3 $\times$ 1 \AA. The resampling was conducted using the ``drizzle'' method in the \textit{Montage} package,\footnote{\href{http://montage.ipac.caltech.edu/}{http://montage.ipac.caltech.edu/}} with a drizzle factor of 0.7. Finally, individually resampled data cubes were weighted by exposure time and averaged, creating the final data cube for each target.

\movetabledown=30mm
\begin{rotatetable*}
\begin{deluxetable*}{l c c c c c c c c c c c}
\tablecaption{\lya\ Measurements}
\label{tab:lya_measurements}
\tablehead{
\colhead{ID} &
\colhead{$F^{\rm tot}_{\lya}$} &
\colhead{Log ($L^{\rm tot}_{\lya}$)} &
\colhead{\lya/\Ha} &
\colhead{(\lya/\Ha)$_{\rm int}$} &
\colhead{$f_{\rm esc}^{\lya}$} &
\colhead{$r^{\rm eff}_{\lya}$} &
\colhead{$W^{\rm 1D}_{\lya}$} &
\colhead{$W^{\rm tot}_{\lya}$} &
\colhead{$(F_{\rm blue}/F_{\rm red})^{\rm 1D}$} &
\colhead{$\Delta v^{\rm 1D}_{\rm peak}$} &
\colhead{$f^{\rm 1D}_{\rm tr}$}  \\
\colhead{\phantom{0}} &
\colhead{($10^{-16}$ \fluxunits)} &
\colhead{(erg s$^{-1}$)} &
\colhead{} &
\colhead{} &
\colhead{} &
\colhead{(kpc)} &
\colhead{(\AA)}  &
\colhead{(\AA)} &
\colhead{} &
\colhead{(\kms)} &
\colhead{} \\
\colhead{(1)} &
\colhead{(2)} &
\colhead{(3)} &
\colhead{(4)}  &
\colhead{(5)} &
\colhead{(6)} &
\colhead{(7)} &
\colhead{(8)} &
\colhead{(9)} &
\colhead{(10)} &
\colhead{(11)} &
\colhead{(12)}}
\startdata
Q0142-BX165 & 3.69 $\pm$ 0.03 & 43.2 & 1.57 $\pm$ 0.22 & 8.4 & 0.19 $\pm$ 0.027 & 22.6 & 54.1 $\pm$ 1.9 & 84.5 $\pm$ 2.7 & 0.17 $\pm$ 0.005 & 658 $\pm$ 10 & 0.021 $\pm$ 0.002 \\
Q0142-BX186 & 0.36 $\pm$ 0.02 & 42.2 & 0.31 $\pm$ 0.04 & 8.4 & 0.04 $\pm$ 0.005 & 16.4 & 30.9 $\pm$ 6.0 & 49.6 $\pm$ 8.0 & 0.32 $\pm$ 0.067 & 687 $\pm$ 31 & 0.019 $\pm$ 0.016 \\
Q0207-BX87 & 3.72 $\pm$ 0.03 & 43.1 & 3.88 $\pm$ 0.69 & 9.0 & 0.43 $\pm$ 0.076 & 25.6 & 73.3 $\pm$ 4.1 & 130.3 $\pm$ 6.9 & 0.23 $\pm$ 0.007 & 522 $\pm$ 13 & 0.074 $\pm$ 0.002 \\
Q0207-BX144 & 4.61 $\pm$ 0.04 & 43.2 & 2.06 $\pm$ 0.22 & 8.5 & 0.24 $\pm$ 0.026 & 27.1 & 44.1 $\pm$ 2.5 & 110.2 $\pm$ 5.1 & 0.18 $\pm$ 0.012 & 631 $\pm$ 7 & $-$0.002 $\pm$ 0.003 \\
Q0449-BX110 & 1.98 $\pm$ 0.03 & 42.9 & 1.10 $\pm$ 0.09 & 9.0 & 0.12 $\pm$ 0.010 & 20.9 & 49.8 $\pm$ 4.0 & 87.6 $\pm$ 5.4 & 0.42 $\pm$ 0.021 & 700 $\pm$ 13 & 0.038 $\pm$ 0.004 \\
Q0449-BX115 & 1.73 $\pm$ 0.03 & 42.9 & 8.57 $\pm$ 2.86\tablenotemark{a} & 8.9 & 0.96 $\pm$ 0.320 & 17.1 & 98.9 $\pm$ 10.6 & 129.7 $\pm$ 12.1 & 0.21 $\pm$ 0.011 & 518 $\pm$ 8 & 0.029 $\pm$ 0.003 \\
Q0821-MD36 & 2.11 $\pm$ 0.03 & 43.1 & 1.13 $\pm$ 0.20 & 9.1 & 0.12 $\pm$ 0.022 & 20.2 & 86.8 $\pm$ 7.9 & 145.8 $\pm$ 12.3 & 0.15 $\pm$ 0.009 & 352 $\pm$ 5 & 0.091 $\pm$ 0.003 \\
Q1549-BX102 & 1.87 $\pm$ 0.03 & 42.9 & 2.71 $\pm$ 1.13 & 8.9 & 0.30 $\pm$ 0.127 & 21.6 & 44.9 $\pm$ 3.8 & 78.0 $\pm$ 5.2 & 0.43 $\pm$ 0.019 & 649 $\pm$ 9 & 0.016 $\pm$ 0.004 \\
Q1700-BX729 & 1.19 $\pm$ 0.02 & 42.7 & 0.53 $\pm$ 0.06 & 8.9 & 0.06 $\pm$ 0.007 & 20.6 & 24.6 $\pm$ 2.0 & 43.6 $\pm$ 2.9 & 0.11 $\pm$ 0.017 & 853 $\pm$ 29 & $-$0.022 $\pm$ 0.007 \\
Q2206-BX151 & 6.92 $\pm$ 0.05 & 43.4 & 10.98 $\pm$ 0.86 & 8.9 & 1.23 $\pm$ 0.096 & 29.8 & 122.5 $\pm$ 8.1 & 195.0 $\pm$ 13.1 & 0.31 $\pm$ 0.007 & 598 $\pm$ 4 & 0.029 $\pm$ 0.002  \\
Q2343-BX418 & 3.97 $\pm$ 0.03 & 43.2 & 2.71 $\pm$ 0.24 & 9.1 & 0.30 $\pm$ 0.026 & 19.8 & 64.3 $\pm$ 2.2 & 80.6 $\pm$ 2.3 & 0.38 $\pm$ 0.009 & 563 $\pm$ 2 & 0.029 $\pm$ 0.001 \\
Q2343-BX660 & 2.51 $\pm$ 0.03 & 43.0 & 1.54 $\pm$ 0.17 & 8.7 & 0.18 $\pm$ 0.020 & 26.0 & 33.5 $\pm$ 1.7 & 110.4 $\pm$ 4.7 & 0.09 $\pm$ 0.011 & 690 $\pm$ 11 & $-$0.016 $\pm$ 0.003 \\
\enddata
\tablenotetext{}{\textbf{Columns:}\ (1) Galaxy ID; (2) Total \lya\ flux in halo; (3) Total \lya\ luminosity of halo; (4) Observed \lya/\Ha\ ratio, with \Ha\ corrected for extinction; (5) Intrinsic \lya/\Ha\ ratio calculated from electron density; (6) \lya\ escape fraction; (7) Effective circular radius of halo, calculated from total area $A$ as $r^{\rm eff}_{\lya} = (A/\pi)^{1/2}$; (8) Rest-frame \lya\ equivalent width from optimally-extracted 1D spectrum; (9) Total rest-frame equivalent width of \lya\ halo; (10) Blue/red peak flux ratio of optimally-extracted 1D spectrum; (11) Peak separation of optimally-extracted 1D spectrum; (12) Fraction of the total \lya\ emission within $\pm100$ \kms\ of the trough between the peaks. }
\tablenotetext{a}{Calculated from \Hb\ flux, assuming $E(B-V)=0.0$ from SED fitting and an intrinsic ratio $\Ha/\Hb=2.79$. }
\end{deluxetable*}
\end{rotatetable*}

\section{Global \lya\ measurements}
\label{sec:global}
In this section we describe the global properties of the \lya\ emission measured by KCWI. We begin with one-dimensional spectra designed to optimize the continuum S/N, for comparison with single slit studies. We define isophotal apertures by running SExtractor \citep{sextractor} in detection mode on the collapsed, white-light images, and then extract spectra from these apertures, weighting by (S/N)$^2$. The resulting spectra are then rescaled to match the total aperture flux. Circularized radii of the apertures range from 0\farcs9 to 2\farcs2, with all but three within 0\farcs3 of the sample median of 1\farcs6. At the median redshift of the sample, these spectra cover rest-frame wavelengths $\sim$1060--1660 \AA, and include a number of interstellar absorption lines which we model along with the \lya\ emission in Section \ref{sec:lya_modeling}.

We show the continuum-subtracted \lya\ profiles from these spectra in Figure \ref{fig:lyaprofiles}, which demonstrates that all objects in the sample have double-peaked profiles with a dominant red peak. The \lya-adjacent continuum is defined as the median flux density in two windows on either side of the line, spanning 1199--1210 \AA\ ($-4000$ to $-1400$ \kms) on the blue side and 1225--1236 \AA\ ($+2300$ to $+5000$ \kms) on the red side. Given the generally high continuum S/N of the optimally extracted spectra and the lack of underlying absorption, these relatively narrow windows provide an effective measurement of the continuum around the line, as can be seen by assessing the continuum subtraction in Figure \ref{fig:lyaprofiles}. 

In order to measure equivalent widths, we integrate the line between the limits at which it reaches the continuum and divide the resulting flux by the continuum level determined above. The resulting rest-frame \lya\ equivalent widths are given in the column labeled $W^{\rm 1D}_{\lya}$ in Table \ref{tab:lya_measurements}. As previously known and by design, all are above the canonical \lya-emitter threshold of $W_{\lya}>20$ \AA. 

We next create pseudo-narrowband continuum-subtracted \lya\ surface brightness images for each object in the sample. We first identify the spatial peak of the \lya\ emission in each data cube, and then extract the summed one-dimensional \lya\ profile of a large (2\farcs4 in diameter) region centered on this peak. We measure the wavelengths at which the \lya\ emission from this large region meets the continuum on either side of the line, typically $\sim -900$ to $+1200$ \kms, and use these as the wavelength limits of a $10\farcs5 \times 10\farcs5$ (i.e.\ $35 \times 35 $ 0\farcs3 pixels) subcube centered on the \lya\ peak. We also extract blue and red subcubes with the same spatial size and spectral widths of 20 \AA\ in the rest frame from either side of the \lya\ emission, from which we measure the continuum level (because we are here measuring the continuum of individual spaxels rather than a spatially integrated region as we did above, we use slightly wider windows to increase the S/N of the measurement). The median of the blue and red subcubes along the wavelength axis results in a continuum image, which we subtract from the \lya\ subcube to create an emission-only cube. Finally, this cube is integrated along the wavelength axis to construct the continuum-subtracted \lya\ surface brightness images shown in Figure \ref{fig:lyaimages}. We note that, in general, accurate modeling and subtraction of the continuum underlying \lya\ emission can be challenging due to the complex nature of the line profiles, which often display a superposition of emission and absorption (e.g.\ \citealt{Shapley2003,Kornei2010}). However, the targets in the current sample have simpler profiles with strong emission and no detectable absorption, enabling effective continuum subtraction with the simple method described here.

The \lya\ emission shown in Figure \ref{fig:lyaimages} is significantly more extended than the underlying UV continuum in all cases, as can be seen by comparing the white (\lya) and black (continuum) contours in Figure \ref{fig:lyaimages}. This comparison also shows that in most cases the \lya\ and continuum peaks are spatially coincident. We measure the total \lya\ flux of each extended halo by summing the largest connected region with S/N~$>1$ (corresponding to a surface brightness of $\sim1$--\expnt{2}{-18} \sbunits\ for targets with approximately 5 hrs of integration), and calculate an effective circular radius $r^{\rm eff}_{\lya} = (A/\pi)^{1/2}$, where $A$ is the area of the region. The total \lya\ fluxes range from 0.4 \edit1{to} \expnt{7}{-16} \fluxunits, corresponding to \lya\ luminosities ranging from \expnt{1.6}{42} to \expnt{2.6}{43} erg s$^{-1}$ with median $10^{43}$ erg s$^{-1}$, while the radii vary between 16 and 30 kpc. 

\lya\ escape fractions are computed by comparing the total \lya\ flux with the predicted flux calculated from the dust-corrected \Ha\ emission\footnote{We use \Hb\ emission for Q0449-BX115, assuming \Ha/\Hb~=2.79.} and the theoretical \lya/\Ha\ ratio. Estimates of the \lya\ escape fraction typically assume an intrinsic ratio $(\lya/\Ha)_{\rm int}=8.7$ (e.g.\ \citealt{Hayes2015}), but the \lya/\Ha\ ratio is density-dependent, increasing at higher electron densities as collisional processes suppress two-photon continuum emission. For the range of densities in our sample, $n_e \sim 100$ -- 700 cm$^{-3}$, $(\lya/\Ha)_{\rm int}$ ranges from 8.4 to 9.1.\footnote{We assume $T_e=15,000$ K. The dependence of the \lya/\Ha\ ratio on temperature is smaller than the dependence on density, $\sim2$\% between 10,000 and 20,000 K.} We therefore adopt an intrinsic ratio for each object that depends on the electron density; these ratios are listed in the column labeled $(\lya/\Ha)_{\rm int}$ in Table \ref{tab:lya_measurements}. For the three galaxies that do not have measurements of $n_e$ (Q0449-BX115, Q1700-BX729 and Q2206-BX151) we adopt the sample median of $n_e\sim450$ cm$^{-3}$, corresponding to $(\lya/\Ha)_{\rm int}=8.9$. 

The resulting escape fractions range from 0.04 to 1.2, with a median of 0.22. We measure $f_{\rm esc}^{\lya}=1.23\pm0.096$ for Q2206-BX151, suggesting $\sim100$\% of the \lya\ emission escapes; however, as discussed in Section \ref{sec:sample}, the \Ha\ flux for this object has significant systematic uncertainties and is likely underestimated, leading to an overestimate of the escape fraction. Nevertheless, Q2206-BX151 does have the largest \lya\ luminosity, halo size, and equivalent width in the sample, so a high escape fraction may also be expected. Similarly, the second highest escape fraction measured, $f_{\rm esc}^{\lya}=0.96\pm0.32$ for Q0449-BX115, may also be overestimated due to the use of $E(B-V)_{\rm cont}$ rather than $E(B-V)_{\rm neb}$ for the dust correction; in practice, $E(B-V)_{\rm cont}=0$ for Q0449-BX115, so no dust correction was applied.

We next calculate the total \lya\ equivalent width by dividing the total \lya\ flux by the median continuum flux density of the optimally extracted one-dimensional spectra, determined as described above. After conversion to the rest frame, these total equivalent widths are larger than the 1D equivalent widths measured above by factors ranging from 1.3 to 3.3, with a median of 1.7. The largest total equivalent width in the sample is $\sim200$ \AA\ (for Q2206-BX151), roughly the maximum value expected from a normal stellar population \citep{Charlot1993}. All \lya\ measurements are given in Table \ref{tab:lya_measurements}. 

Finally, we note that in this work we focus on the spectral properties of the extended \lya\ emission in this strongly \lya-emitting subset of the KBSS sample; an analysis of the structural and spectral properties of \lya\ emission in the full KBSS-KCWI sample will be presented elsewhere.

\begin{figure*}[t]
\centerline{\epsfig{angle=00,width=\hsize,file=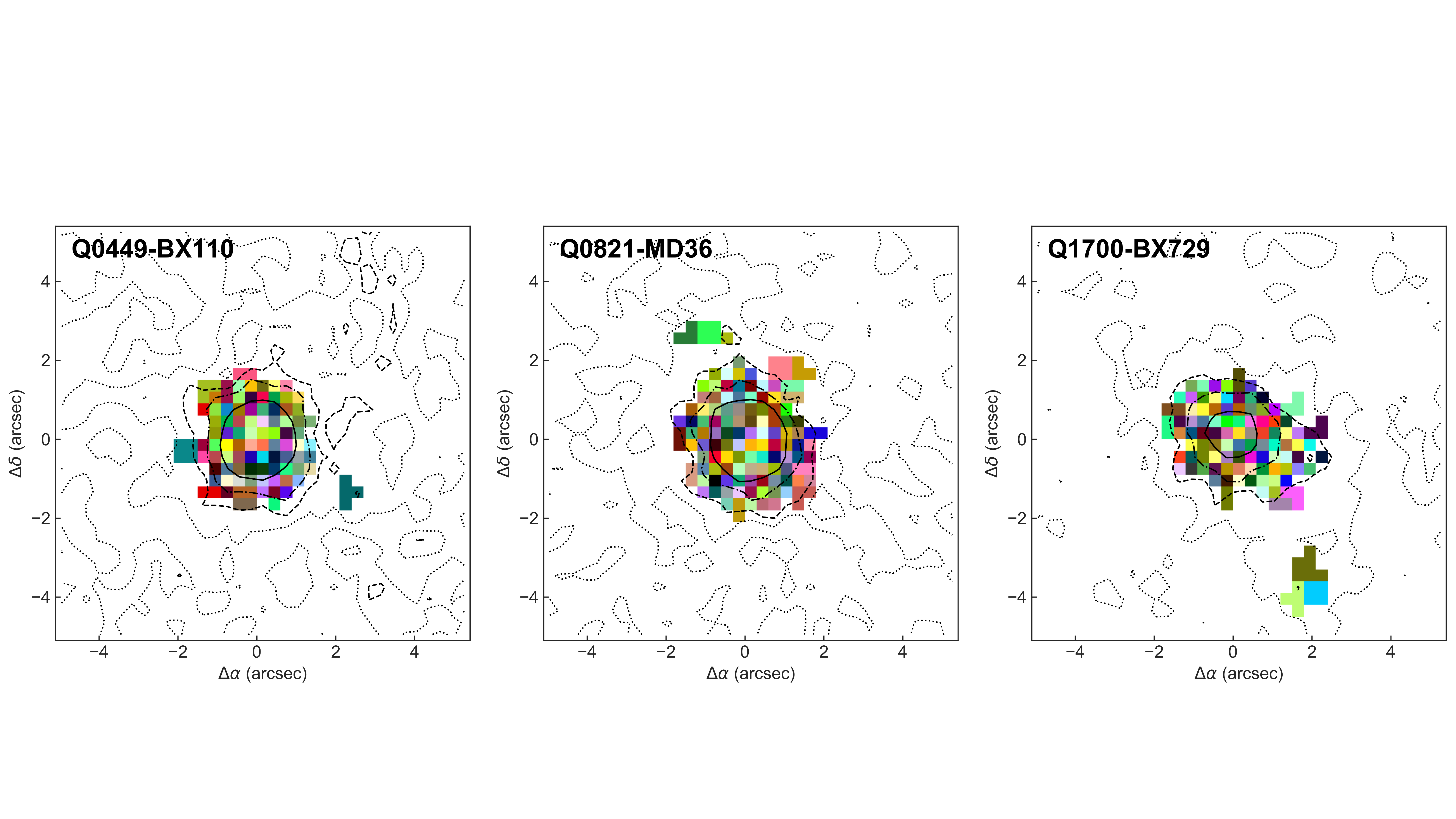}}
\caption{Representative results of the Voronoi binning of the \lya\ surface brightness images. The bins include only individual pixels with S/N~$\geq1$, and are constructed to achieve a target S/N~$\geq3$. Bins generally consist of single pixels, except in the outer halo where each object has a few bins of $\sim2$--5 pixels. Contours showing \lya\ surface brightness are the same as in Figure \ref{fig:lyaimages}, and show that regions reaching the target S/N generally have surface brightness $\gtrsim5\times10^{-18}$ \sbunits\ (dashed contour).}
\label{fig:vorbins}
\end{figure*}

\section{The spectral profiles of spatially extended \lya\ emission}
\label{sec:spaxelspectra}
In this section we analyze the spatial variations of the spectral \lya\ profiles across the extended halos using individual spaxels and small bins, focusing on the flux ratio and velocity separation of the two peaks. We first apply an adaptive two-dimensional Voronoi binning procedure to increase the S/N in the outer regions of the halos using the python package \texttt{vorbin}, an implementation of the method described in detail by \citet{Cappellari2003}. Beginning with the highest S/N pixel, this routine works its way outward to lower S/N regions until a pixel with S/N lower than a specified target threshold is reached; a bin is then constructed from adjacent pixels, seeking to match the target S/N. We run the binning on the \lya\ surface brightness images described in Section \ref{sec:global} above, avoiding the inclusion of pure noise by using only individual pixels with S/N~$>1$ and setting a target for each bin of S/N~$\geq3$. In practice, most ($\sim90$\%) of the resulting bins consist of single pixels, and very few (0--3 per object) contain four or more pixels. Three representative examples of the results of the Voronoi binning are shown in Figure \ref{fig:vorbins}, with only bins with S/N~$\geq3$ shown and each bin indicated by a different color.

\begin{figure*}[htbp]
\centerline{\epsfig{angle=00,width=\hsize,file=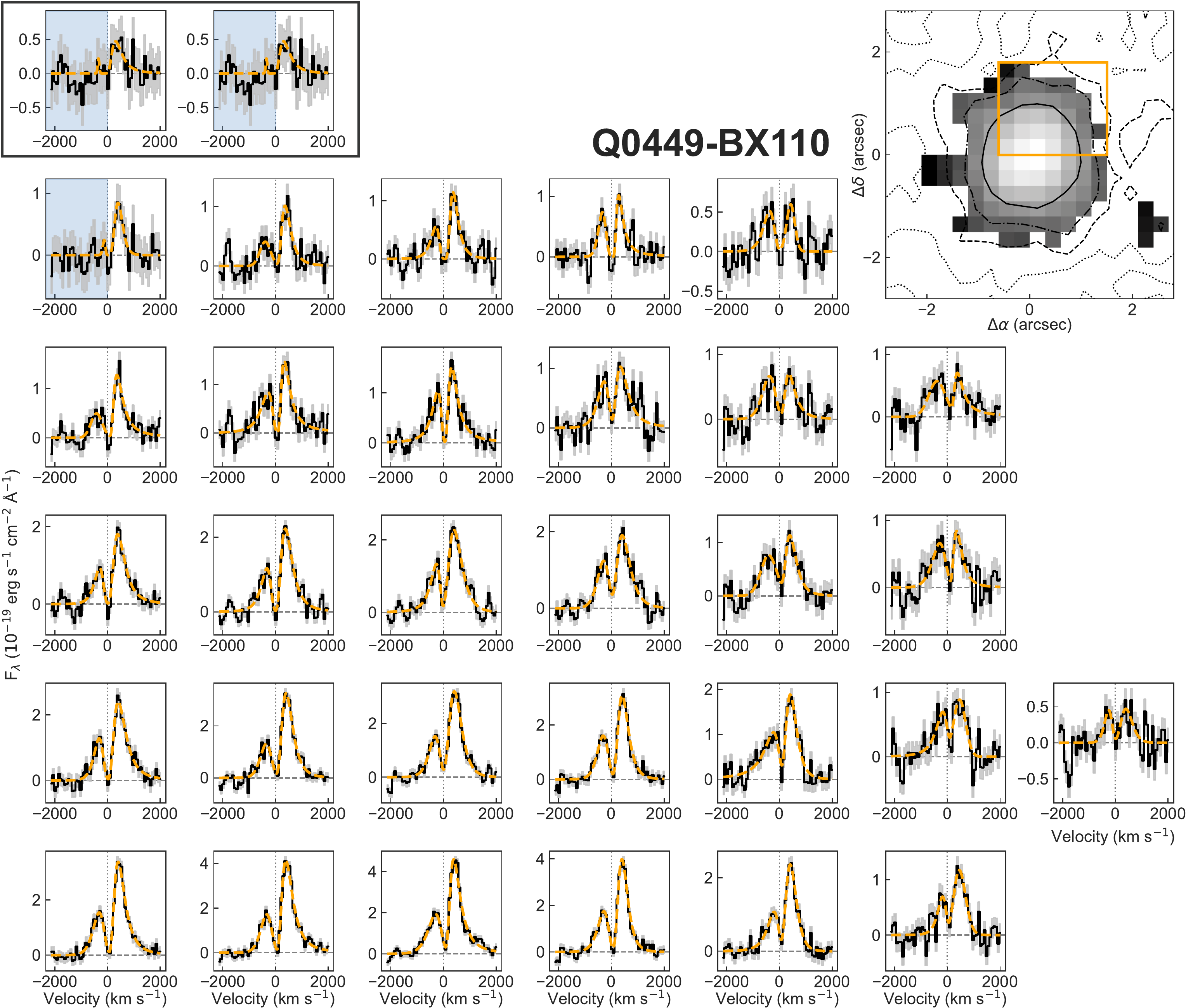}}
\caption{The \lya\ profiles (black) and asymmetric Gaussian fits (orange dashed lines) for individual spaxels and Voronoi bins in the NW portion of the halo of Q0449-BX110. The two spaxels in the grey box in the top row comprise a single bin, and the two spectra plotted within the box are identical. The blue shaded regions indicate that the blue side of the line is undetected. The image at upper right shows the \lya\ surface brightness of the full modeled region of Q0449-BX110 on a logarithmic scale, with the portion shown here outlined in orange. The \lya\ profiles are shown in the order of their corresponding spaxels in the orange box. Contours of \lya\ surface brightness are the same as in Figure \ref{fig:lyaimages}.}
\label{fig:spaxelspecs}
\end{figure*}

Once the bins are identified, we extract the spectrum and variance of each bin from the continuum-subtracted \lya\ datacube, using the mean of the individual spaxels for bins consisting of more than one spaxel. We then fit the \lya\ profile of each bin with a double asymmetric Gaussian function in velocity space, defined as
\begin{equation}
\begin{aligned}
    f(v) ={}  A_{\rm blue}\, &\exp \left( \frac{-(v-v_{\rm 0, blue})^2}{2\sigma_{\rm blue}^2} \right) + \\
     A_{\rm red}\, &\exp \left( \frac{-(v-v_{\rm 0, red})^2}{2\sigma_{\rm red}^2} \right),
\end{aligned}
\end{equation}
where $A_{\rm blue}$ and $A_{\rm red}$ and $v_{\rm 0,blue}$ and $v_{\rm 0,red}$ are the amplitudes and peak velocities of the blue and red components respectively. The asymmetric line width $\sigma$ is further defined as $\sigma = a(v-v_0)+d$, where $a$ and $d$ describe the asymmetry and width of the profile respectively. A single asymmetric Gaussian has been previously used to fit \lya\ profiles at high redshift by \citet{Shibuya2014} and \citet{Leclercq2020}; here we introduce separate components for the blue and red parts of the lines, given the strong double-peaked nature of our sources. Double asymmetric Gaussian fits to the spatially integrated \lya\ spectra are shown in Figure \ref{fig:lyaprofiles}, demonstrating that they generally provide an excellent representation of the line profile.

For each spaxel we measure the flux ratio of the blue and red peaks by integrating each side of the line between the trough between the peaks (or zero velocity, in the rare cases in which the trough is not present) and the point at which the S/N drops below unity. For spaxels with significant detections of both peaks, we use the parameters of the fit to define the \lya\ peak separation for each spaxel as $\Delta v_{\rm peak} = v_{\rm 0,red} - v_{\rm 0,blue}$. Flux uncertainties are determined via error propagation of the variance cube, while uncertainties on the peak separation result from the covariance matrix of the Gaussian fit. We also tested a Markov Chain Monte Carlo approach to the fitting and uncertainties, but found that it did not change the results significantly while being very time-consuming given the number of fits involved. 

Figure \ref{fig:spaxelspecs} shows an example of the spectra of individual spaxels and Voronoi bins for a portion of the halo of Q0449-BX110, chosen because it is near the sample median in both total \lya\ flux and halo size. The asymmetric Gaussian fits to each spaxel are overplotted in orange. 

\begin{figure*}[htbp]
\centerline{\epsfig{angle=00,width=\hsize,file=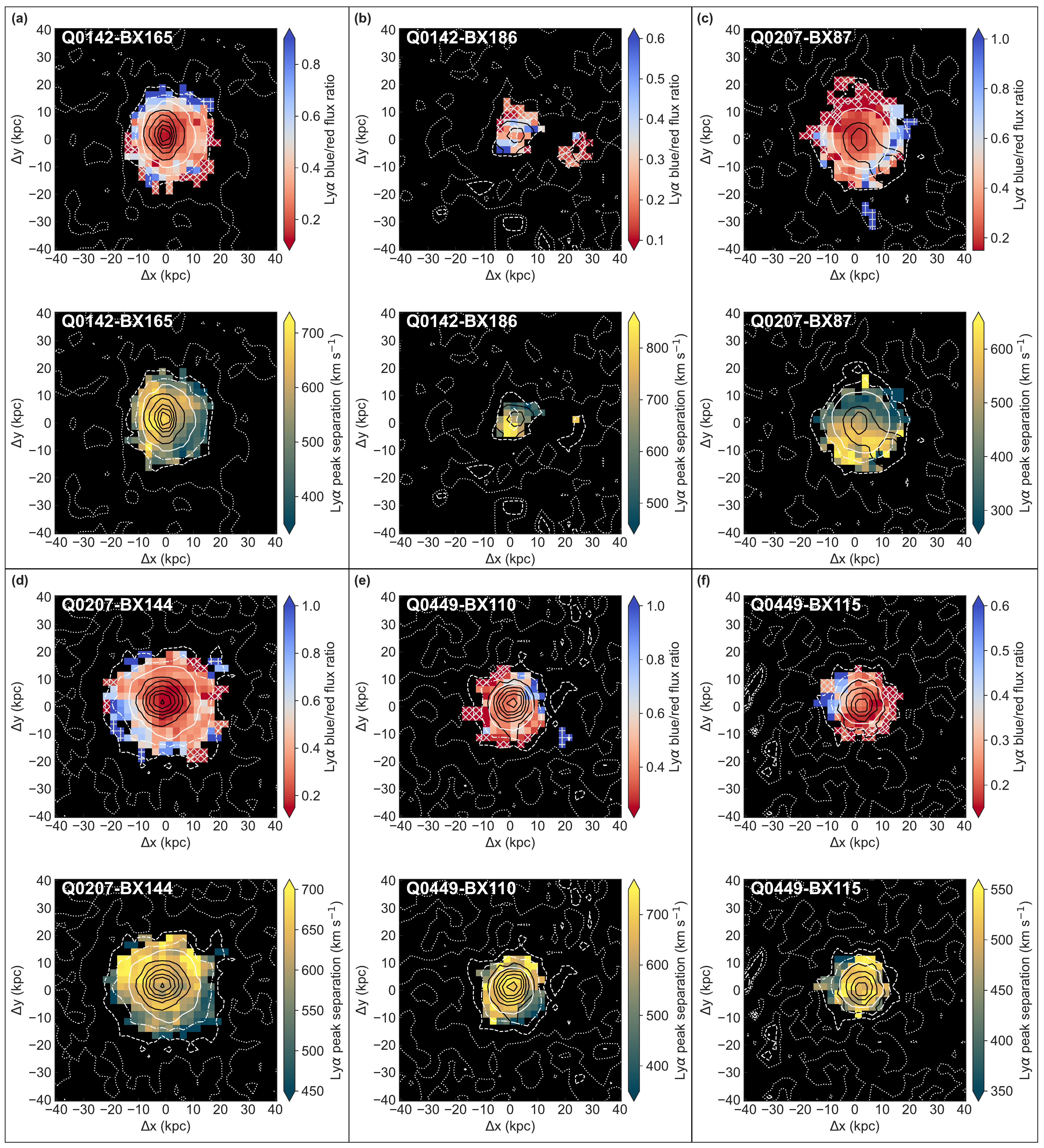}}
\caption{Maps of the \lya\ peak ratio and separation for the first six objects in the sample. Each lettered panel (a) through (f) shows the peak ratio at the top and the peak separation at bottom for a given object. The blue peak is undetected for spaxels marked with an $\times$, the red peak is undetected for spaxels marked with a $+$, and the color of these spaxels indicates the 1$\sigma$ upper limit in the case of blue non-detections and the 1$\sigma$ lower limit in the case of red non-detections. We measure the peak separation only for spaxels for which both peaks are detected. White and black contours show the \lya\ and continuum surface brightnesses respectively, as in Figure \ref{fig:lyaimages}. }
\label{fig:peakmaps1}
\end{figure*}

\begin{figure*}[htbp]
\centerline{\epsfig{angle=00,width=\hsize,file=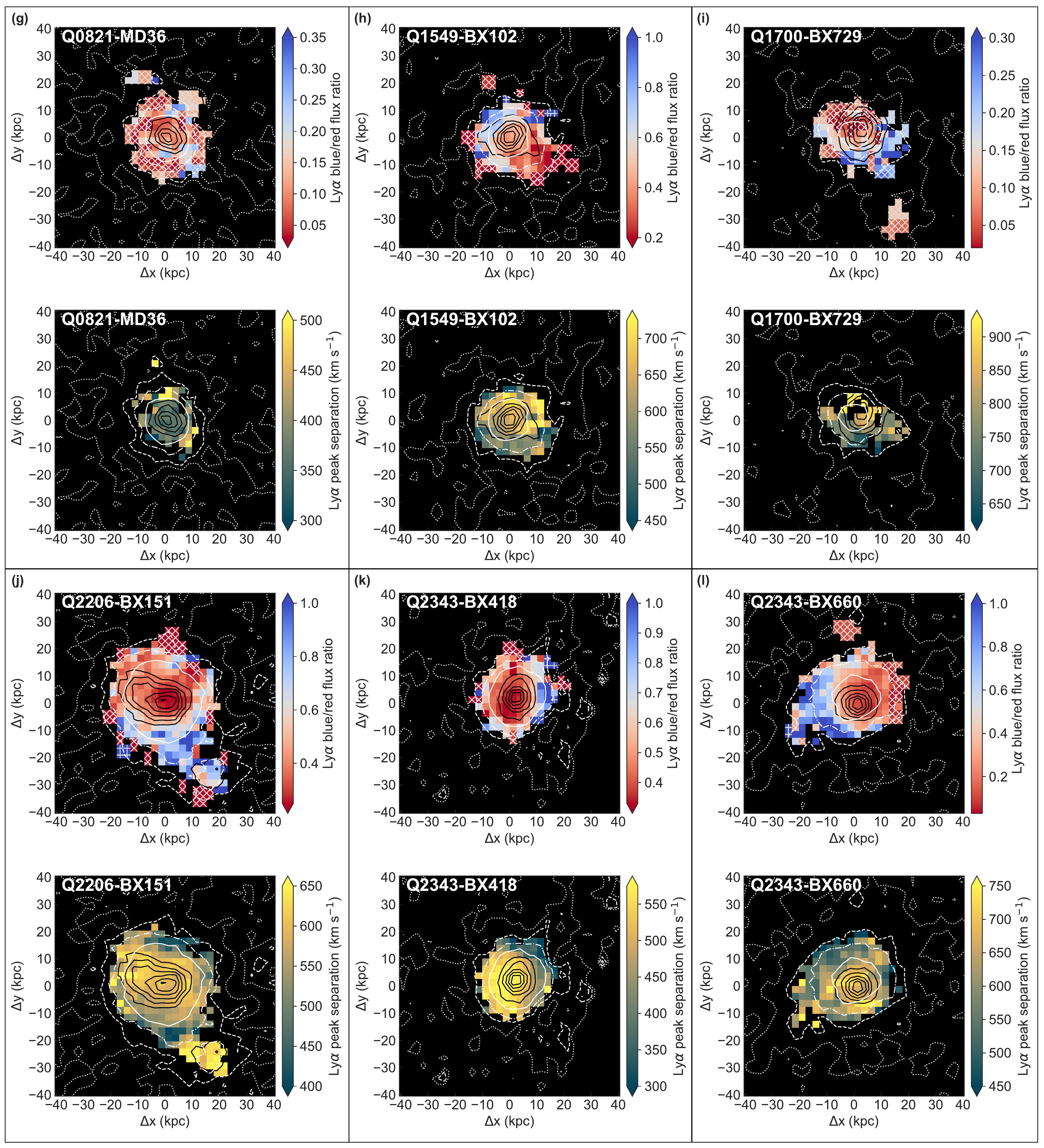}}
\caption{Maps of the \lya\ peak ratio and separation for objects 7--12. Each lettered panel (g) through (l) shows the peak ratio at the top and the peak separation at bottom for a given object. The blue peak is undetected for spaxels marked with an $\times$, the red peak is undetected for spaxels marked with a $+$, and the color of these spaxels indicates the 1$\sigma$ upper limit in the case of blue non-detections and the 1$\sigma$ lower limit in the case of red non-detections. We measure the peak separation only for spaxels for which both peaks are detected. White and black contours show the \lya\ and continuum surface brightnesses respectively, as in Figure \ref{fig:lyaimages}. }
\label{fig:peakmaps2}
\end{figure*}

Maps of the \lya\ blue/red flux ratio and peak separation resulting from these measurements for the 12 galaxies are shown in Figures \ref{fig:peakmaps1} and \ref{fig:peakmaps2}, where each lettered panel shows the peak ratio at the top and the peak separation at bottom. Higher flux ratios (i.e.\ those with a stronger blue peak) and smaller peak separations are indicated in blue. Most objects have peak ratios ranging from $\sim0.1$--$1$ and peak separations ranging from $\sim300$ to $\sim700$ \kms, although a few have lower peak ratios (Q0449-BX115, Q0821-MD36, Q1700-BX729) or larger separations (Q1700-BX729). The maps also indicate that regions of higher peak ratio and lower peak separations are generally found in the outer parts of the halos, although not always in the same regions. 

\begin{figure*}[htbp]
\centerline{\epsfig{angle=00,width=\hsize,file=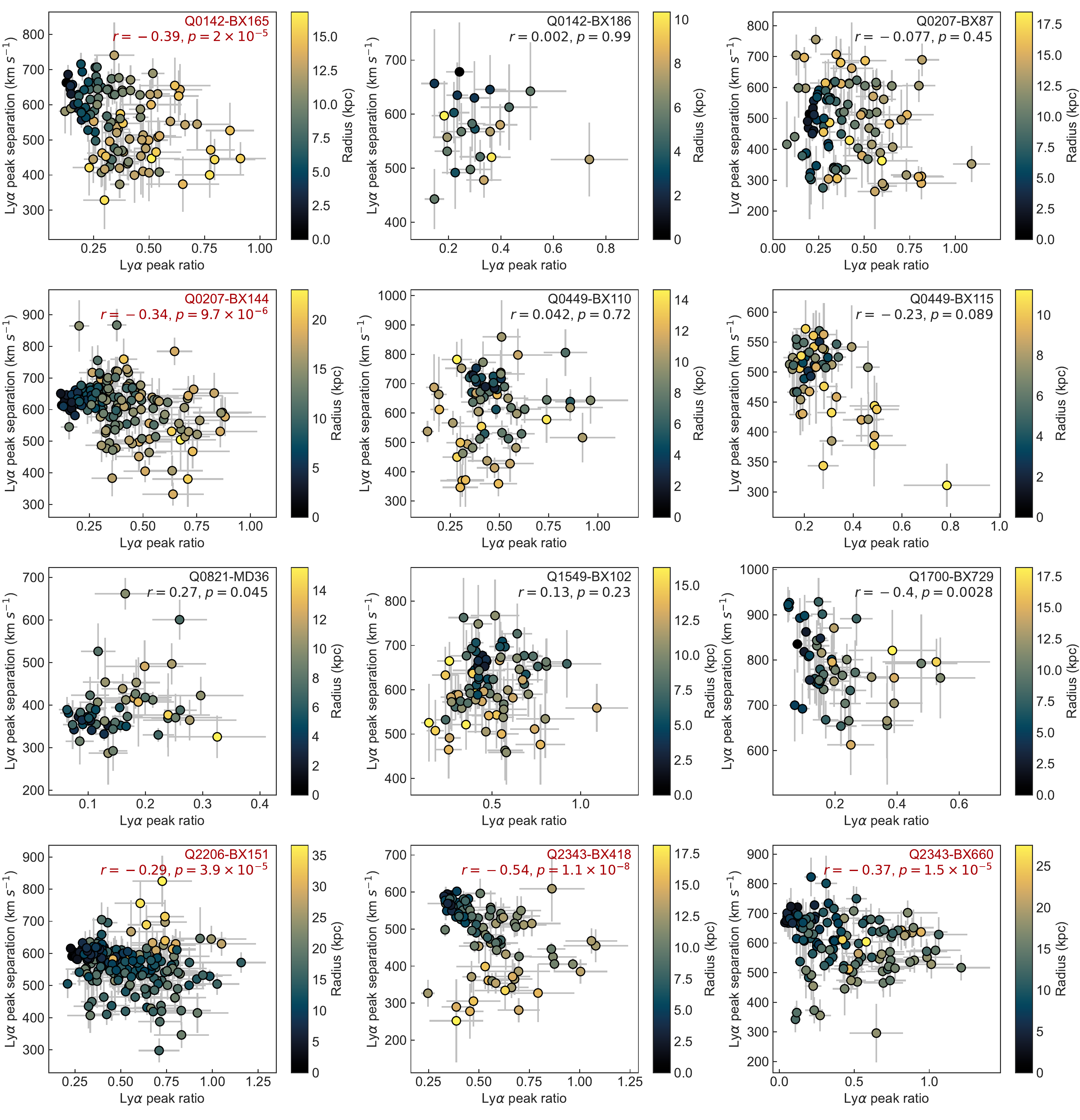}}
\caption{\lya\ peak ratio vs.\ separation. Each point represents a single spaxel or Voronoi bin, and is color-coded by its distance from the peak of the \lya\ surface brightness image. Spearman $r$ and $p$ values are given in the upper right of each panel, with significant ($>3\sigma$) correlations labeled in red. With $p=\expnt{2.83}{-3}$, Q1700-BX729 has 2.99$\sigma$ significance.}
\label{fig:ratio_vs_sep}
\end{figure*}

\begin{figure}[htbp]
\centerline{\epsfig{angle=00,width=\hsize,file=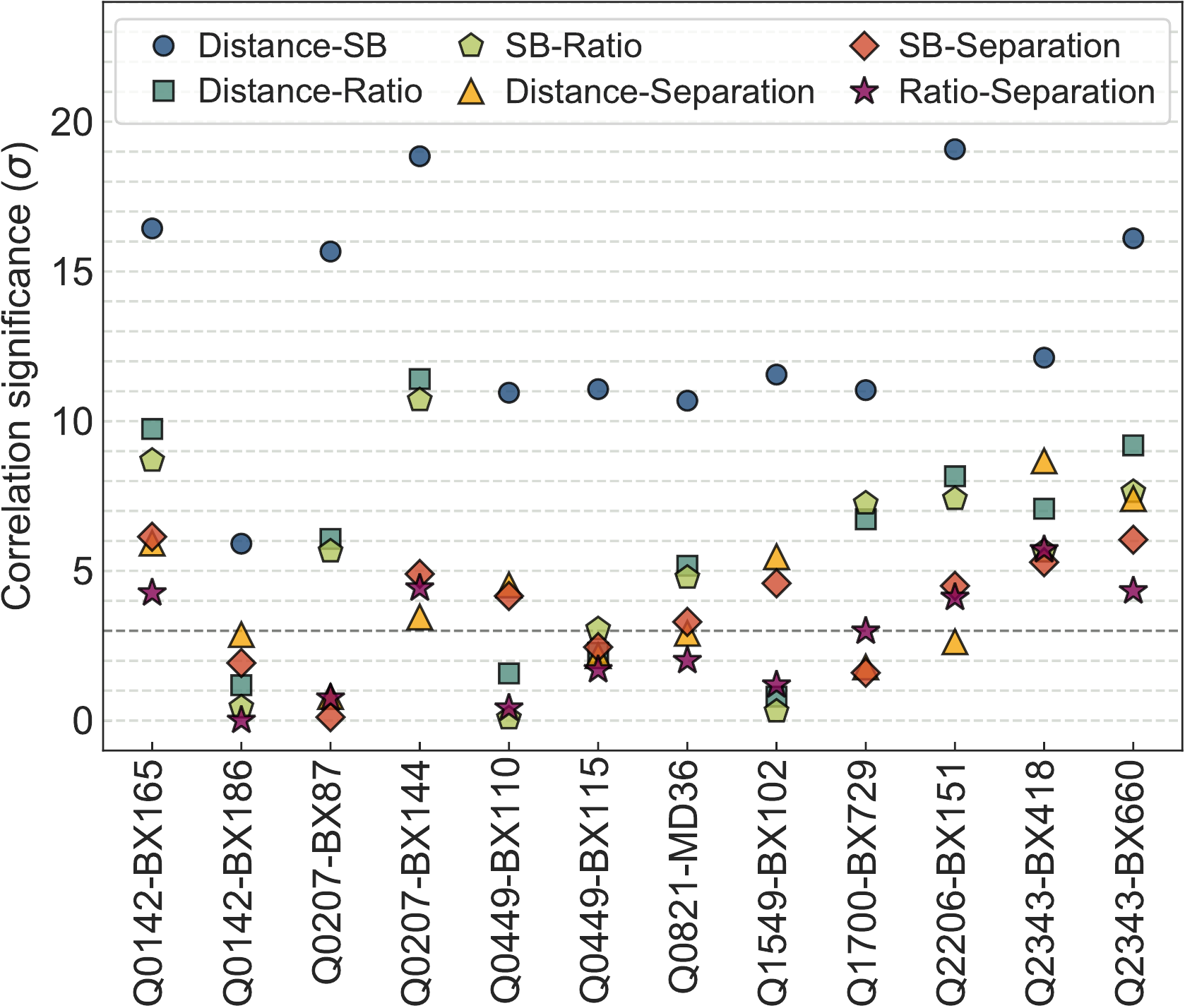}}
\caption{Significance levels from pairwise Spearman correlation tests of distance from the peak of the \lya\ surface brightness image, \lya\ surface brightness, peak ratio and peak separation.}
\label{fig:correlations}
\end{figure}

We study the relationship between the peak ratio and separation further in Figure \ref{fig:ratio_vs_sep}, in which we plot the ratio vs.\ separation for each spaxel or bin of each of the 12 objects, with each point color-coded by its distance from the center, defined as the \lya\ surface brightness peak. This color-coding again shows that higher ratios and lower separations tend to be found at larger radii. For each object we measure the Spearman correlation coefficient $r$ and the probability $p$ of the null hypothesis that the peak ratio and separation are uncorrelated; these results are reported in the upper right of each panel, with significant correlations ($>3\sigma$, $p<0.0027$) labeled in red. Five of the twelve objects in the sample show significant anti-correlations, such that a higher peak ratio is associated with a lower peak separation; one additional object, Q1700-BX729, shows a 2.99$\sigma$ correlation.  

To gain further insight into the connection between peak ratio and separation, we perform pairwise Spearman correlation tests on the four quantities we measure for each spaxel or Voronoi bin:\ distance from the center, \lya\ surface brightness, peak ratio, and peak separation. This results in six correlation coefficients for each object, and we show the significance levels of these correlations in Figure \ref{fig:correlations}. In addition to the expected extremely strong correlation between distance and surface brightness, we see strong ($>5\sigma$) correlations between peak ratio and distance or surface brightness for most (8/12) of the sample. Correlations between peak separation and distance or surface brightness are usually weaker, but are present at $>3\sigma$ in 8/12 objects. For two galaxies in the sample (Q0449-BX110 and Q1549-BX102) the separation is more strongly correlated with distance and surface brightness than the ratio. All objects except Q0142-BX186 have at least one $>3\sigma$ correlation in addition to that of distance and surface brightness, and there is no obvious preference for either distance or surface brightness to be more strongly correlated with the line profile properties. The galaxy showing no correlations with \lya\ peak properties, Q0142-BX186, is the faintest object in the sample, with total \lya\ flux more than three times lower than the second faintest (Q1700-BX729), demonstrating that high S/N over a relatively large area is needed to detect these trends.  

These results suggest that the correlations between peak ratio and separation are largely driven by the underlying tendencies of these quantities to increase and decrease respectively in the outer, fainter parts of the halos. We will discuss the relationship between the ratio and separation in more detail in the following sections.

\begin{figure*}[htbp]
\centerline{\epsfig{angle=00,width=0.9\hsize,file=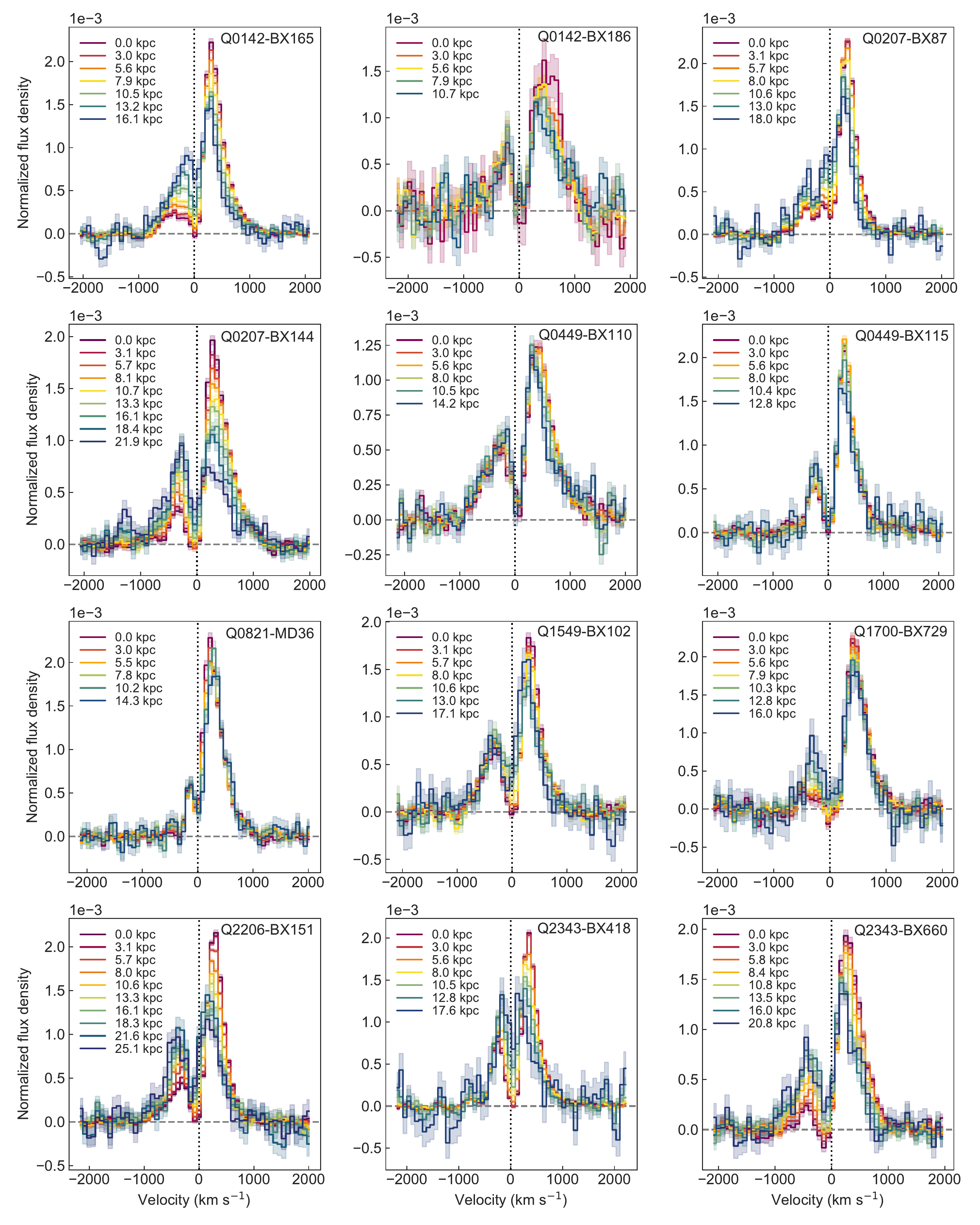}}
\caption{Normalized annular \lya\ profiles, constructed by binning all spaxels that have S/N~$>2$ in the continuum-subtracted \lya\ images in single spaxel (0\farcs3) radial increments. The spectra are color-coded by radius, with the inner portions of the halo in red and the outer portions in blue. The legend in each panel gives the median radius of each bin. For most of the sample, the  blue-to-red peak ratio increases and the depth of the trough between the peaks decreases with increasing radius. \edit1{In nearly all cases, the optimally extracted spectra shown in Figure \ref{fig:lyaprofiles} are statistically indistinguishable from the annular profiles at $r\approx3$ kpc.}}
\label{fig:annularspectra}
\end{figure*}

\section{Spatially averaged \lya\ profiles}
\label{sec:averagespectra}
In order to determine general trends and study the spectral properties of the \lya\ halos to larger radii than can be measured with individual spaxels and the small Voronoi bins, we also construct binned spectra of larger regions, using both the entire halo and smaller regions chosen based on their spectral properties.

\subsection{Annular Ly$\alpha$ profiles}
\label{sec:annularspectra}
We first study the average variation of the \lya\ profile as a function of radius by making annular spectra binned by radius for all objects in the sample. Beginning with the central, highest surface brightness spaxel and including all spaxels with S/N~$>2$ in the continuum-subtracted \lya\ images, we bin each halo in single spaxel (0\farcs3) radial increments. The spectra of all the spaxels in each bin are then summed, and the resulting \lya\ profile is  normalized to a total flux of 1. This normalization enables a straightforward visual examination of changes in the shape of the profile with radius, and is also used to format the spectra for the radiative transfer modeling discussed in Section \ref{sec:modeling}. The normalized spectra are shown in Figure \ref{fig:annularspectra}, color-coded by radius with the central portions of each halo in red and the outer portions in blue.

The increasing strength of the blue peak relative to the red peak with increasing radius is clearly apparent for most of the objects in the sample. It is also clear that the depth of the trough between the two peaks decreases with increasing radius for most of the sample. Although generally less obvious to the eye, the trend of decreasing peak separation with increasing radius is also apparent in many of the sources. We quantify these trends by fitting double asymmetric Gaussian profiles to the binned annular spectra as described in Section \ref{sec:spaxelspectra} above, measuring the average peak ratio and separation as a function of both radius and average surface brightness. We also quantify the depth of the trough between the peaks by measuring $f_{\rm tr}$, defined as the fraction of the total emission within $\pm100$ \kms\ of the trough.\footnote{$f_{\rm tr}$ differs slightly from the quantity $f_{\rm cen}$ defined by \citet{Naidu2022}, who measure the fraction of flux escaping within $\pm100$ \kms\ of the systemic velocity; we instead measure the flux on either side of the trough to account for the fact that the trough is occasionally slightly offset from zero velocity (e.g.\ Q2343-BX660).}

\begin{figure*}[ht!]
\centerline{\epsfig{angle=00,width=\hsize,file=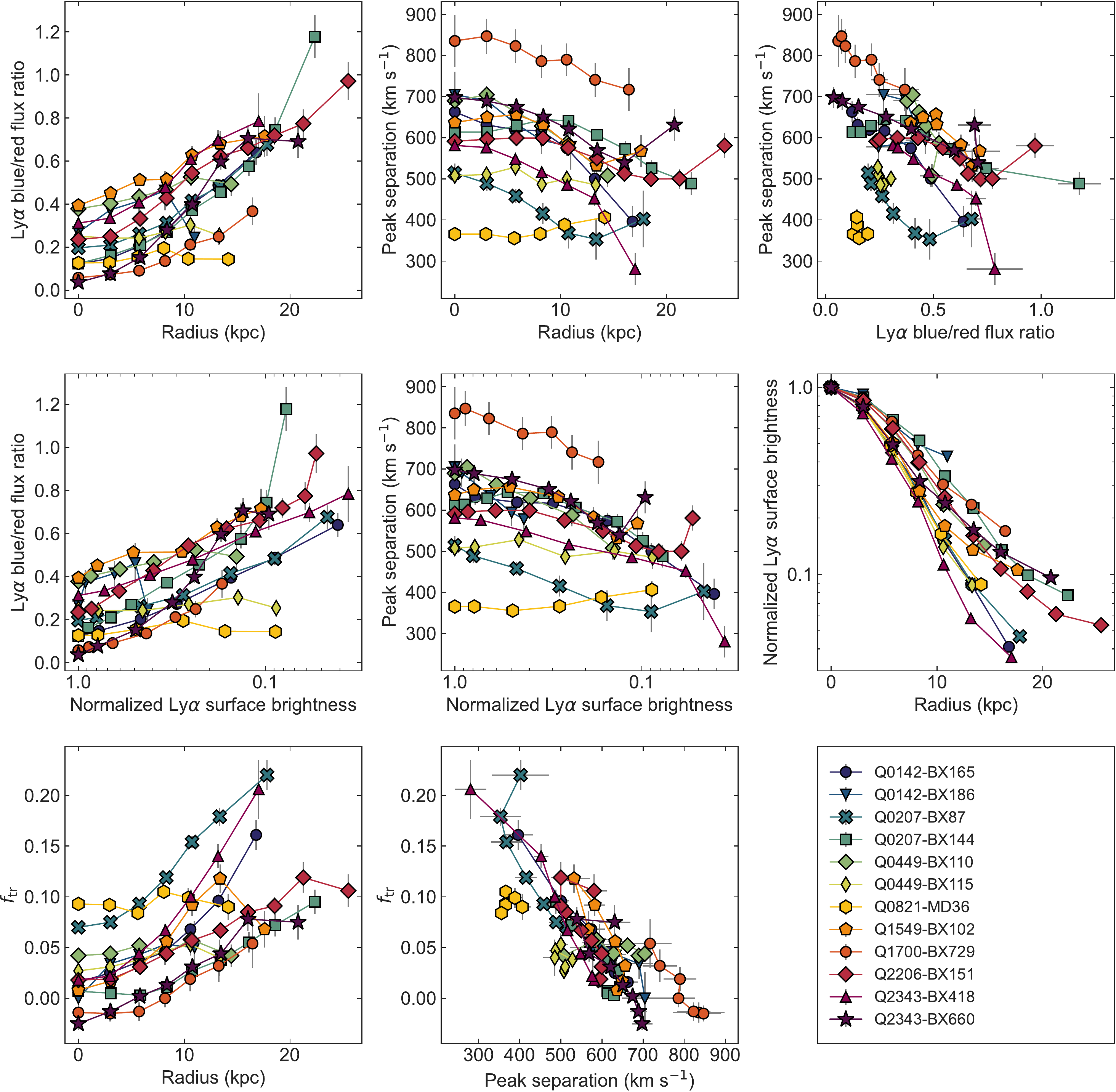}}
\caption{Results of the line profile measurements of the annular spectra described in Section \ref{sec:annularspectra}. \textit{Top row:} \lya\ peak ratio vs.\ radius; \lya\ peak separation vs.\ radius; and peak separation vs.\ ratio. \textit{Middle row:} \lya\ peak ratio vs.\ normalized \lya\ surface brightness; \lya\ peak separation vs.\ surface brightness; and surface brightness vs.\ radius. \textit{Bottom row:} The fraction of total flux within $\pm100$ \kms\ of the trough between the peaks $f_{\rm tr}$ vs.\ radius; $f_{\rm tr}$ vs.\ peak separation.}
\label{fig:annularmeasurements}
\end{figure*}

The results are shown for all objects in Figure \ref{fig:annularmeasurements}, and generally confirm expectations from visual inspection of the spectra. Central blue-to-red flux ratios are $\sim0.2\pm0.2$, and the average ratio increases consistently with radius for most of the galaxies in the sample; all objects that can be measured at a radius beyond $\sim16$ kpc have flux ratios $>0.6$ at that radius. The trough flux fraction $f_{\rm tr}$ ranges from $<0$ to $\sim0.1$ at the center of the halos, and rises consistently with radius for most objects. The largest measured values are $f_{\rm tr}\sim0.2$, found in the outer halos of Q0207-BX87 and Q2343-BX418.  

Trends with the average peak separation are somewhat more complicated. Most (10/12) of the halos have a central peak separation of $\sim500$--700 \kms, with the exceptions of Q0821-MD36 (365 \kms) and Q1700-BX729 (835 \kms). In most cases the average peak separation decreases with radius, with a typical change of $\sim-100$ \kms\ such that peak separations in the outer halo are $\sim400$--600 \kms; however, a few objects (e.g.\ Q0142-BX165 and Q2343-BX418) show steeper gradients. Two galaxies in the sample (Q2206-BX151 and Q2343-BX660) also show an \textit{increase} in the peak separation at the largest radius; in both cases these increases are due to small regions with large separations at large radius, as can be seen in Figure \ref{fig:peakmaps2}. Unsuprisingly, however, the peak separation is closely related to the trough depth $f_{\rm tr}$, decreasing as $f_{\rm tr}$ increases.

\begin{figure*}[t]
\centerline{\epsfig{angle=00,width=\hsize,file=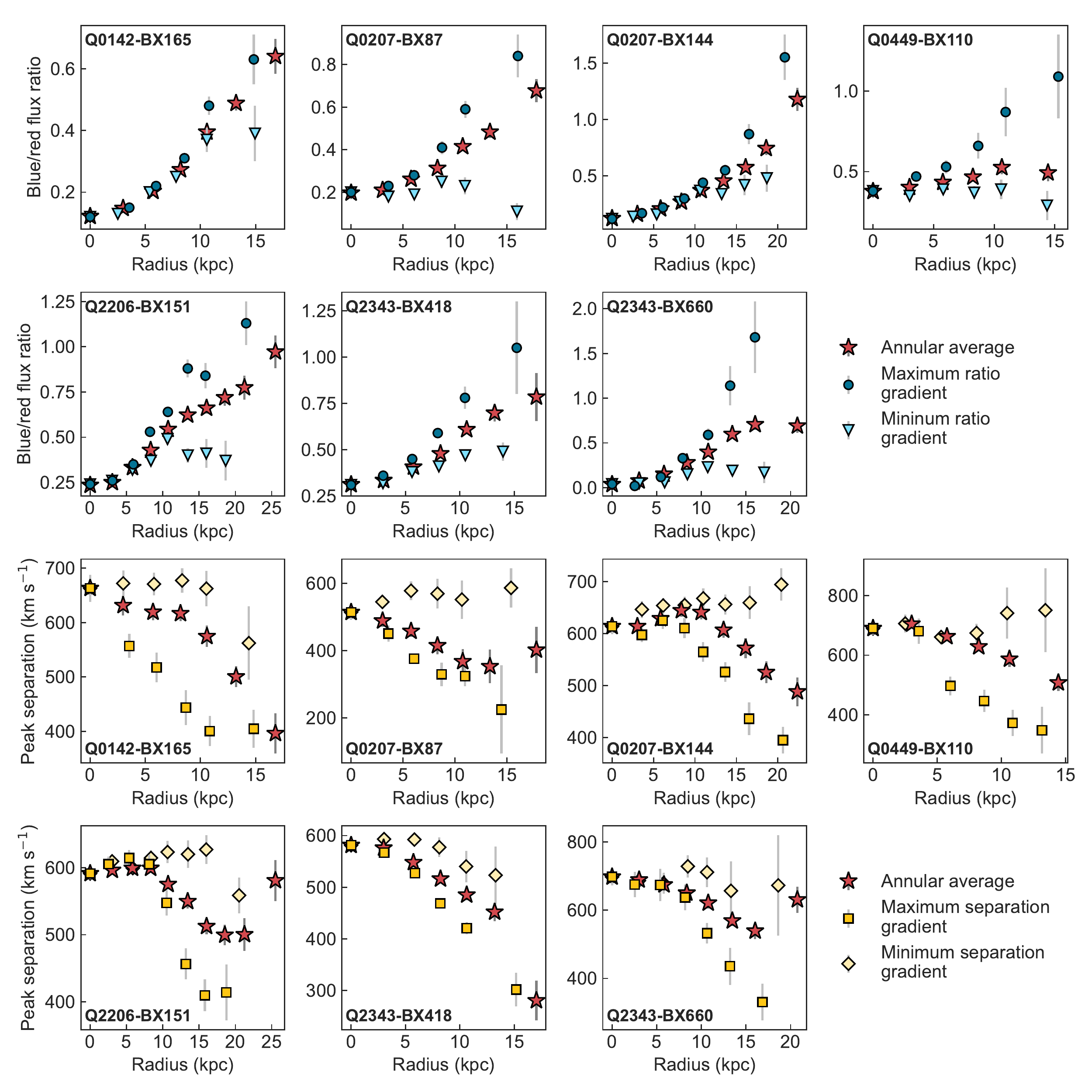}}
\caption{Top two rows:\ Blue-to-red \lya\ flux ratios from radially binned spectra of the 60$^{\circ}$ angular regions that maximize (dark blue circles) and minimize (light blue triangles) the gradient in peak ratio with radius. Second two rows:\ Same as first two rows, for measurements of the \lya\ peak separation, with maximum gradients indicated by gold squares and minimum gradients by light yellow diamonds. The annular averages from Figure \ref{fig:annularmeasurements} are also shown as red stars in all panels. See Section \ref{sec:peakgradients} for details.}
\label{fig:gradients}
\end{figure*}

\subsection{Gradients in Ly$\alpha$ peak ratio and separation}
\label{sec:peakgradients}
While the binned, annular profiles described above are useful to characterize general trends in the extended \lya\ emission, they also wash out the spectral variations seen in different parts of individual halos. As is readily apparent from the maps of peak ratio and separation in Figures \ref{fig:peakmaps1} and \ref{fig:peakmaps2}, the \lya\ profiles across the halos are not radially symmetric, and there are significant differences in both peak ratio and separation at different position angles in a given halo. We therefore characterize the variations in the \lya\ profile within individual halos by binning smaller regions, using seven of the eight brightest sources in the sample (we do not include Q0821-MD36, for which the blue peak is too weak to obtain useful measurements from binning smaller regions).

Our goal is to construct a series of binned spectra that maximize or minimize the gradients in peak ratio or separation from the center to the outskirts of the halo. Again beginning with all spaxels with S/N~$>2$ in the \lya\ images, we then take a subset of each halo corresponding to a 60$^{\circ}$ angular region (chosen to encompass a large enough region to increase the S/N by binning while still isolating different parts of the halos). As with the annular spectra, we radially bin the datacube in this region in single spaxel annular increments and measure the peak ratio and separation of each of the resulting \lya\ profiles. We then rotate the 60$^{\circ}$ region by 10$^{\circ}$ and repeat the process until the entire halo has been covered. 

We next measure the peak ratio and separation for each of the resulting 36 spectra, and locate the regions of maximum and minimum gradients in peak ratio and separation by identifying the two regions for which the difference in peak ratio with radius is maximized, and the two regions for which the difference in separation is maximized. For the peak ratio, the maximum gradient corresponds to the largest \textit{increase} from the center to the outskirts, while for the separation it is the largest \textit{decrease}. In other words, the steepest peak ratio gradient is found in the direction of the highest blue-to-red flux ratio, and the steepest peak separation gradient is found in the direction of the narrowest peak separation.

The results of this process are shown in Figure \ref{fig:gradients}, in which we plot the maximum and minimum ratio gradients in the top two rows and the maximum and minimum separation gradients in the bottom two rows, along with the annular averages from Figure \ref{fig:annularmeasurements}. Although the sample for which these measurements are feasible is small, this exercise shows that all of the halos have a region for which the peak ratio increases with radius, and a region for which the separation decreases with radius. Notably, however, in all cases the angular regions corresponding to these two maximum gradients do not overlap; this result is consistent with the finding in Section \ref{sec:spaxelspectra} above that the correlation between peak ratio and separation is largely due to the underlying relationship of both with radius.

Turning to the minimum gradients, most of the halos also have at least one sightline for which the increase in peak ratio with radius is small or nonexistent, and at least one sightline for which the peak separation is relatively flat with radius (or even rising, in the case of Q0207-BX144). Unlike the maximum gradients, there is some overlap between the regions of minimum gradient; for four of the seven objects, the minumum gradient regions overlap by 10--30$^{\circ}$. The minimum gradients show that most halos have regions for which the \lya\ line profile does not follow the average trends. We discuss the implications of this observation further in Section \ref{sec:discuss}, informed by the results of spatially resolved modeling of the \lya\ emission. 

\section{Modeling Ly$\alpha$ emission and low ionization interstellar absorption lines}
\label{sec:modeling}
In the previous sections we have shown that the spectral morphology of \lya\ emission changes significantly across the extended halos. On average, the blue-to-red peak flux ratio increases, the peak separation decreases, and the fraction of the total flux emerging between the two peaks increases with increasing radius; there are, however, variations in these patterns with azimuthal angle within a given halo. In this section we further examine both the spatially resolved \lya\ profiles and the ``down-the-barrel'' rest-frame UV low ionization interstellar metal absorption lines using physical models. This analysis will help us construct a consistent picture of the ISM and CGM of the galaxies in our sample. 

\subsection{Ly$\alpha$ radiative transfer modeling}
\label{sec:lya_modeling}

To extract physical properties of the gas in the halos from the observed \lya\ profiles, we perform Monte Carlo radiative transfer (MCRT) modeling of the \lya\ line. In contrast to the majority of previous studies in which spatially integrated \lya\ spectra are modeled, in this work we attempt to fully leverage the power of KCWI and reproduce the spatially varying trends of the observed \lya\ profiles.

Following a similar methodology to \citet{Li22a}, we model the spatially resolved \lya\ profiles using the multiphase, clumpy model. Each model is a 3D spherically symmetric region that emulates a galactic halo with a \lya\ emitting source located at its center and two phases of gas:\ cool ($\sim 10^4$\,K) \HI\ clumps and a hot ($\sim 10^6$\,K), highly ionized inter-clump medium (ICM). As we will show below, such a hot, diffuse, low-density \HI\ component is necessary to reproduce the observed \lya\ profiles, primarily by producing additional absorption near the \lya\ line center.\footnote{In the multiphase, clumpy model, the flux at line center of the emergent \lya\ spectra is predominantly controlled by the residual ${\rm H\,{\textsc {i}}}$ number density in the static ICM component. Without the ICM, a significant number of photons will escape at the line center.} In reality, such a component may correspond to the low column density absorbers (${\rm log}\,N_{\rm HI} \lesssim 10^{17} \rm cm^{-2}$) that provide additional \lya\ scatterings in a galactic outflow (see e.g.\ Section 7.3 of \citealt{Dijkstra12}). After interacting with these two phases of gas, \lya\ photons that escape from different impact parameters can be separated into different spatial bins and the emergent spectra can then be compared to the corresponding observed spatially resolved \lya\ profiles. 

In practice, we construct a grid of multiphase, clumpy models for fitting the \lya\ spectra by varying the five most important physical parameters:\ $F_{\rm V}$, the volume filling factor of the clumps; ${\rm log}\,N_{\rm HI,\,{\rm cl}}$, the \HI\ column density of the clumps; $\sigma_{\rm cl}$, the random velocity dispersion of the clumps; $v_{\rm cl}$, the radial outflow velocity of the clumps; and ${\rm log}\,n_{\rm HI,\,{\rm ICM}},$ the residual \HI\ number density in the ICM.\footnote{We have also experimented with a varying $v_{\rm ICM}$ (the radial outflow velocity of the ICM), but found that in almost all cases $v_{\rm ICM} \simeq 0$ is preferred by the fitting. This is due to the prominent trough in most of the \lya\ profiles that requires significant absorption at the line center. Therefore, we have fixed $v_{\rm ICM}$ to zero to reduce the dimensionality of our model grid. \edit2{Similarly, varying the ICM temperature may have a minor effect on the emergent \lya\ spectra, but we chose to fix it to 10$^6$K to keep the computational cost affordable.}} An additional parameter, $\Delta v$, is used in post-processing to determine the deviation between the best-fit systemic redshift of the \lya\ emitting source and the observed systemic redshift inferred from non-resonant nebular emission lines. The parameter values of the model grid are summarized in Table \ref{tab:params}. Note that the range of $F_{\rm V}$ corresponds to a cloud covering factor $f_{\rm c}^{\lya}$ (the average number of clumps per line of sight) of $\sim 1-10$, which is similar to or moderately larger than the critical threshold $f_{\rm c,\,crit}^{\lya}$. Here $f_{\rm c,\,crit}^{\lya}$ denotes the critical average number of clumps per line of sight, above which the clumpy medium starts to transition to a homogeneous medium. In other words, we are exploring a unique physical regime ($f_{\rm c}^{\lya} \simeq f_{\rm c,\,crit}^{\lya}$) where the \lya\ RT in a multiphase, clumpy medium does not fully converge to the homogeneous shell model \citep{Gronke2016b,Gronke17,Li22b}.

\begin{table}
    \scriptsize \caption{Parameter values of the multiphase, clumpy model grid.}
    \label{tab:params}
    \setlength{\tabcolsep}{2pt}
    \begin{tabular}{ccc}
    \hline\hline
    Parameter & Definition & Values\\ 
     (1)  & (2) & (3)\\
    \hline
    $F_{\rm V}$ & Clump volume filling factor & (0.01, 0.04, 0.08, 0.16)\\
     ${\rm log}\,N_{\rm HI,\,cl}$ & Clump {\rm H\,{\textsc {i}}} column density & (17, 17.5, 18, 18.5, 19) log cm$^{-2}$\\
     $\sigma_{\rm cl}$ & Clump velocity dispersion & (0, 25, 50, ..., 150) km\,s$^{-1}$ \\
     $v_{\rm cl,\,\infty}$ & Clump asymptotic outflow velocity & (500, 600, 700, 800, 900) km\,s$^{-1}$ \\
     ${\rm log}\,n_{\rm HI,\,{\rm ICM}}$ & ICM \HI\ number density & (-8, -7.5, -7, -6.5) log cm$^{-3}$\\
     $\Delta v$ & Velocity shift relative to systemic $z$ & [-120,\,120] km\,s$^{-1}$ \\
    \hline\hline
    \end{tabular}
    \tablenotetext{}{\textbf{Notes.} The parameter values of the model grid that we used for fitting the \lya\ profiles. The columns are: (1) parameter name; (2) parameter definition; (3) parameter values on the grid.}
\end{table}

Previous work \citep{Li22a} assumed constant radial outflow velocities, but here we adopt a more physically realistic radially varying clump outflow velocity profile. Our choice is inspired by \citet{Dijkstra12}, who find that a radially varying velocity profile is able to better reproduce the surface brightness (SB) profiles of \lya\ halos. Specifically, the momentum equation of an \HI\ clump can be written as \citep{Murray05, Martin05}:
\begin{equation}
\frac{\mathrm{d}v(r)}{\mathrm{d}t}=-\frac{GM(r)}{r^2}+Ar^{-\alpha}
\label{eq:momentum}
\end{equation}
where $r$ is the clump's radial position, $v(r)$ is the clump radial outflow velocity at $r$, $M(r)$ is the total gravitational mass within $r$, and $A$ is a constant that characterizes the amplitude of the power-law acceleration $r^{-\alpha}$. The acceleration of the clump is determined by two competing terms on the right hand side, the first of which is due to gravitational deceleration and the second of which is an empirical power-law acceleration term \citep{Steidel10}. Major acceleration mechanisms for the cool clumps may include radiation pressure, ram pressure from a hot wind, and shock-accelerated cosmic rays, which all may correspond to an $r^{-2}$ force (see \citealt{Chisholm16}, and note that the radiation pressure should be in the optically thin regime). However, in reality, the clumps may suffer from extra deceleration (and acceleration, see \citealp{Gronke20}) due to their interaction with other phases of gas, which yields an effective $\alpha$ less than 2. 

Assuming the gravitational potential is an isothermal sphere, we have $M(r)=2 \sigma_{\rm cl}^2\,r/G$, where $\sigma_{\rm cl}$ is the velocity dispersion of the clumps. Equation \ref{eq:momentum} can then be analytically solved as:
\begin{align}
v(r) = \sqrt{4\,\sigma_{\rm cl}^2\,{\rm ln}\Big{(}\frac{r_{\rm min}}{r}\Big{)} + v_{\rm cl,\,\infty}^2 \Big{(}1-\Big{(}\frac{r}{r_{\rm min}}\Big{)}^{1-\alpha}\Big{)}}
\label{eq:solution}
\end{align}
where $r_{\rm min}$ is the inner cutoff (or ``launching'') radius that satisfies $v(r_{\rm min})=0$, and $v_{\rm cl,\,\infty}$ = $\sqrt{2Ar_{\rm min}^{1-\alpha}/(\alpha-1)}$ is the asymptotic maximum outflow velocity if there were no gravitational deceleration. Note that in general the actual $v(r)$ does not reach $v_{\rm cl,\,\infty}$ due to the gravitational deceleration term; even the maximum radial $v(r)$ is usually several hundred km\,s$^{-1}$ smaller than $v_{\rm cl,\,\infty}$. Following \citet{Dijkstra12}, we have fixed $\alpha$ = 1.4 and left $\sigma_{\rm cl}$ and $v_{\rm cl,\,\infty}$ as the free parameters in this model. We set $r_{\rm min}$ \edit1{to be 1\% of the simulated halo radius $r_{\rm h}$}, so that $\frac{r}{r_{\rm min}} \in [1, 100]$. The model is  intrinsically rescalable (i.e.\ increasing the size of every component in the model by any factor with all column densities unchanged would yield an identical model) and constrains only the ratio $\frac{r}{r_{\rm min}}$, so the following analysis applies to \lya\ halos of varying physical sizes. 

For each multiphase, clumpy model on the grid, MCRT has been performed on 10$^4$ \lya\ photon packages emitted at the center of the simulation sphere in the form of a normalized Gaussian intrinsic spectrum $\mathcal{N}(v, \mu = 0, \sigma = \sigma_{\rm i,\,cl})$, where $\sigma_{\rm i,\,cl}$ = 12.85\,km\,s$^{-1}$ is the canonical thermal velocity dispersion of the \HI\ gas in the clumps at $T$ = 10$^4$\,K.\footnote{In the multiphase, clumpy model, the width of the intrinsic spectrum is always assumed to be small and the clump velocity dispersion is responsible for broadening the spectrum. Such a choice has the advantage of avoiding obtaining unphysically large intrinsic line widths from fitting the spectrum (e.g.\ using the shell models, see \citealp{Li22b}).} The \HI\ clumps with a constant column density $N_{\rm HI,\,cl}$ are placed uniformly radially, so that their number density $n_{\rm cl} \propto r^{-2}$ (i.e.\ mass conserving if the radial outflow velocity is constant).

Each model on the grid is further used to generate three spatially binned \lya\ profiles by separating all the photons into three spatial bins according to their last-scattering impact parameters: $b/b_{\rm max} \in$ (0, 0.25], (0.25, 0.50] and (0.50, 0.75], where $b_{\rm max}$ is the largest impact parameter of the scattered \lya\ photons (see Figure 5 of \citet{Li22a} for an illustrative schematic), and the impact parameter $b$ is measured orthogonal to the direction of the photon's escape trajectory. The difference between $b_{\rm max}$ and the halo radius $r_{\rm h}$ is negligible, and we simply \edit1{fix $b_{\rm max} = r_{\rm h}$.}

We only include the photons within 75\% of $b_{\rm max}$ (or equivalently, within the inner $\sim$ 56\% of the total area) in our fitting in order to ensure a direct comparison between the model and the data, because the S/N~$>2$ regions of the halos used for the spectra (see Section \ref{sec:annularspectra}) contain on average 58\% of the total halo area. The spectra to be modeled are constructed in the same way as the annular spectra described in Section \ref{sec:annularspectra}, except that the spaxels are divided into three radial bins with $0 < r \leq 0.33r_{\rm max}$, $0.33r_{\rm max} < r \leq 0.67r_{\rm max}$, and $0.67r_{\rm max} < r \leq r_{\rm max}$, where $r_{\rm max}$ is the radius of the most distant spaxel in the modeled area.  When we present our modeling results later in \S\ref{sec:mod_results}, we consider only the photons included in the modeling and renormalize the halo to 0.75\,$b_{\rm max}$, so that $b/b_{\rm max} \in (0, \frac{1}{3}]$, $(\frac{1}{3}, \frac{2}{3}]$ and $(\frac{2}{3}, 1]$.  

Our fitting pipeline employs the \texttt{python} nested sampling package \texttt{dynesty} \citep{Skilling04, Skilling06, Speagle20}. At each visited point of the parameter space, the pipeline executes the following three steps:

(1) calculate three binned \lya\ model spectra via linear flux interpolation on the model grid (to circumvent doing computationally expensive RT ``on the fly''), where the flux density of the model spectrum at each wavelength is calculated by a parameter-weighted multidimensional linear interpolation\footnote{Such an interpolation is carried out based on the distance between the visited point in the parameter space and its adjacent points on the grid (realized by the \texttt{PYTHON} function \texttt{scipy.interpolate.interpn}).} of the flux densities of the adjacent grid model spectra at the corresponding wavelength. The three binned \lya\ model spectra are then convolved with a Gaussian function with $\sigma = 65\,\rm km\,s^{-1}$ (the KCWI line spread function [LSF]) to mimic the finite instrumental resolution;

(2) compare each binned model spectrum to an observed \lya\ spectrum at the corresponding impact parameter range and calculate the likelihood;

(3) sum the likelihoods of these three binned models as the likelihood of the current set of parameters.

Each fitting run yields a posterior probability distribution (PDF) of the model parameters. The parameter uncertainties can be further determined as certain quantiles (e.g. 16\%--84\%, or 1$\sigma$ confidence intervals) of the samples in the marginalized PDF.

\subsection{Metal absorption line modeling}\label{sec:UV_modeling}
In addition to the \lya\ profiles observed at both $b = 0$ and $b > 0$, the rest-UV, low-ionization metal absorption lines observed ``down-the-barrel'' (i.e., at $b = 0$) also encode rich information on the physical properties of the cool gas. These metal absorption line profiles are typically ``sawtooth'' shaped (e.g.\ \citealt{Weiner09}), where the part blueward of the absorption trough (the location of the minimum flux density) gradually decreases with velocity while the part redward of the absorption trough increases with velocity relatively rapidly. The blueshifted absorption at negative velocities is produced by gas clumps with radially varying outflow velocities along the line-of-sight, whereas the red part is mainly produced by a group of non-outflowing, randomly moving clumps. In this work, we focus on modeling the portion blueward of the absorption trough for the average line profile of the {\ion{Si}{2}} $\lambda$1260 and {\ion{C}{2}} $\lambda$1334 transitions,\footnote{We did not fit {\ion{Si}{2}} $\lambda$1260 and {\ion{C}{2}} $\lambda$1334 separately as many of the individual lines have fairly low S/N ratios.} as we are most interested in constraining the clump outflow kinematics. Our model is similar to the kinematic model used by \citet{Steidel10}, but with a different clump radial velocity profile.

In our model, we first assume that the clump radial outflow velocity is described by the same model we use for the \lya\ emission, i.e.\ Equations \ref{eq:momentum} and \ref{eq:solution} with two free parameters: the clump velocity dispersion $\sigma_{\rm cl}$ and the asymptotic maximum clump outflow velocity $v_{\rm cl,\,\infty}$. Assuming that the absorption lines are saturated\footnote{The assumption of saturation comes from the fact that in our sample, {\ion{Si}{2}} $\lambda$1260 and {\ion{Si}{2}} $\lambda$1526 have similar equivalent widths (see e.g.\ footnote 27 of \citealt{Steidel2018}).} (i.e.\ the column densities of the absorbing gas are so high that the depth of absorption simply reflects the gas covering fraction), the down-the-barrel absorption line profile $I(v)$ (the normalized, residual flux density as a function of velocity) is simply given by 
\begin{equation}
I(v) = 1-f_c(v)
\label{eq:residual_flux}
\end{equation}
where $f_c(v)$ is the (clumpy) gas geometric covering fraction as a function of velocity, which is the fraction of the total lines of sight of the rest-UV emission that are intercepted by the absorbing gas. We further assume that the gas covering fraction decreases as a function of radius, in the form of a power law:
\begin{equation}
f_c (r) = f_{c, \rm max} \left(\frac{r}{r_{\rm min}}\right)^{-\gamma} 
\label{eq:fc}
\end{equation}
where $r_{\rm min}$ is the launching radius and $f_{c, \rm max}$ is the maximum gas covering fraction that corresponds to the deepest part of the absorption trough. $f_c (r)$ can then be translated into $f_c (v)$ using the $v(r)$ dictated by Equation \ref{eq:solution}. Note that the gas geometric covering fraction in the \lya\ RT models, which is a function of the number density and the physical size of the clumps (both of which may vary as a function of velocity or radius; see Equation 2 in \citealt{Dijkstra12}), may not be fully consistent with the power law $f_c (r)$ assumed here. One may match them by using clumps with radially-varying sizes in the RT model; we plan to explore this option in future work.

To be consistent with the \lya\ modeling in \S\ref{sec:lya_modeling}, we fix $\alpha$ = 1.4 in the clump radial velocity profile and set $r_{\rm min}$ = 0.1 kpc with $\frac{r}{r_{\rm min}} \in [1, 100]$. Note again that only the ratio $\frac{r}{r_{\rm min}}$ (rather than $r$ or $r_{\rm min}$ individually) is constrained by the absorption line modeling. We then fit the observed absorption line profiles with \texttt{dynesty} to determine the PDF of the four parameters in this model: $\sigma_{\rm cl}$, $v_{\rm cl,\,\infty}$, $f_{c,\,\rm max}$ and $\gamma$. We use flat priors for the fitted parameters (which can vary continuously): $\sigma_{\rm cl} \in [0, 120]$ \kms, $v_{\rm cl,\,\infty} \in [100, 1500]$ \kms, $f_{c,\,\rm max} \in [0, 1]$, and $\gamma \in [0.1, 2.0]$. We restrict $\gamma$ to be no larger than 2, as otherwise it suggests that the clumps are destroyed rapidly as they move outwards, contradictory to the observation of metal absorption at large impact parameters ($b \sim 100$ kpc, see Figure 21 of \citealt{Steidel10} and \citealt{Rudie19}).

\begin{figure*}[htbp]
\centerline{\epsfig{angle=00,width=0.9\hsize,file=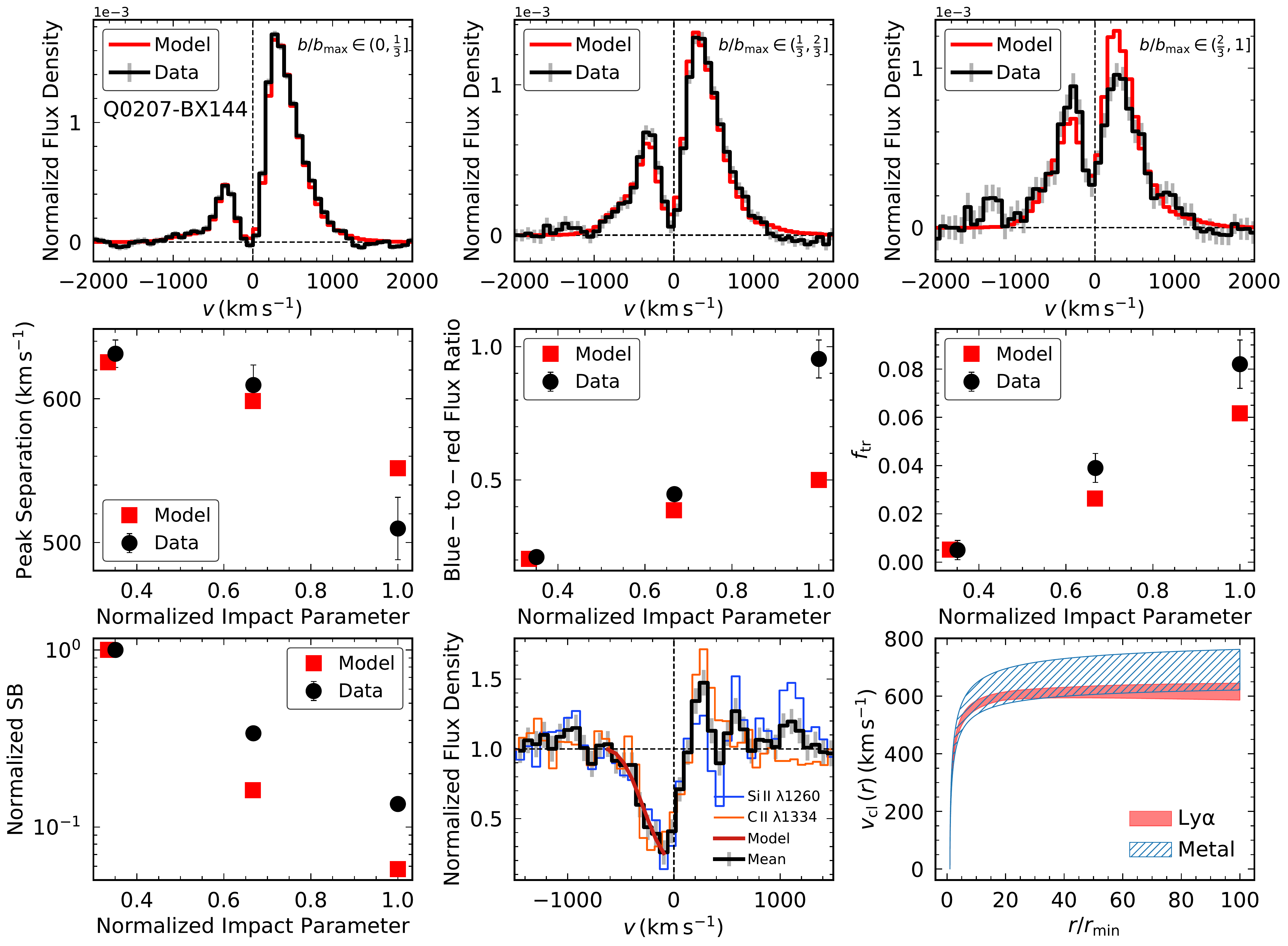}}
\caption{Modeling results of the annular-averaged, spatially resolved \lya\ spectra and of the average line profile of {\ion{Si}{2}} $\lambda$1260 and {\ion{C}{2}} $\lambda$1334 observed down the barrel for Q0207-BX144 (see Appendix \ref{sec:appendix} for the rest of the sample). The top row shows the best-fit models (red) to the spatially resolved \lya\ spectra (black, with 1-$\sigma$ uncertainties shown in grey) from the inner to the outer halo. In each subpanel of the top row, the vertical and horizontal black dashed lines indicate the systemic redshift (determined from nebular emission lines) and zero flux density, respectively. The middle row and the first panel of the bottom row show a comparison between the radial trends of peak separation, blue-to-red flux ratio, trough flux fraction, and normalized SB versus the normalized impact parameter predicted by the best-fit models (red squares) and measured from observation (black points, with 1-$\sigma$ uncertainties). Note that the impact parameters may be slightly different for the model and the data:\ the models are binned consistently as $b/b_{\rm max} \in (0, \frac{1}{3}]$, $(\frac{1}{3}, \frac{2}{3}]$ and $(\frac{2}{3}, 1]$, and while the data are binned in the same way, the halos are asymmetric with the result that the median distance to the spaxels included in each bin varies from object to object. The rest of the bottom row shows the best-fit models (red) to the average line profile (black, with 1-$\sigma$ uncertainties shown in grey) of {\ion{Si}{2}} $\lambda$1260 (blue) and {\ion{C}{2}} $\lambda$1334 (orange) profiles, as well as a comparison of clump radial outflow velocity profiles inferred from \lya\ RT modeling (red) and metal absorption line fitting (blue hatched patch). The shaded regions represent the velocity ranges spanned by 50 points in the parameter space after convergence has been achieved for the fitting.}
\label{fig:fits_example}
\end{figure*}

\begin{table*}
\begin{center}
\setlength{\tabcolsep}{4pt}
\renewcommand\arraystretch{1.5}
  \footnotesize \caption{Best-fit parameters from modeling \lya\ emission and the rest-UV low ionization metal absorption lines}
  \label{tab:Fitting_results}
  \begin{tabular}{c|ccccccc|ccccc}
    \hline \hline 
    \multicolumn{1}{c|}{}&\multicolumn{7}{c|}{Best-fit Parameters (\lya)}&\multicolumn{5}{c}{Best-fit Parameters (Absorption)}\\
    \hline 
    ID&$F_{\rm V}$&${\rm log}\,N_{\rm HI,\,{\rm cl}}$&$\sigma_{\rm cl}$&$v_{\rm cl,\,\infty}$&$v_{\rm cl,\,max}$&${\rm log}\,n_{\rm HI,\,{\rm ICM}}$&$\Delta v$&$\sigma_{\rm cl,\,abs}$&$v_{\rm cl,\,\infty,\,abs}$&$v_{\rm cl,\,max,\,abs}$&$f_{c,\,\rm max}$&$\gamma$\\%
&&(cm$^{-2})$&(km\,s$^{-1})$&(km\,s$^{-1})$&(km\,s$^{-1})$&(cm$^{-3})$&(km\,s$^{-1})$&(km\,s$^{-1})$&(km\,s$^{-1})$&(km\,s$^{-1})$&&\\%
(1)&(2)&(3)&(4)&(5)&(6)&(7)&(8)&(9)&(10)&(11)&(12)&(13)\\
    \hline
Q0142{-}BX165&0.06$^{+0.01}_{-0.01}$&18.3$^{+0.1}_{-0.1}$&142$^{+4}_{-6}$&748$^{+24}_{-25}$&392$^{+26}_{-27}$&-7.13$^{+0.05}_{-0.07}$&9$^{+4}_{-4}$&58$^{+36}_{-33}$&835$^{+69}_{-69}$&707$^{+63}_{-46}$&0.8$^{+0.1}_{-0.1}$&1.7$^{+0.2}_{-0.3}$\\%
Q0142{-}BX186&0.12$^{+0.02}_{-0.02}$&18.8$^{+0.1}_{-0.1}$&117$^{+14}_{-4}$&533$^{+48}_{-23}$&235$^{+17}_{-4}$&-7.28$^{+0.16}_{-0.13}$&55$^{+10}_{-10}$&72$^{+32}_{-41}$&429$^{+400}_{-205}$&423$^{+432}_{-157}$&0.6$^{+0.3}_{-0.3}$&1.2$^{+0.6}_{-0.7}$\\%
Q0207{-}BX87&0.08$^{+0.01}_{-0.00}$&18.0$^{+0.0}_{-0.0}$&98$^{+0}_{-1}$&617$^{+11}_{-10}$&389$^{+13}_{-12}$&-7.14$^{+0.02}_{-0.02}$&-4$^{+5}_{-4}$&61$^{+40}_{-34}$&1029$^{+311}_{-264}$&891$^{+292}_{-262}$&0.2$^{+0.1}_{-0.1}$&1.1$^{+0.6}_{-0.6}$\\%
Q0207{-}BX144&0.08$^{+0.01}_{-0.01}$&18.4$^{+0.1}_{-0.1}$&106$^{+17}_{-21}$&842$^{+40}_{-47}$&622$^{+17}_{-17}$&-6.69$^{+0.04}_{-0.04}$&-47$^{+3}_{-3}$&53$^{+37}_{-29}$&807$^{+79}_{-76}$&684$^{+72}_{-50}$&0.8$^{+0.1}_{-0.1}$&1.7$^{+0.2}_{-0.3}$\\%
Q0449{-}BX110&0.09$^{+0.02}_{-0.01}$&18.3$^{+0.1}_{-0.1}$&9$^{+11}_{-6}$&892$^{+5}_{-9}$&815$^{+5}_{-9}$&-6.89$^{+0.04}_{-0.04}$&33$^{+5}_{-5}$&62$^{+36}_{-37}$&1051$^{+130}_{-102}$&913$^{+121}_{-84}$&0.8$^{+0.1}_{-0.1}$&1.6$^{+0.3}_{-0.4}$\\%
Q0449{-}BX115&0.13$^{+0.02}_{-0.02}$&18.1$^{+0.1}_{-0.1}$&18$^{+20}_{-13}$&550$^{+19}_{-14}$&495$^{+9}_{-9}$&-6.93$^{+0.04}_{-0.04}$&1$^{+3}_{-3}$&58$^{+39}_{-35}$&636$^{+307}_{-155}$&502$^{+319}_{-131}$&0.7$^{+0.2}_{-0.2}$&1.3$^{+0.5}_{-0.6}$\\%
Q0821{-}MD36&0.15$^{+0.00}_{-0.01}$&18.1$^{+0.0}_{-0.0}$&60$^{+4}_{-3}$&506$^{+10}_{-4}$&386$^{+9}_{-10}$&-7.47$^{+0.03}_{-0.02}$&6$^{+4}_{-7}$&61$^{+36}_{-37}$&870$^{+383}_{-269}$&735$^{+373}_{-249}$&0.3$^{+0.3}_{-0.2}$&1.2$^{+0.5}_{-0.7}$\\%
Q1549{-}BX102&0.07$^{+0.01}_{-0.01}$&17.7$^{+0.1}_{-0.1}$&134$^{+6}_{-11}$&882$^{+11}_{-16}$&578$^{+24}_{-19}$&-6.67$^{+0.03}_{-0.04}$&-15$^{+4}_{-4}$&58$^{+40}_{-33}$&633$^{+170}_{-125}$&497$^{+166}_{-100}$&0.7$^{+0.1}_{-0.2}$&1.4$^{+0.4}_{-0.6}$\\%
Q1700{-}BX729&0.16$^{+0.00}_{-0.00}$&18.5$^{+0.0}_{-0.0}$&123$^{+1}_{-3}$&606$^{+13}_{-6}$&287$^{+16}_{-11}$&-6.60$^{+0.07}_{-0.09}$&67$^{+7}_{-5}$&44$^{+35}_{-24}$&563$^{+74}_{-65}$&468$^{+44}_{-60}$&0.9$^{+0.1}_{-0.1}$&1.7$^{+0.2}_{-0.3}$\\%
Q2206{-}BX151&0.10$^{+0.02}_{-0.02}$&17.6$^{+0.1}_{-0.2}$&142$^{+5}_{-8}$&864$^{+17}_{-23}$&530$^{+14}_{-15}$&-6.73$^{+0.03}_{-0.03}$&-73$^{+3}_{-3}$&57$^{+39}_{-33}$&743$^{+258}_{-137}$&612$^{+251}_{-127}$&0.5$^{+0.1}_{-0.1}$&1.4$^{+0.4}_{-0.6}$\\%
Q2343{-}BX418&0.07$^{+0.01}_{-0.01}$&18.0$^{+0.0}_{-0.1}$&6$^{+14}_{-4}$&606$^{+19}_{-14}$&553$^{+11}_{-12}$&-6.99$^{+0.03}_{-0.02}$&34$^{+2}_{-2}$&59$^{+39}_{-35}$&857$^{+165}_{-107}$&725$^{+154}_{-96}$&0.6$^{+0.2}_{-0.2}$&1.4$^{+0.4}_{-0.5}$\\%
Q2343{-}BX660&0.07$^{+0.01}_{-0.01}$&18.4$^{+0.1}_{-0.1}$&134$^{+8}_{-19}$&876$^{+17}_{-47}$&571$^{+21}_{-20}$&-6.62$^{+0.04}_{-0.05}$&-112$^{+5}_{-5}$&55$^{+37}_{-32}$&697$^{+78}_{-78}$&569$^{+69}_{-44}$&0.9$^{+0.1}_{-0.1}$&1.5$^{+0.3}_{-0.3}$\\%

    \hline \hline
  \end{tabular}
  \end{center}
  \tablenotetext{}{\textbf{Notes.} Best-fit parameters (averages and 16\% -- 84\% quantiles, i.e., 1$\sigma$ confidence intervals) from the \lya\ and low ionization metal absorption line (the average of {\ion{Si}{2}} $\lambda$1260 and {\ion{C}{2}} $\lambda$1334) modeling. The columns are: (1) the object ID; (2) the clump volume filling factor; (3) the clump \HI\ column density; (4) the clump velocity dispersion; (5) the clump asymptotic outflow velocity; (6) the actual maximum clump radial outflow velocity; (7) the residual \HI\ number density of the ICM; (8) the velocity shift relative to the systemic redshift of the source. (9) - (13) are determined from the average metal absorption line profile. (9) the clump velocity dispersion; (10) the clump asymptotic outflow velocity; (11) the actual maximum clump radial outflow velocity; (12) the maximum clump covering fraction; (13) the power-law index of the clump covering fraction function.}
\end{table*}

\subsection{Modeling results \& interpretation}\label{sec:mod_results}
Our modeling of both the spatially resolved \lya\ emission and the UV absorption lines has achieved the following principal results:

(1) reproducing the radially varying, spatially resolved \lya\ profiles;

(2) reproducing the radial trends of several important physical quantities of the \lya\ profiles, including the peak separation, peak flux ratio, trough flux fraction, and SB versus the impact parameter;

(3) reconciling the clump outflow velocities inferred from \lya\ emission and metal absorption lines.

We present the modeling results for the spatially resolved \lya\ spectra and the average line profile of {\ion{Si}{2}} $\lambda$1260 and {\ion{C}{2}} $\lambda$1334 for our sample in Figure \ref{fig:fits_example} (using Q0207-BX144 as an example) and Appendix \ref{sec:appendix}. In each panel, the top row shows the best-fit RT models (red) to the spatially resolved \lya\ spectra (black); the middle row and the first panel of the bottom row show a comparison between the radial trends of peak separation, peak flux ratio, trough flux fraction, and SB predicted by the best-fit models and measured from observations; and the rest of the bottom row shows the best-fit models (red) to the average metal absorption line profile (black), as well as a comparison of clump radial outflow velocity profiles inferred from \lya\ emission and the average metal absorption line. The best-fit parameters are summarized in Table \ref{tab:Fitting_results}, \edit1{and we present the posterior distribution of Q0207-BX144 as an example in Appendix \ref{sec:appendix2}}. In Section \ref{sec:radial_trends} below we describe the relationships between impact parameter, the properties of the model \lya\ profiles, and the parameters of the model, and in Section \ref{sec:bestfit_parameters} we further discuss the best-fit parameters and relationships between them. Section \ref{sec:resolved_vs_integrated} provides a comparison of spatially integrated vs.\ spatially resolved \lya\ modeling, and we discuss caveats to the models in Section \ref{sec:caveats}.

\subsubsection{Radial trends}\label{sec:radial_trends}

\begin{figure}[htbp]
\centerline{\epsfig{angle=00,width=\hsize,file=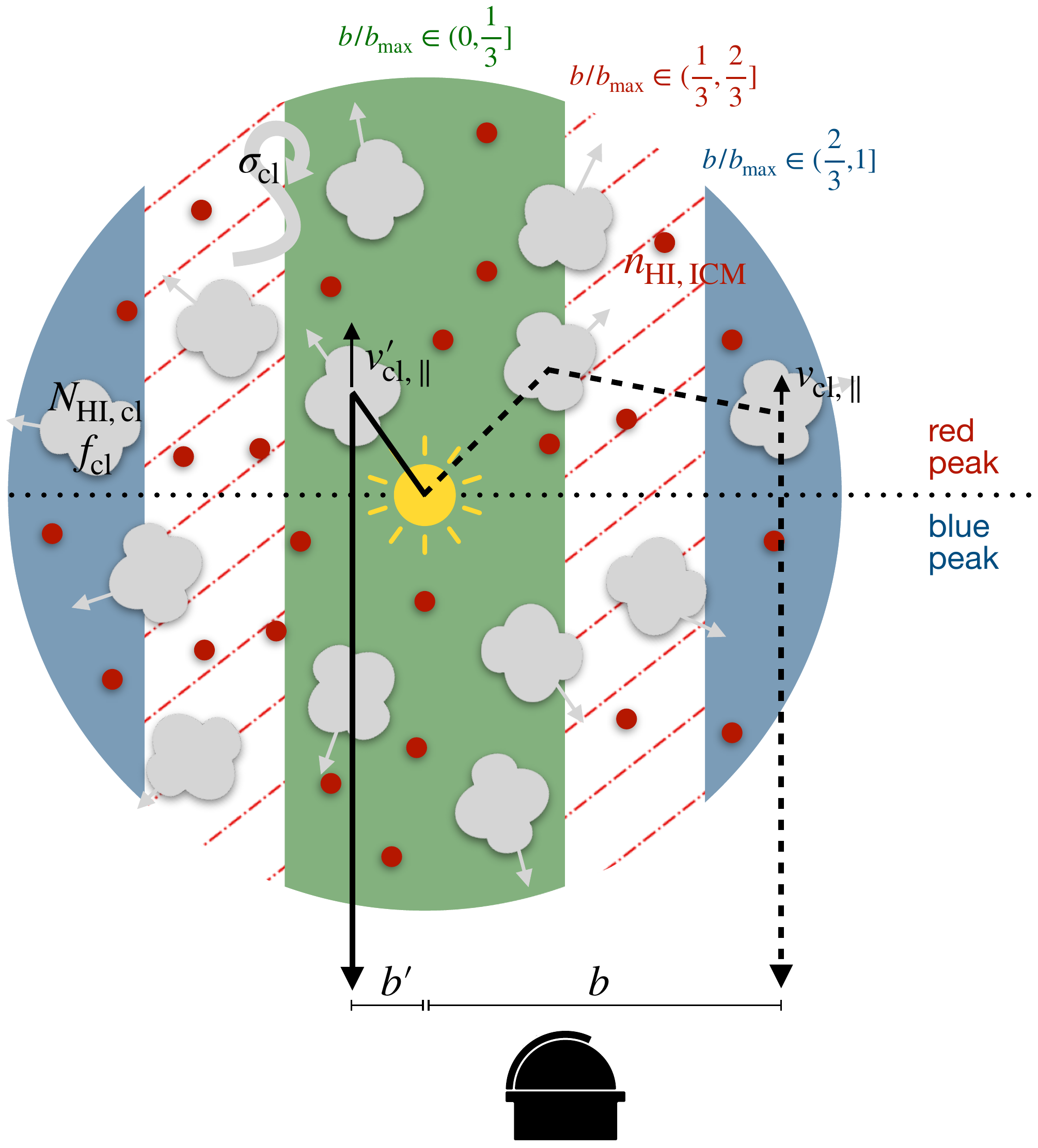}}
\caption{Schematic of the escape of two \lya\ photons at low and high impact parameters in the multiphase, clumpy model. The large circle represents the boundary of the simulated spherical region, divided into shaded green, red hatched, and blue regions indicating the three ranges of impact parameters modeled. The location of the observer is indicated by the telescope dome at the bottom, and the dotted horizontal line indicates that photons in the blue peak arise from the near side of the halo while those in the red peak predominantly come from the far side. The gold sun symbol represents the \lya\ emitting source at the center, the grey clouds represent \HI\ clumps with random motions and radial outflows, and the small red circles represent the diffuse, hot ICM. The impact parameters $b$ and $b'$ are defined as the orthogonal distance from the center to the direction of the photon escape trajectories shown by the black solid and dashed lines. The photon that escapes at a higher $b > b'$ will experience several differences before it escapes: (1) it will scatter with lower \HI\ column densities from the clumps, due to the decrease in the clump covering fraction at large radii; (2) it will experience (on average) a lower projected component of the clump outflow velocity along its traveling direction ($v_{\rm cl,\,\parallel} < v'_{\rm cl,\,\parallel}$, as indicated by the black arrows near the last clump that scatters each photon); (3) it will suffer from less absorption at line center from the ICM, due to its lower traveling distance at the outskirts of the halo. Also note that the photon escaping at $b'$ passes through a clump on the near side of the halo unimpeded, because it is out of resonance with the clump due to its previous scattering.}
\label{fig:schematic}
\end{figure}

The modeling results show that our multiphase, clumpy model is able to reproduce the spatially resolved \lya\ spectra fairly well, especially for the innermost two spatial bins. In a number of cases (e.g.\ Q0142-BX165, Q0207-BX87, Q0207-BX144, Q1549-BX102 and Q2343-BX660) there is a noticeable mismatch between the model and data in the outermost bin, which may be because the gas in the outer halo does not fully follow the outflowing kinematics of the gas in the inner halo (e.g.\ due to external forces). In general, as the impact parameter increases, the best-fit \lya\ RT model predicts a decrease in the peak separation, an increase in the blue-to-red peak flux ratio, and an increase in the trough flux fraction. These predicted radial trends of peak separation and peak flux ratio are broadly consistent with the observational data, although the exact values differ in some cases. The increase in the trough flux fraction is also evident in almost all objects, especially from a comparison between the innermost two spatial bins. 

From a \lya\ RT perspective, the peak separation, which reflects the most likely frequencies at which the \lya\ photons escape, is directly related to the \lya\ optical depth of the system. The optical depth, which is the product of the \lya\ cross-section\footnote{Strictly speaking, the peak separation is also related to the gas outflow velocity, since the \lya\ cross section depends on the photons' apparent frequencies in the gas frame. However, our tests have shown that such an effect is minor compared to the one that the \HI\ column density has on peak separation.} and the \HI\ column density of the absorber, can therefore be expressed as a function of the temperature and column density of the absorber.\footnote{For example, the peak separation of \lya\ photons that escape from an opaque, static \HI\ sphere is $\Delta v_{\rm peak} \simeq 320 \Big{(}\frac{N_{\rm HI}}{10^{20}\,{\rm cm^{-2}}}\Big{)}^{1/3} \Big{(}\frac{T_{\rm HI}}{10^{4}\,{\rm K}}\Big{)}^{1/6}\,\rm km\,s^{-1}$ \citep{Dijkstra14}.} The blue-to-red peak flux ratio, however, is negatively correlated with the \HI\ gas outflow velocity as seen by the \lya\ photons, as the blue photons are less likely to escape since they appear closer to resonance in the reference frame of the outflowing gas. Finally, as the absorption at the line center is mainly produced by the ICM, the trough flux fraction is mostly set by the ICM column density.

One can then imagine that the \lya\ photons that escape at large impact parameters (i.e.\ the directions of their escape trajectories are almost orthogonal to the radial direction) will experience the following differences relative to photons from smaller impact parameters before they escape: (1) experience lower \HI\ column densities from the clumps, as the area covering fraction of the clumps decreases at large radii due to the increase of the physical volume of the halo; (2) encounter (on average) a lower projected component of the clump outflow velocity along their traveling directions in the portion of the outer halo that they pass through before they escape;\footnote{This is a purely geometrical effect; assuming the clump outflow is nearly isotropic, at high impact parameters ($b \simeq b_{\rm max}$) the maximum projected component of the clump outflow velocity along the traveling direction of a photon goes as $v_{\rm cl,\,\parallel,\,max}(r) = \sqrt{1 - (b / b_{\rm max})^2}\,v_{\rm cl}(r) \simeq 0$ \citep{Li22a}.} (3) suffer from lower absorption (or equivalently, ``see'' a lower optical depth) at line center from the ICM in the outer halo, as on average the distance a photon travels within the halo before it escapes at large impact parameters is smaller than that at small impact parameters.\footnote{Considering the spherical geometry of the halo, the largest distance that a photon can travel through without changing direction at impact parameter $b$ is $\sim 2\sqrt{R^2 - b^2}$.} These three effects are presumably responsible for the observed radial variation of the spatially binned \lya\ profiles, and we illustrate them in Figure \ref{fig:schematic}.

\begin{figure*}
\includegraphics[width=0.245\hsize]{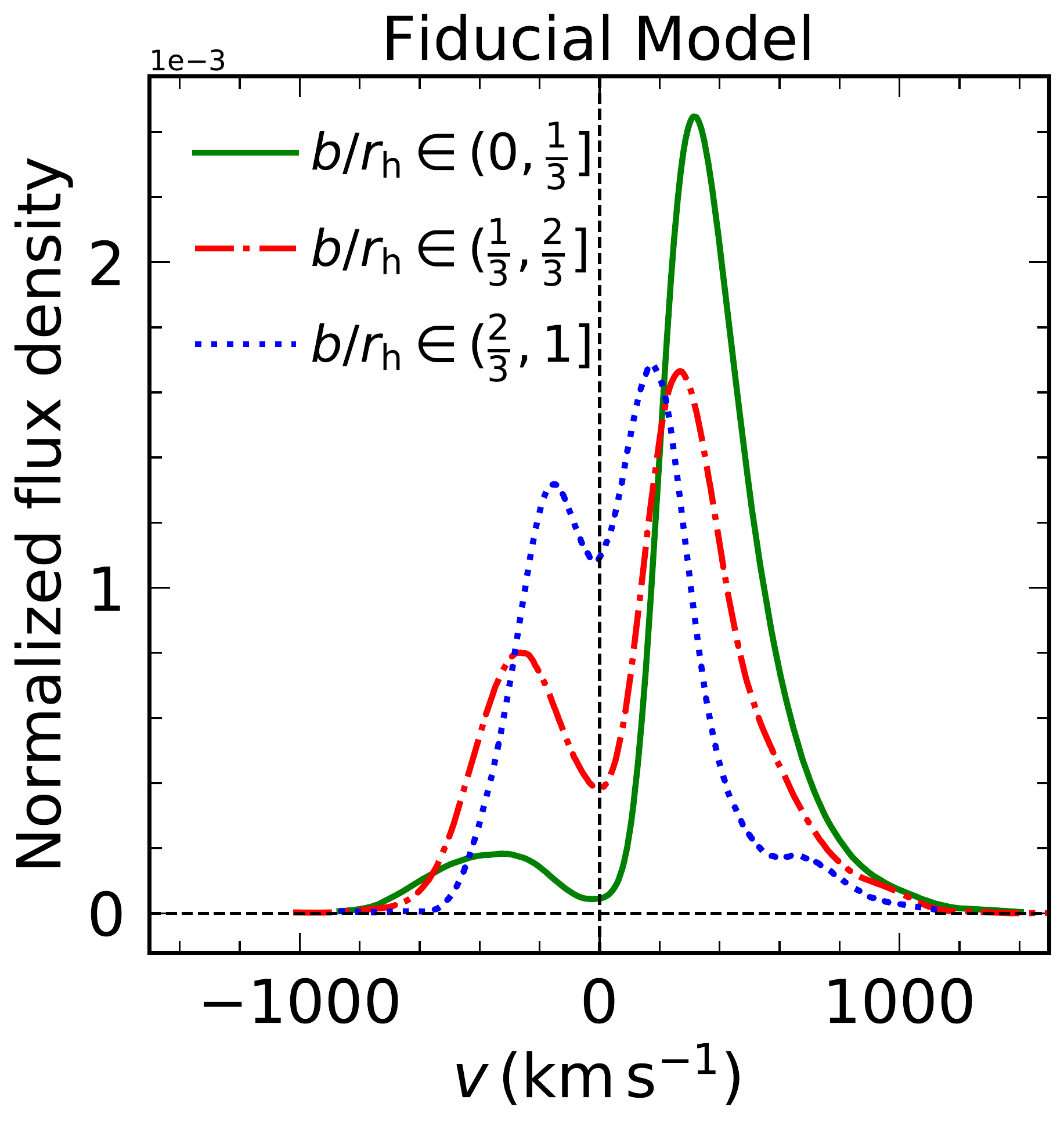}
\includegraphics[width=0.245\hsize]{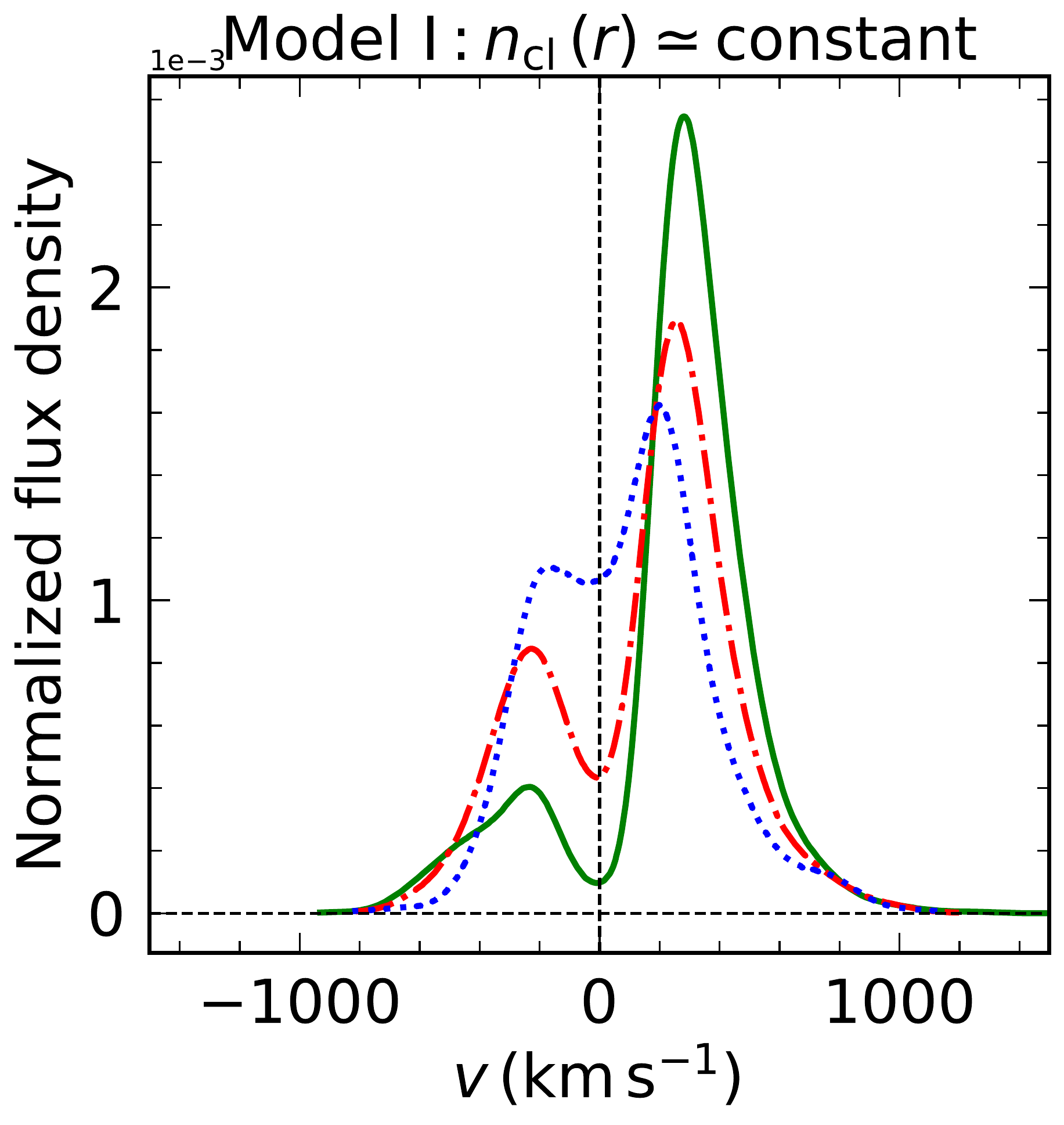}
\includegraphics[width=0.245\hsize]{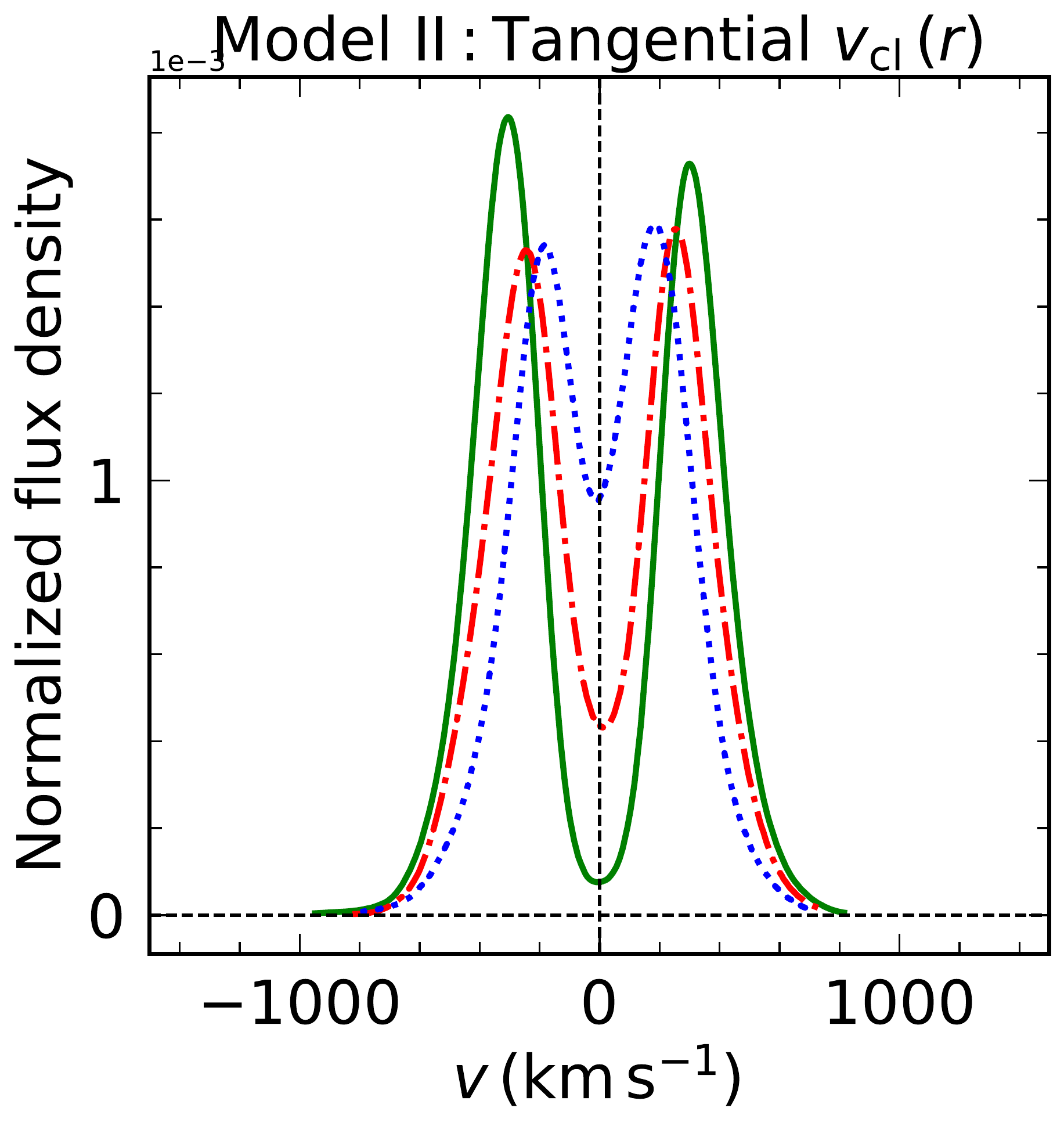}
\includegraphics[width=0.245\hsize]{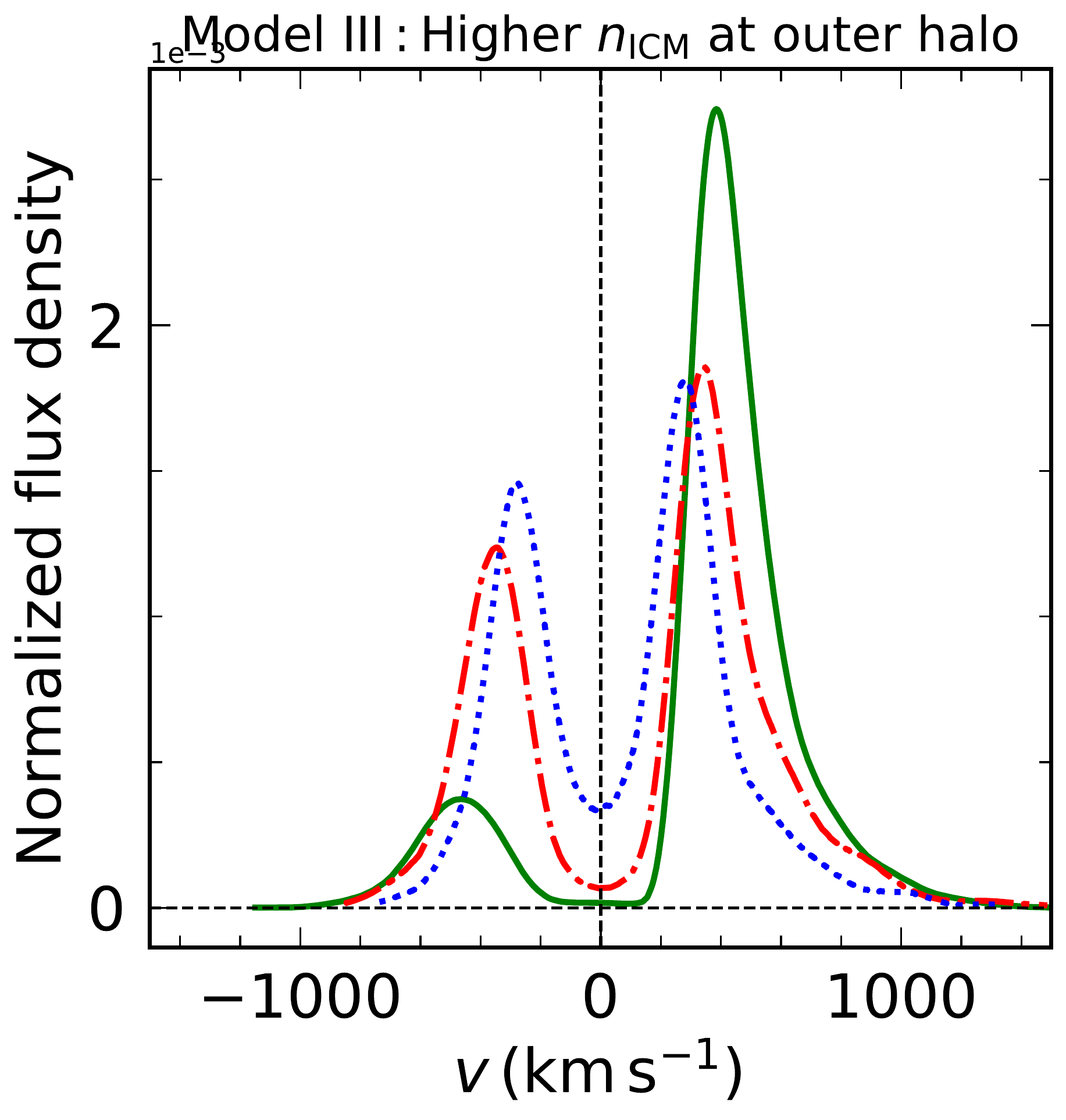}
\caption{Experiments designed to test our hypotheses for the differences between \lya\ photons that escape at low and high impact parameters. In each of the four subpanels, three binned model \lya\ spectra are shown according to their last-scattering impact parameters: $b/r_{\rm h} \in (0, \frac{1}{3}]$ (green solid), $(\frac{1}{3}, \frac{2}{3}]$ (red dash-dotted) and $(\frac{2}{3}, 1]$ (blue dotted), where $r_{\rm h}$ is the radius of the modeled halo. \emph{Left}: The fiducial model with ($F_{\rm V}$, ${\rm log}\,N_{\rm HI,\,cl}$, $\sigma_{\rm cl}$, $v_{\rm cl,\,\infty}$, ${\rm log}\,n_{\rm HI,\,ICM}$) = (0.05, 18.5, 80, 500, -7.0). \emph{Second from left}: Model I, in which more clumps are placed at large radii so that the number density of the clumps $n_{\rm cl}(r) \simeq$ constant. \emph{Third from left}: Model II, in which the clump radial velocity is set to be tangential, so that the projected component of the clump outflow velocity along the traveling direction is no longer preferentially small for photons that escape at high impact parameters. \emph{Right}: Model III, in which the number density of the ICM is increased by a factor of 20 in the outer 60\% of the halo radius. In each of the three test models, the change in the model configuration offsets the corresponding spatial variation of the \lya\ spectral morphology (i.e. peak separation, peak flux ratio and trough flux fraction), hence supporting our explanation.}
\label{fig:models_test}
\end{figure*}

To test these hypotheses, we have designed several experiments and present them in Figure \ref{fig:models_test}. We first generate our fiducial model by setting ($F_{\rm V}$, ${\rm log}\,N_{\rm HI,\,cl}$, $\sigma_{\rm cl}$, $v_{\rm cl,\,\infty}$, ${\rm log}\,n_{\rm HI,\,ICM}$, $\Delta v$) = (0.05, 18.5, 80, 500, -7.0, \edit1{0}). Such a choice roughly corresponds to the median parameter values of the model grid and proves to clearly demonstrate the radial variation of the peak separation, peak flux ratio, and the trough flux fraction of the radially binned \lya\ spectra. We then generate three test models for comparison by modifying the configuration of the fiducial model in specific ways. In Model I, we adjust the spatial distribution of the clumps:\ instead of placing the clumps radially uniformly, we place more clumps at large radii so that the number density of the clumps $n_{\rm cl}(r) \simeq$ constant. In Model II, we change the direction of the clumps' outflow velocity from radial to tangential \edit1{by rotating the clumps' velocity vector by 90 degrees}, so that the projected component of the clump outflow velocity along the traveling direction is no longer preferentially small for photons that escape at high impact parameters. In Model III, we increase the number density of the ICM by a factor of 20 in the outer 60\% of the halo radius in order to offset the shorter photon traveling distance at large radii. As shown in Figure \ref{fig:models_test}, in Model I, the peak separation of the three binned \lya\ model spectra is now roughly constant; in Model II, the significant increase in the blue-to-red peak flux ratio is no longer present, yet a slight decrease towards the outskirts is seen; and in Model III, the trough flux fractions are all much closer to zero. Therefore, we conclude that these experiments strongly support our above explanation  for the radial trends of the peak separation, peak flux ratio and trough flux fraction.

Incidentally, our model has also reproduced the decreasing trend of \lya\ SB versus impact parameter, with only a few exceptions (e.g.\ Q0142-BX186 and Q1700-BX729). These two objects, which have a more gradual decline in SB, are the faintest objects in the sample, with the smallest fraction of the total halo area used for the spatially resolved \lya\ modeling. This overall consistency adds further credence to our multiphase, clumpy RT model.

\subsubsection{Best-fit parameters}
\label{sec:bestfit_parameters}
\begin{figure}[htbp]
\begin{tabular}{p{0.48\textwidth} p{0.40\textwidth}}
  \vspace{0pt} \includegraphics[width=0.42\textwidth]{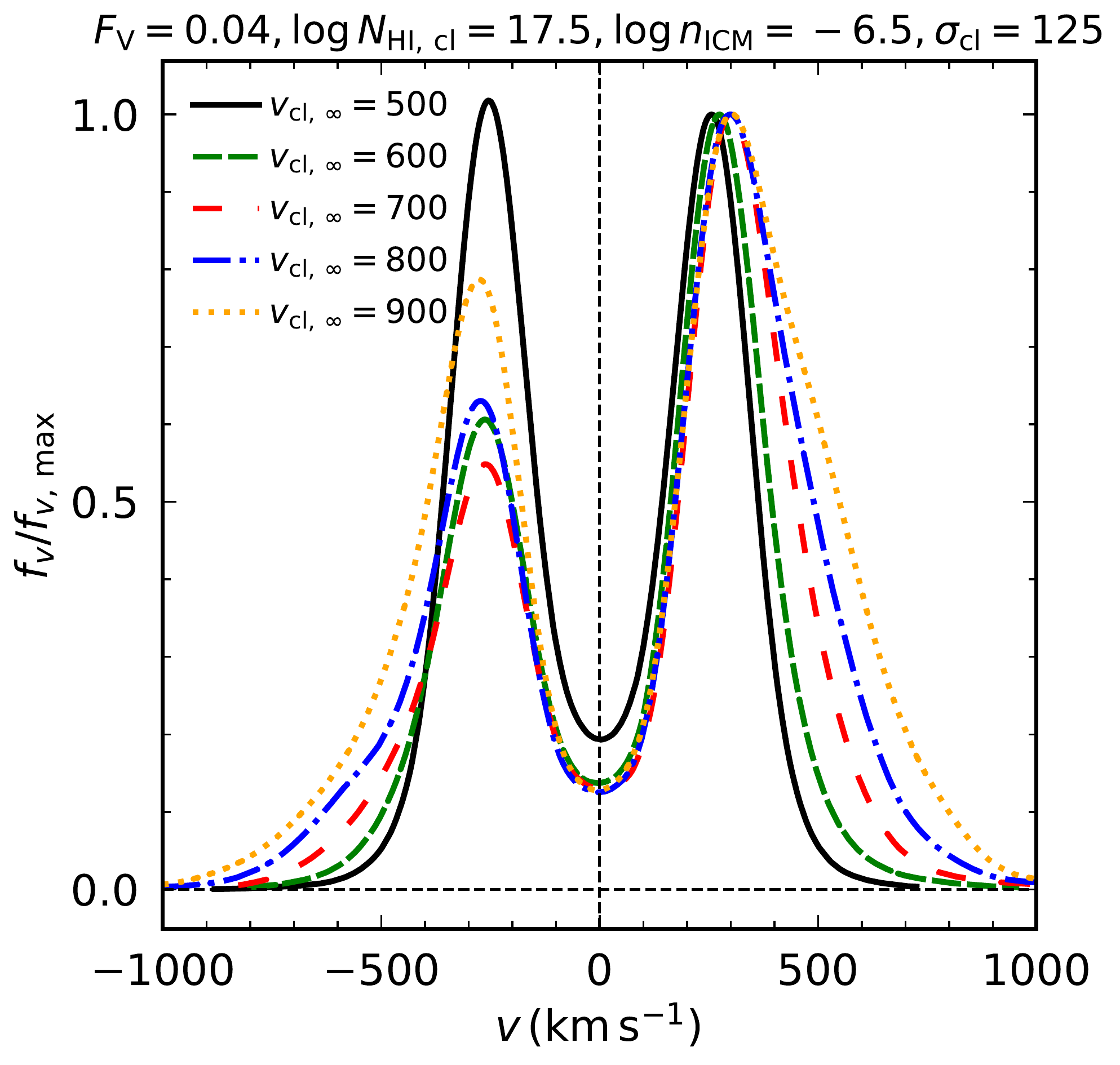} 
  \vspace{0pt} \includegraphics[width=0.40\textwidth]{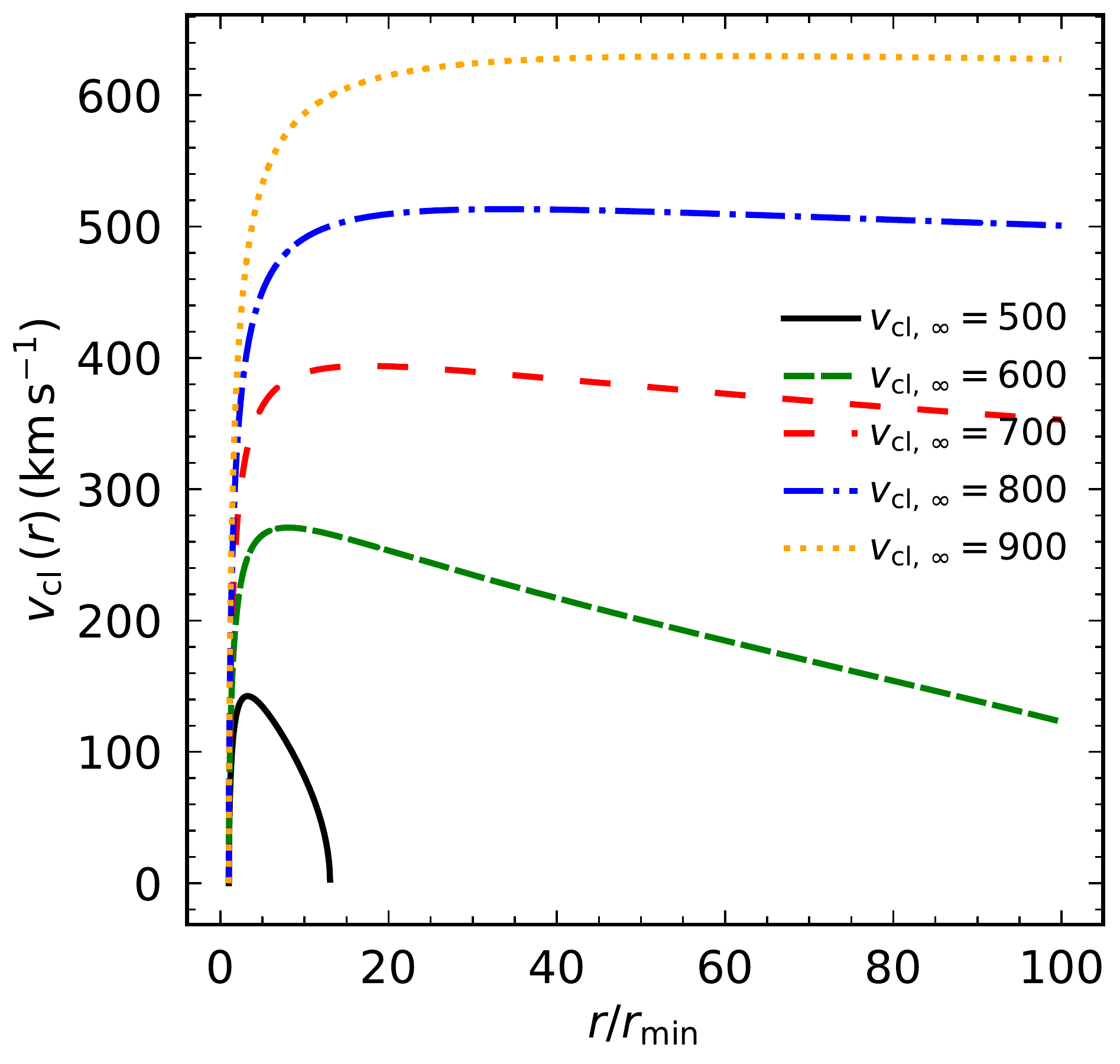}
\end{tabular}
\caption{Examples of \lya\ model spectra with different clump outflow velocities showing the pattern in the change of the blue-to-red peak flux ratio. Five models with ($F_{\rm V}, {\rm log}\,N_{\rm HI,\,cl}, {\rm log}\,n_{\rm ICM}, \sigma_{\rm cl}) = (0.04, 17.5, -6.5, 125)$ and $v_{\rm cl,\,\infty} = (500, 600, 700, 800, 900)$ are shown with different colors and linestyles. \emph{Upper}: spatially integrated \lya\ model spectra with different $v_{\rm cl,\,\infty}$ values. As $v_{\rm cl,\,\infty}$ increases, the average clump radial outflow velocity increases, and the blue-to-red peak flux ratio first decreases (comparing the black and green curves) and then increases (comparing the red, blue and orange curves). \edit1{For visual convenience, we have normalized all the model spectra so that the maximum flux density of the red peak is one.} \emph{Lower}: The corresponding clump radial velocity profiles for different $v_{\rm cl,\,\infty}$. Note that $v_{\rm cl}(r)$ and $v_{\rm cl,\,\infty}$ are positively correlated, but $v_{\rm cl}(r)$ is always smaller than (typically by several hundred km\,s$^{-1}$) $v_{\rm cl,\,\infty}$ due to the effect of gravitational deceleration.}
\label{fig:vinf}
\end{figure}

One of the most interesting discoveries from our modeling is that the clump outflow velocities inferred from \lya\ emission and the low ionization metal absorption lines can be mutually consistent, with typical values of $\sim 400 - 600$ \kms\ obtained for both \edit1{(see Table \ref{tab:Fitting_results})}. The mismatch between the gas outflow velocities inferred from \lya\ and from metal absorption lines has been a long-standing problem. For example, it is reported that the $\lesssim 150$ \kms\ outflow velocities of the shell model required to match the \lya\ profiles of local starburst and green pea galaxies are {\it{much lower}} than the $\gtrsim 300$ \kms\ characteristic velocities of the metal absorption lines (e.g.\ \citealt{Leitherer13, Orlitova2018}). The high outflow velocity regime of \lya\ RT models has been little explored, possibly due to the belief that the \lya\ photons will be seen as out of resonance by the fast moving gas and will therefore not scatter (e.g.\ \citealt{Verhamme2015}). However, we observe an interesting pattern in our multiphase, clumpy model:\ for a typical double-peaked \lya\ profile, as the clump outflow velocity increases, the blue-to-red peak flux ratio (or the ``level of symmetry'') first decreases and then increases, until the clump outflow velocity is so large that all the photons are completely shifted out of resonance as seen by the gas. This pattern, as shown by an example in Figure \ref{fig:vinf}, suggests the possibility of matching the observed asymmetric \lya\ profiles in the high outflow velocity regime ($v_{\rm cl,\,max} \gtrsim 400 - 600 \rm\;km\,s^{-1}$). 

In our sample, consistency (accounting for uncertainties) between the clump outflow velocities inferred from \lya\ and metal absorption lines is achieved in 8 / 12 objects.\footnote{We define two velocity profiles as being consistent if they have a non-negligible overlap at $r > r_{\rm min}$.} Such a high success rate demonstrates the feasibility of matching both the observed \lya\ and metal absorption line profiles simultaneously with one clump radial velocity profile. Among the four inconsistent cases, two (Q0207-BX87 and Q2343-BX418) have relatively irregular and noisy absorption line profiles that yield a broad range of velocities, whereas in the other two cases (Q0142-BX165 and Q1700-BX729), \edit1{the inconsistency may come from the unusual asymmetry of either their \lya\ halo or stellar continuum:\ \citet{Chen2021} found that Q0142-BX165 is a significant outlier that has particularly asymmetric \lya\ emission as a function of azimuthal angle, whereas Q1700-BX729 is one of the 5 galaxies in their sample of 40 that requires more than one source for a successful Sersic profile fit to its stellar continuum}. We also note that exact matches between the \lya\ and absorption-line-inferred outflow velocities are not necessarily expected because the transitions probe somewhat different gas:\ the absorption lines are purely a line-of-sight measurement that probes the gas only on the near side of the halo, while the \lya\ results incorporate gas on the far side \edit1{of the galaxy} and at large impact parameters that is not seen in absorption. \edit1{An exact match between the velocities would therefore be seen only in the case of perfect angular symmetry.}

The best-fit radial velocity profile of the clumps in the multiphase, clumpy model typically exhibits a rapid acceleration phase to $v_{\rm cl} = v_{\rm cl,\,max}$ within $1 \lesssim \frac{r}{r_{\rm min}} \lesssim 10$ followed by a gradual deceleration\footnote{Note, however, that such a deceleration phase is not preferred by the absorption line modeling as it will break the one-to-one relation between $r$ and $v_{\rm cl}$ and yield a pathological absorption line profile $I(v)$.} (or $v_{\rm cl} \simeq $ constant) phase at $\frac{r}{r_{\rm min}} \gtrsim 10$. The decline in the outflow velocity and possible transition to an inflow are physically expected due to the increasing importance of gravitational deceleration at large radii, and have been explored in previous works (e.g.\ \citealt{Chen2020}); however, the exact location of the transition is model-dependent and may need additional observational constraints.

We also note that there is a significant velocity difference between the outflowing cool clumps and the static hot ICM in the best-fit models, which is at odds with the traditional ``hot wind entrains (and co-outflows with) the cold gas'' paradigm (see e.g.\ \citealt{Gronke18, Gronke20}, and references therein). It is possible, however, that the interaction between the hot phase and the \lya\ photons is dominated by the decelerated, semi-static hot gas, as suggested by the deep troughs at line center in the observed \lya\ profiles. A larger sample with more diverse \lya\ morphologies will be helpful in assessing the impact of an outflowing hot gas component in the future.

We next turn to the other best-fit parameters of the models. For the \lya\ modeling, the best-fit clump volume filling factors ($F_{\rm V}$) range from 0.06 to 0.16 (corresponding to $\sim 5 - 10$ clumps on average per line-of-sight\footnote{Note that the number of clumps per line-of-sight and the associated gas covering fraction both decrease with $r$ due to the increase of the halo volume at large $r$ (cf.\ Figures 15 and 16 of \citealt{Rudie2012}).}), and the best-fit clump column densities ($N_{\rm HI,\,cl}$) range from $\sim 10^{17.6}$ to $10^{18.8}\,\rm cm^{-2}$. The total \HI\ column densities ($N_{\rm HI,\,total} \simeq \frac{4}{3} f_{\rm cl} N_{\rm HI,\,cl} = (r_{\rm h} / r_{\rm cl})\,F_{\rm V} N_{\rm HI,\,cl}$, \citealt{Gronke2016b}) of the best-fit models range from $\sim 10^{18.5}$ to $10^{19.9}\,\rm cm^{-2}$. Here $N_{\rm HI,\,total}$ represents the inferred total \HI\ column density of the modeled halo that a \lya\ photon typically interacts with, either via scattering or free-streaming; the scattered, out-of-resonance \lya\ photons may stream through the high-velocity, outflowing clumps without scattering \citep{Gronke17}.

The residual \HI\ column densities of the hot, diffuse ICM ($N_{\rm HI,\,ICM} \simeq n_{\rm HI,\,ICM}\,r_{\rm h}$) range from $\sim 10^{15}$ to $10^{16}\,\rm cm^{-2}$. Such column densities are much smaller than those within the clumps, but are necessary to produce the absorption trough at line center, and may serve as optically-thin channels for LyC escape along lines of sight that have relatively few \HI\ clumps. The best-fit systemic redshifts of the \lya\ sources are mostly consistent with the systemic redshifts determined from nebular emission lines ($|\Delta v| < 50\;\rm km\,s^{-1}$ for 8 / 12 objects). The best-fit clump velocity dispersions ($\sigma_{\rm cl}$) are all smaller than $150\;\rm km\,s^{-1}$ and span a similar range to the observed nebular emission line widths ($\sim 50 - 120\;\rm km\,s^{-1}$). We compared the best-fit $\sigma_{\rm cl}$ values with the MOSFIRE $H$-band (\OIII\ and \Hb) and $K$-band (\Ha) nebular emission line widths (corrected for instrumental LSF), but did not find any significant correlation.

For the metal absorption line modeling, the best-fit clump velocity dispersions\footnote{Note that here the velocity dispersions are determined independently from the \lya\ modeling. In fact, they are not very well constrained (i.e.\ flat posterior) by the absorption line data, as the acceleration term is preferred to be dominant.} are all smaller than $75\;\rm km\,s^{-1}$, suggesting that the gravitational deceleration only plays a minor role compared to the acceleration forces. The clump outflow velocities are high, mostly $\gtrsim 500 \rm\;km\,s^{-1}$, and generally correspond to the velocity where the blue side of the absorption line profile meets the continuum. The maximum clump covering fractions $f_{\rm c, max}$ range from $\sim$ 0.2 to 0.9, depending on the minimum flux density of the absorption line profile. The power-law indices of the clump covering fraction function ($\gamma$) range from $\sim$ 1.1 to 1.7, corresponding to a mass-conserving-like (or more gradual) decrease in the number density of the clumps. A $\gamma$ smaller than 2 may suggest that the clumps expand as they move outwards (e.g.\ due to the decrease of thermal/radiation pressure at large radii), because if the clumps are uniformly distributed radially and their sizes remain constant at different radii, $\gamma$ will be exactly 2 due to the geometric volume increase at large radii ($dV \propto 4\pi r^2 dr$). 

We have also checked if any correlations exist between the best-fit clump outflow velocities and the host galaxy properties such as stellar mass and SFR, as these are expected to be correlated due to the causal relation between stellar feedback and galactic outflows (e.g.\ \citealt{Martin05, Rupke05, Weiner09, Chen10, Martin12, Rubin14, Chisholm15, Heckman2015, Trainor15}). Specifically, we tested for correlations between the \emph{actual} maximum clump radial velocities $v_{\rm cl,\,max}$ inferred from the \lya\ and absorption line modeling versus the stellar masses, SFRs and sSFRs of the host galaxies. We find that all three correlations are insignificant and have considerable scatter. Such a null result is unsurprising, however, as our sample is intentionally restricted to low-mass galaxies with high SFR and sSFR values and therefore has a limited dynamic range by design. We will revisit these correlations with larger and more well-rounded samples in future work. 

\subsubsection{Advantages of spatially resolved Ly$\alpha$ modeling}
\label{sec:resolved_vs_integrated}
In this section, we demonstrate the advantages of spatially resolved \lya\ modeling by comparing it to the spatially integrated \lya\ modeling that has typically been carried out in previous works. Assuming that for a \lya-emitting source of interest, only a spatially integrated \lya\ spectrum within a certain aperture can be obtained (e.g.\ due to the unavailability of IFU observations), we consider the following two scenarios: (1) the spatially integrated spectrum corresponds to the \lya\ emission from only the central region, typical of observations using a slit or other small aperture (the spectra we extracted in Section \ref{sec:global} and showed in Figure \ref{fig:lyaprofiles} belong to this category); (2) the spatially integrated spectrum is extracted from a larger aperture that also includes the \lya\ emission from a significant portion of the extended halo. For exploratory purposes, we model scenario (1) with spatially integrated multiphase, clumpy models in which all the emitted photons are included in the emergent spectra, assuming that we are completely unaware of any spatial variation of the \lya\ emission. We model scenario (2) with spatially integrated models that include the photons with $b/r_{\rm h} \lesssim 75\%$, assuming that we are aware that the data only represent part of the extended halo and should be compared to a corresponding fraction of the modeled halo. This is equivalent to merging the 3-bin spectra for both the data and the models in the spatially resolved modeling routine that we described in Section \ref{sec:lya_modeling}.

For scenario (1), we find that the best-fit clump outflow kinematics (namely the $\sigma_{\rm cl}$ and $v_{\rm cl,\,\infty}$ values) are similar in both the spatially integrated and resolved modeling, but the required clump volume filling factors (and hence the covering factor) and ICM column densities are higher, on average, in the spatially integrated \lya\ modeling. This is mainly because in the observed spatially integrated spectra, the trough depth at line center is similar to that of the innermost binned spectra used in the spatially resolved modeling, as they correspond to similar regions of the halos. In contrast, the trough depth at the line center of a spatially integrated \lya\ model spectrum lies between that of its corresponding innermost and outermost binned model spectra due to the radial variation of the profile (see Section \ref{sec:radial_trends}). Therefore, larger clump volume filling factors (which contribute to the total \HI\ column densities) and ICM column densities are required to reproduce the deep troughs in the spatially integrated \lya\ profiles. 

\begin{figure*}[htbp]
\centerline{\epsfig{angle=00,width=\hsize,file=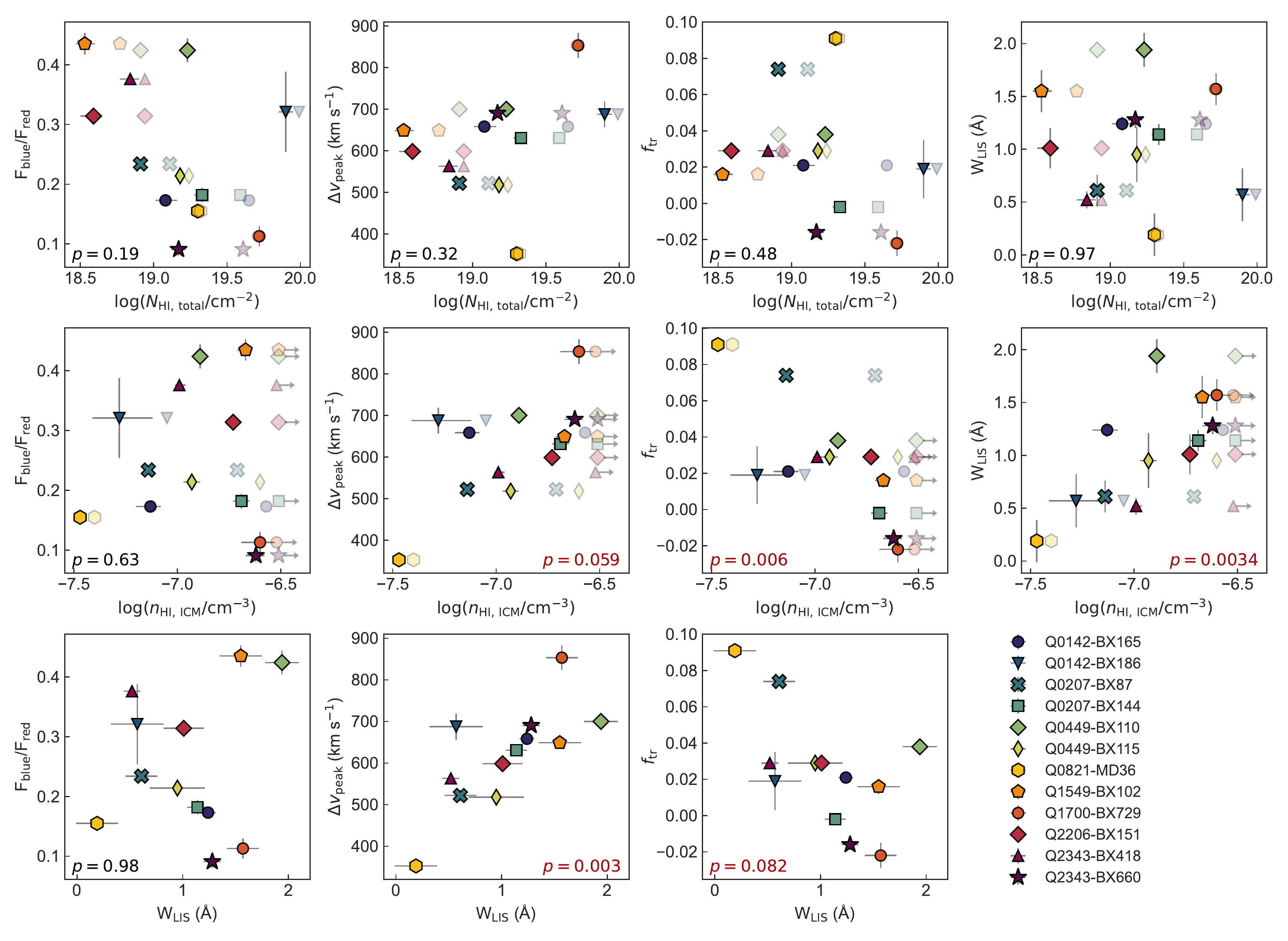}}
\caption{Comparison of results from the radiative transfer models with properties of the spatially integrated spectra of the central regions of the galaxies shown in Figure \ref{fig:lyaprofiles}. \textit{Top row, left to right:}\ the \lya\ blue-to-red flux ratio, peak separation, trough flux fraction $f_{\rm tr}$ and mean low ionization absorption line equivalent width vs.\ the total \ion{H}{1} column density. \textit{Middle row:}\ the same four spectral quantities vs.\ the residual \ion{H}{1} density in the ICM. \textit{Bottom row, left to right:}\ \lya\ blue-to-red flux ratio, peak separation, and $f_{\rm tr}$ vs.\ mean low ionization absorption line equivalent width. In the top two rows the darker points show the results of our best-fit spatially resolved modeling, while the fainter points show the results of modeling the single, spatially integrated line profiles. The lower corner of each panel gives the $p$-value resulting from a Spearman correlation test using the spatially resolved models only. Values with $p<0.1$ are highlighted in red.}
\label{fig:modelresults}
\end{figure*}

A quantitative comparison of the best-fit total \HI\ column densities from the clumps and the \HI\ densities in the ICM for the spatially resolved and scenario (1) models is shown in Figure \ref{fig:modelresults}, with the darker and fainter points indicating the resolved and spatially integrated models from scenario (1) respectively. We plot the total $N_{\rm HI}$ and $n_{\rm HI,\,ICM}$ versus properties measured from the integrated spectra, and discuss the comparison further in Section \ref{sec:discuss} below. We find that values of total $N_{\rm HI}$ from spatially integrated modeling of the central region are larger on average by a factor of 1.5, while $n_{\rm HI,\,ICM}$ is larger by at least a factor of 1.9, and likely significantly more because more than half of the sample requires values of $n_{\rm HI,\,ICM}$ higher than the maximum value allowed by the model grid. The overestimation of $n_{\rm HI,\,ICM}$ in the spatially integrated models manifests as an overestimation of the depth of the trough between the peaks, which is due to the omission of spatial information on the outer halo.

For scenario (2), we find that the best-fit parameters of the spatially resolved and integrated modeling are fully consistent with each other. This result is probably unsurprising, as a reasonable match between all three bins of model and data should still hold if the bins are merged for both the models and the data. \edit1{The tightness of the constraints on the model parameters is also similar in both cases due to the similar average S/N ratio of the observational data.} However, we stress that this result does not indicate that the spatially resolved modeling is no longer necessary, as we would not have found that the radial trends of peak separation, peak ratio, trough flux ratio, and SB can all be reasonably well-matched by the same best-fit model if we had not separated the photons into different spatial bins and modeled the \lya\ profiles in a spatially resolved manner.

In short, our experiments in this section suggest that although spatially integrated modeling may be used to crudely extract certain global properties of the CGM, it tends to either lose information about the outer regions of the halos and overestimate the neutral hydrogen content encountered by \lya\ photons, or fail to account for the radial variation of the \lya\ morphological properties. In comparison, spatially resolved \lya\ modeling has the advantage of fully leveraging the spatial variation in the \lya\ halo as observed by integral field unit spectrographs such as KCWI and quantifying the corresponding spatial changes of the physical parameters of the CGM. The overall good match in radial trends between the spatially resolved data and models provides a reassuring check on the validity of the multiphase, clumpy model.

\subsubsection{Caveats}
\label{sec:caveats}
There are several important caveats to this work. First, we did not include the effect of dust (but note that the dust extinction of our sample is typically small), which means that all of the emitted \lya\ photons will eventually escape from the simulation region and contribute to the emergent model spectra. Considering that the actual \lya\ escape fraction is always smaller than one (when it is robustly measured; see Table \ref{tab:lya_measurements} and discussion in Section \ref{sec:global}), we essentially assumed that the observed frequency distribution of \lya\ photons is representative of the \lya\ photons that escape in all directions. The validity of such an assumption requires further scrutiny. 

Second, we used spherically symmetric RT models to model the angularly averaged \lya\ profiles of asymmetric halos, so the results should be interpreted as average parameters within the modeled region. We have also experimented with modeling the spatially resolved \lya\ profiles along the directions of maximum and minimum peak ratio and peak separation gradients (see Section \ref{sec:peakgradients}), but did not find any significant dependence of the model parameters on these higher order spatial variations. This is mainly because the best-fit model is primarily constrained by the spectra of the two innermost bins, which have higher S/N, whereas the spectrum of the outermost bin may contribute strongly to the measured gradients but does not put strong constraints on the model parameters. Development of anisotropic RT models may shed light on this problem, \edit1{as future observational facilities will likely improve the S/N of the spectra of the outer halo, and eventually the higher-order spatial variations should be able to put extra constraints on the model parameters.}

Last but not least, some of the assumptions in our models are inevitably over-simplified. For example, we assumed a two-component model with temperatures of 10$^4$ K and 10$^6$ K, whereas in reality \HI\ absorbers at intermediate temperatures should exist \citep{Rudie19}. The \HI\ column densities and the physical sizes of the clumps are also simplistically assumed to be constant in the multiphase, clumpy model. Moreover, the actual motion of the clumps in the CGM may be more complicated than the idealistic kinematic model we employed (see e.g.\ \citealt{Fielding22}). We plan to upgrade our models in future work.

\section{Summary and discussion}
\label{sec:discuss}
We have presented KCWI integral field spectroscopy and radiative transfer modeling of spatially extended \lya\ emission in a sample of 12 relatively low mass ($M_{\star} \sim 10^9$ \msun), extreme emission line galaxies at median redshift $z=2.3$. As described in Section \ref{sec:sample}, the targets are primarily selected based on nebular emission line ratios indicating high ionization and low metallicity, and all are previously known \lya-emitters. The sample galaxies have specific star formation rates $\sim4$ times larger than than that of their  $z\sim2$ parent sample, and may more closely resemble galaxies at earlier epochs of cosmic history. Our primary results are as follows:

\begin{enumerate}
\item{All of the galaxies show strong, double-peaked \lya\ emission (see Section \ref{sec:global} and Figure \ref{fig:lyaprofiles}) and spatially extended \lya\ halos, with luminosities ranging from \expnt{3}{42} to \expnt{3}{43} erg s$^{-1}$ and radii between 16 and 30 kpc (Figure \ref{fig:lyaimages}).}

\item{We fit double asymmetric Gaussian profiles to the \lya\ emission of individual spaxels and small Voronoi bins in each halo, as described in Section \ref{sec:spaxelspectra} and shown in Figure \ref{fig:spaxelspecs}, and measure the flux ratio of the blue and red peaks and the peak separation velocity $\Delta v_{\rm peak}$ for each spaxel or bin. Maps of the peak ratio and separation are presented in Figures \ref{fig:peakmaps1} and \ref{fig:peakmaps2}. The halos show significant azimuthal variation, but the blue-to-red flux ratio tends to increase at larger radii and regions of narrower peak separation are usually found in the outer halo. The peak ratio and separation are anti-correlated for half of the sample, but this is likely driven by the underlying tendency of both to change with radius (Figures \ref{fig:ratio_vs_sep} and \ref{fig:correlations}).}

\item{We also construct spatially averaged \lya\ profiles, in order to identify general trends and measure the profiles to larger radii. We first construct azimuthally averaged spectra binned as a function of radius (Section \ref{sec:annularspectra} and Figure \ref{fig:annularspectra}), and again measure the peak ratio and separation in each annular region as well as $f_{\rm tr}$, the fraction of total flux escaping within $\pm100$ \kms\ of the trough between the peaks (Figure \ref{fig:annularmeasurements}). The blue-to-red flux ratio increases consistently with radius for most objects in the sample, with a typical central value of $\sim0.2$; all objects that can be measured at a radius $\gtrsim16$ kpc have peak flux ratios $>0.6$ at that radius. $f_{\rm tr}$ also increases with radius for most of the sample. Trends with peak separation are more complex, but the typical central peak separation is $\sim600$ \kms, with a moderate decrease toward the outer halo. }

\item{Because the annular binned spectra wash out the significant azimuthal variations in the line profiles, we also construct binned spectra of 60$^{\circ}$ angular regions designed to maximize the gradients in peak ratio and separation from the center to the outer halo, using seven of the brightest galaxies in the sample (Section \ref{sec:peakgradients} and Figure \ref{fig:gradients}). These spectra show that all of the halos have sightlines for which the peak ratio increases (typically from $\sim0.2$ to $\sim1$) or the peak separation decreases (typically from $\sim600$ -- 700 to $\sim300$ -- 400 \kms) with radius. In all cases, however, the regions of maximum peak ratio increase and maximum peak separation decrease do not overlap. We also construct spectra designed to minimize the gradients in peak ratio and separation, finding that most halos also have regions for which the changes in peak ratio and separation with radius are relatively small.}

\item{Using a new suite of \lya\ radiative transfer simulations, we model the spatially resolved \lya\ profiles in three radial bins with multiphase, clumpy models with radially-varying outflow velocities (Section \ref{sec:lya_modeling}). These models are broadly successful in reproducing the observed line profiles, as well as the radial trends of peak flux ratio, peak separation, and trough flux fraction (Figures \ref{fig:fits_example} and \ref{fig:fits1}--\ref{fig:fits6}). The clumps reach a typical maximum velocity of $\sim500$ \kms\ and have \ion{H}{1} column densities of $\sim10^{17.6}$ to $10^{18.8}$ cm$^{-2}$, while the total $N_{\rm HI}$ of the best-fit models ranges from  $\sim10^{18.5}$ to $10^{19.9}$ cm$^{-2}$. The clumps are embedded in a hot inter-clump medium with residual $N_{\rm HI,\,ICM}\sim10^{15}$ -- $10^{16}$ cm$^{-2}$. Best-fit parameters of the models are given in Table \ref{tab:Fitting_results}.}

\item{We find that the trend in \lya\ peak separation with radius is primarily governed by the \ion{H}{1} column density, as photons that escape at larger radii are able to do so with a smaller velocity shift because they experience lower \HI\ column densities from the clumps before they escape due to the decrease in clump covering fraction with radius. The \lya\ peak ratio depends on the line-of-sight velocity, with the result that the variation in peak ratio with radius is largely a geometric effect as the projected component of the outflow velocity along the line of sight decreases with increasing impact parameter (Figure \ref{fig:schematic}). The depth of the trough (or the trough flux fraction, $f_{\rm tr}$) between the two peaks primarily depends on the residual neutral \HI\ density of the ICM. We show the results of experiments designed to test these conclusions in Figure \ref{fig:models_test}, and further explore the relationship between outflow velocity and peak ratio in Figure \ref{fig:vinf}. }

\item{We self-consistently model the mean low ionization absorption line profile of each object, employing the same radially varying velocity model used for the \lya\ emission and a radially decreasing gas covering fraction (Section \ref{sec:UV_modeling} and Figures \ref{fig:fits_example} and \ref{fig:fits1}--\ref{fig:fits6}). Typical clump maximum outflow velocities inferred from the absorption line profiles are $\gtrsim 500$ \kms, in broad agreement with the velocities inferred from \lya; exact matches may not be expected because the down-the-barrel UV spectra and the radially binned \lya\ emission are not probing entirely the same regions of the halos. This agreement alleviates a long-standing discrepancy between outflow velocities inferred from \lya\ shell models and the UV absorption lines.}

\item{Finally, we compare the results of the spatially resolved \lya\ modeling with those obtained from applying the same model to single, spatially integrated \lya\ profiles, using both a small aperture capturing only the brightest region (scenario 1) and a larger aperture encompassing most of the halo (scenario 2). We find that modeling the integrated central profile (scenario 1) results in higher inferred values for both the total \ion{H}{1} column density and the neutral component of the ICM, largely because the spatially integrated modeling does not account for the decrease in the depth of the trough between the peaks at larger radii; this decrease in depth reflects the lower neutral hydrogen content experienced by photons that escape from larger radii and indicates that some photons may escape at line center in the outer halo. The best-fit parameters obtained from modeling a larger aperture in scenario (2) are consistent with those from the spatially resolved  modeling, but fail to capture the trends in the \lya\ profile with radius and the physical insights these variations provide.}
\end{enumerate}

Our observations and modeling suggest a self-consistent physical picture of the CGM of this sample of $z \sim 2$ star-forming galaxies:\ a multiphase, clumpy medium in which cool ($\sim 10^4$\,K), outflowing gas clumps are embedded in a hot ($\sim 10^6$\,K), highly ionized, diffuse medium with low-density residual \HI. The clumps typically have \HI\ column densities of $\sim 10^{18}\,\rm cm^{-2}$ and provide a total column density of $\sim 10^{19}\,\rm cm^{-2}$, and the \lya\ photons ``solve the maze'' by being resonantly scattered by and free-streaming through the clumps until they escape. The cool clumps also have random velocity dispersions of $\sim 100\,\rm km\,s^{-1}$, and are accelerated to high radial outflow velocities of $\gtrsim 500\rm \,km\,s^{-1}$ at large impact parameters, which give rise to both the asymmetric \lya\ profiles and broad low ionization metal absorption lines. The hot ICM is nearly static and has a low total \HI\ column density ($\sim 10^{15}$ -- $10^{16}\,\rm cm^{-2}$), but is essential to shaping the emergent double-peaked \lya\ profiles as it provides additional scattering that produces the absorption trough at line center.

\subsection{Central Ly$\alpha$ profiles and LyC escape}
With this physical model of the CGM in mind, we revisit the spatially integrated central \lya\ profiles shown in Figure \ref{fig:lyaprofiles} and assess how (or if) quantities measured from these profiles relate to the properties of the CGM inferred from the spatially resolved modeling; such a comparison may aid in the interpretation of \lya\ profiles when information from the outer halo is unavailable. In Figure \ref{fig:modelresults} we compare the total $N_{\rm HI}$ and $n_{\rm HI,\,ICM}$ from the models with the peak ratio, peak separation and trough flux fraction $f_{\rm tr}$ and the mean low ionization absorption equivalent width $W_{\rm LIS}$ measured from the spatially integrated one-dimensional spectra, as well as the equivalent width vs. the \lya\ profile properties in the bottom row. Darker points indicate the results of the spatially resolved \lya\ modeling, while the fainter points are the result of modeling the central spatially integrated profiles (scenario 1 in Section \ref{sec:resolved_vs_integrated}). The lower corner of each panel gives the $p$-value resulting from a Spearman correlation test, with values of $p<0.1$ highlighted in red. While none of the correlations are formally ($>3\sigma$) significant, the strongest trends ($\sim2.75$--$3\sigma$) relate to the \HI\ density in the ICM, which tends to be higher for larger peak separations, lower $f_{\rm tr}$, and larger low ionization equivalent width. We also find that smaller peak separations and higher values of $f_{\rm tr}$ tend to be associated with lower $W_{\rm LIS}$. All correlations involving the total $N_{\rm HI}$ or the blue-to-red peak ratio have significance levels $\leq1.3\sigma$.

These results broadly support our conclusion in Section \ref{sec:radial_trends} that the trough flux fraction can be understood as an indication of low $N_{\rm HI}$ in the ICM. Note, however, that the potential relationship between the central $f_{\rm tr}$ and modeled $n_{\rm HI,\,ICM}$ relies on the results inferred from spatially resolved modeling of the extended halo; modeling the central profiles alone results in significantly higher values of $n_{\rm HI,\,ICM}$, half of which are higher than the upper limit of the current model grid. 

Previous work has suggested that significant \lya\ flux at the systemic velocity may be an indication of LyC escape (e.g.\ \citealt{Rivera-Thorsen2019,Naidu2022}); if ionizing photons emerge through optically thin channels between clumps, then the transparency of the ICM is a key property governing LyC escape. A low covering fraction of neutral gas and significant residual intensity in the low ionization absorption lines are also likely related to LyC escape (e.g.\ \citealt{Heckman2011,Reddy2016,Chisholm2018}), so the potential relationship between $n_{\rm HI,\,ICM}$ and $W_{\rm LIS}$ is also unsurprising. 

Given the results of the spatially resolved \lya\ modeling, we expect the peak separation to be most closely related to the total \HI\ column density; however, there is no significant correlation between the central peak separation and $N_{\rm HI, \, total}$ from the spatially resolved models. This lack of correlation may be due to the small sample size and the lack of dynamic range in peak separation, as 10 of the 12 objects in the sample have central peak separations between 500 and 700 \kms. These peak separations are also larger than the $\sim200$--500 \kms\ \edit1{range over} which $\Delta v_{\rm peak}$ is observed to correlate with the LyC escape fraction in local galaxies \citep{Izotov2021}. We do observe potential relationships between the peak separation and both $n_{\rm HI,\,ICM}$ and $W_{\rm LIS}$; these may be due to the strong correlation between the peak separation and $f_{\rm tr}$. Modeling of a larger sample with a wider range of central peak separations will clarify the relationship between $\Delta v_{\rm peak}$ and $N_{\rm HI, \, total}$.

There are no observations covering wavelengths below the Lyman break for the galaxies in our sample, so we have no constraints on their LyC emission. However, based on the criteria discussed above involving the peak separation or central flux fraction, we would not expect most of the galaxies in the sample to have significant LyC emission. Possible exceptions are the two most likely LyC candidates, Q0821-MD36 and Q0207-BX87, which have the highest trough flux fractions $f_{\rm tr}\sim0.1$, relatively narrow peak separations, and the second and third highest \lya\ equivalent widths in the sample (after Q2206-BX151).

\subsection{Future prospects}
Although the inclusion of spatially resolved information increases the power of the radiative transfer modeling, we are still limited by the assumption of symmetry:\ we fit radially binned spectra with spherically symmetric models, but as we have shown, real halos show significant azimuthal variation (Figure \ref{fig:gradients}). However, insights obtained from the modeling can aid in the interpretation of the variations across a given halo, at least qualitatively. Because the increase in blue-to-red peak ratio with radius is largely a geometric effect due to the decrease in the line-of-sight component of the outflow velocity, portions of the halos for which there is little change in the peak ratio with radius likely correspond to regions for which the velocity still has a significant component along the observer's line of sight even in the outskirts of the halo. More broadly, azimuthal variations in the peak ratio are indicative of velocity asymmetries and non-radial gas motions at large radii. Similarly, variations in the peak separation in the outer halo suggest varying \HI\ column densities in the CGM, with regions for which $\Delta v_{\rm peak}$ does not decrease with radius likely having higher $N_{\rm HI}$. Future modeling that does not assume azimuthal symmetry is needed in order to quantify these conclusions.  

Finally, while the objects in this sample are likely to be more typical of galaxies at higher redshifts than of the general $z\sim2$ population, extending the analysis of double-peaked \lya\ profiles to more distant galaxies will be challenging. For example, the median redshift of the MUSE sample studied by \citet{Leclercq2020} is $z=3.8$, while that of our KCWI sample is $z=2.3$, and this difference in redshift results in a median decrease in surface brightness of a factor of 4.5 for the higher redshift sample. In addition, the blue-to-red \lya\ peak ratio decreases with increasing redshift due to \lya\ absorption by the IGM \citep{Laursen2011,Hayes2021}, and the mean IGM transmission of \lya\ drops strongly from  $\gtrsim80$\% at $z\approx2.3$ to $\sim45$\% at $z\approx3.8$ \citep{Rudie2013,Inoue2014}. The combination of these effects results in a typical factor of $\gtrsim6$ decrease in the surface brightness of the blue peak at $z=3.8$ relative to $z=2.3$. These effects will of course be even more significant at $z>4$.  

Given the power of the double-peaked \lya\ profile for constraining the kinematics and column density of the CGM, we therefore expect that integral field observations of galaxies at $z\sim2$ will only grow in importance. As new observations from the James Webb Space Telescope precisely measure the properties of galaxies at both $z\sim2$ and in the reionization era, it will be increasingly possible to robustly identify $z\sim2$ analogs of reionization-era sources and quantify their CGM via spatially extended \lya\ emission.

\begin{acknowledgements}
The authors thank the referee for a detailed and constructive report, and Crystal Martin, Peng Oh and Tim Heckman for helpful discussions. D.K.E.\ is supported by the US National Science Foundation (NSF) through Astronomy \& Astrophysics grant AST-1909198. C.C.S, Z.L.\ and Y.C.\ are supported in part by NSF grant AST-2009278. R.F.T. is a Cottrell Scholar receiving support from the Research Corporation for Science Advancement under grant ID 28289 and from the Pittsburgh Foundation under grant ID UN2021-121482. Some data presented herein were obtained at the W. M. Keck Observatory from telescope time allocated to the National Aeronautics and Space Administration through the agency's scientific partnership with the California Institute of Technology and the University of California. The Observatory was made possible by the generous financial support of the W. M. Keck Foundation. The authors wish to recognize and acknowledge the very significant cultural role and reverence that the summit of Maunakea has always had within the indigenous Hawaiian community. We are most fortunate to have the opportunity to conduct observations from this mountain. D.K.E.\ acknowledges in Milwaukee that we are on traditional Potawatomi, Ho-Chunk and Menominee homeland along the southwest shores of Michigami, North America's largest system of freshwater lakes, where the Milwaukee, Menominee and Kinnickinnic rivers meet and the people of Wisconsin's sovereign Anishinaabe, Ho-Chunk, Menominee, Oneida and Mohican nations remain present.

\end{acknowledgements}

\appendix
\section{Modeling results of the full sample}
\label{sec:appendix}
\begin{figure*}[htbp]
\centerline{\epsfig{angle=00,width=0.8\hsize,file=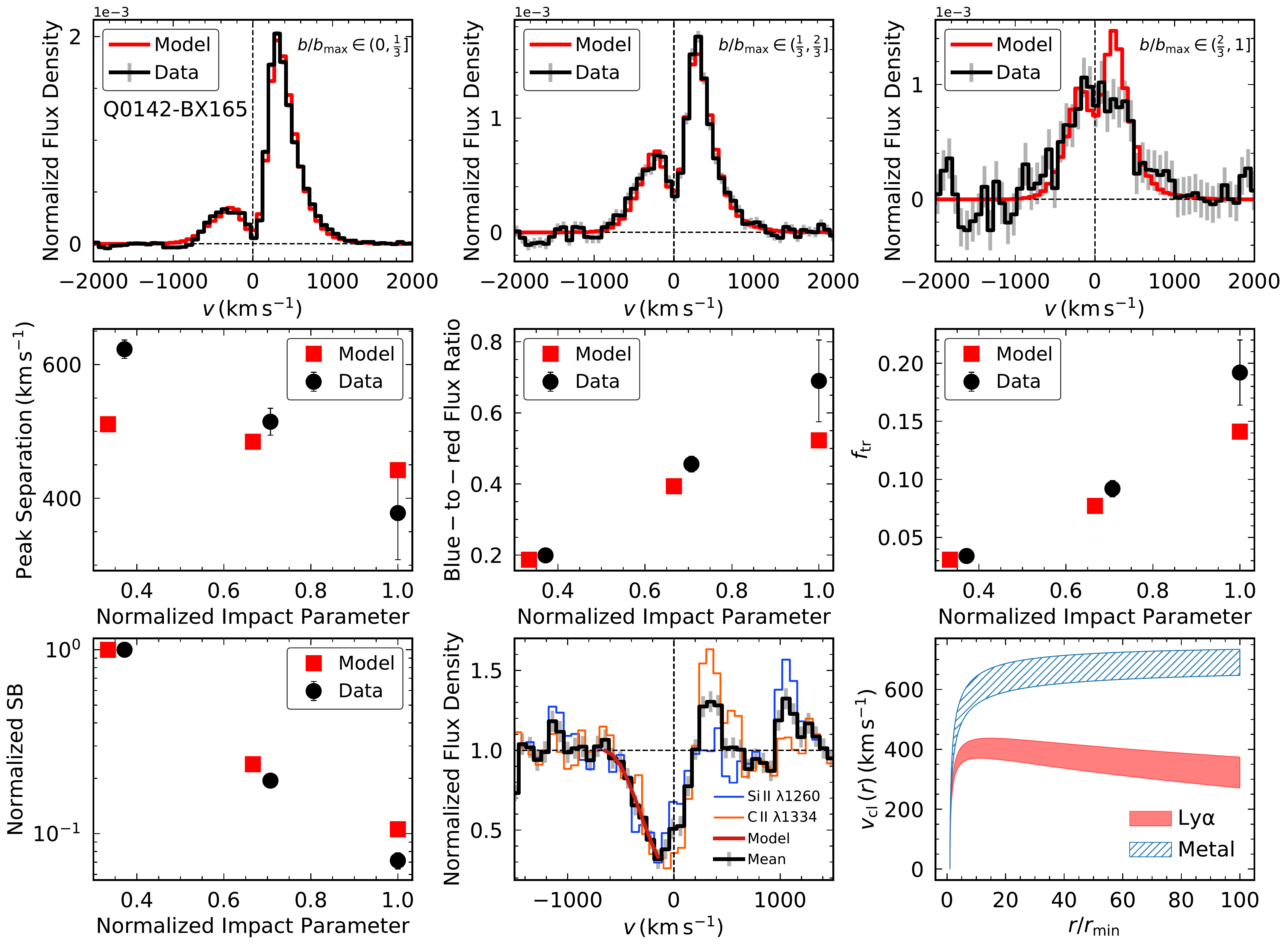}}
\vspace{0.3cm}
\centerline{\epsfig{angle=00,width=0.8\hsize,file=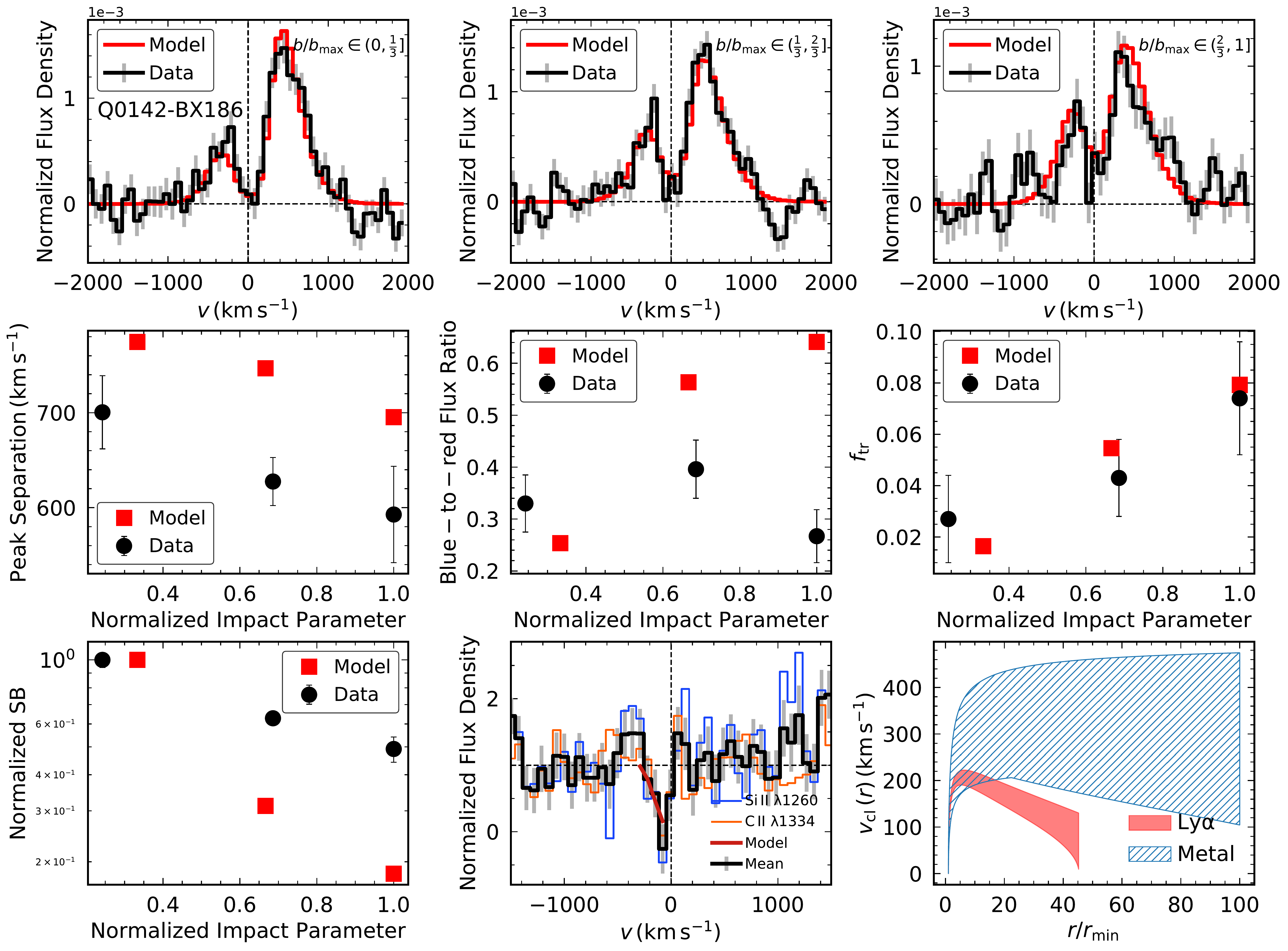}}
\caption{Same as Figure \ref{fig:fits_example}, but for Q0142-BX165 and Q0142-BX186.}
\label{fig:fits1}
\end{figure*}

\begin{figure*}[htbp]
\centerline{\epsfig{angle=00,width=0.8\hsize,file=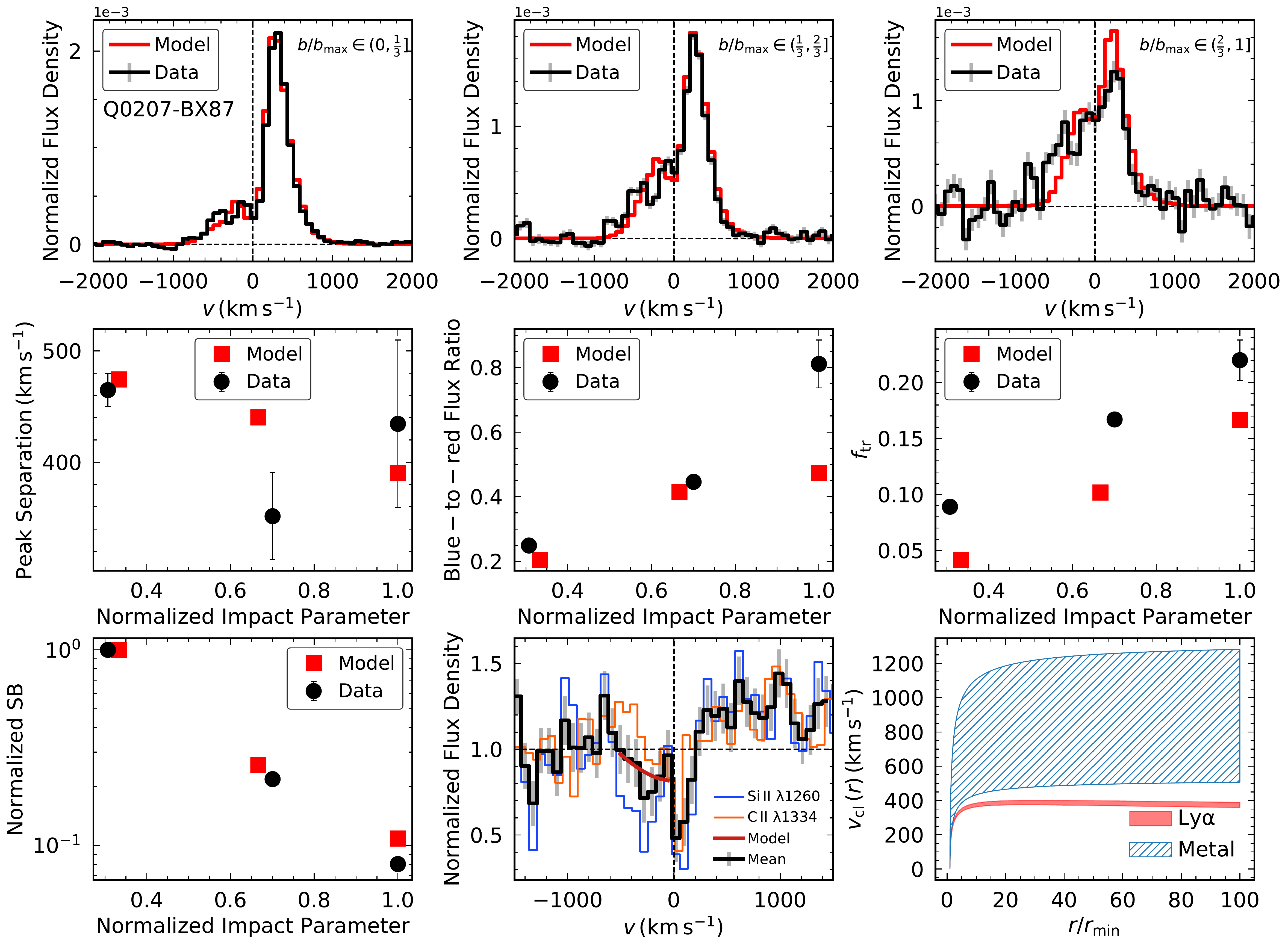}}
\vspace{0.3cm}
\centerline{\epsfig{angle=00,width=0.8\hsize,file=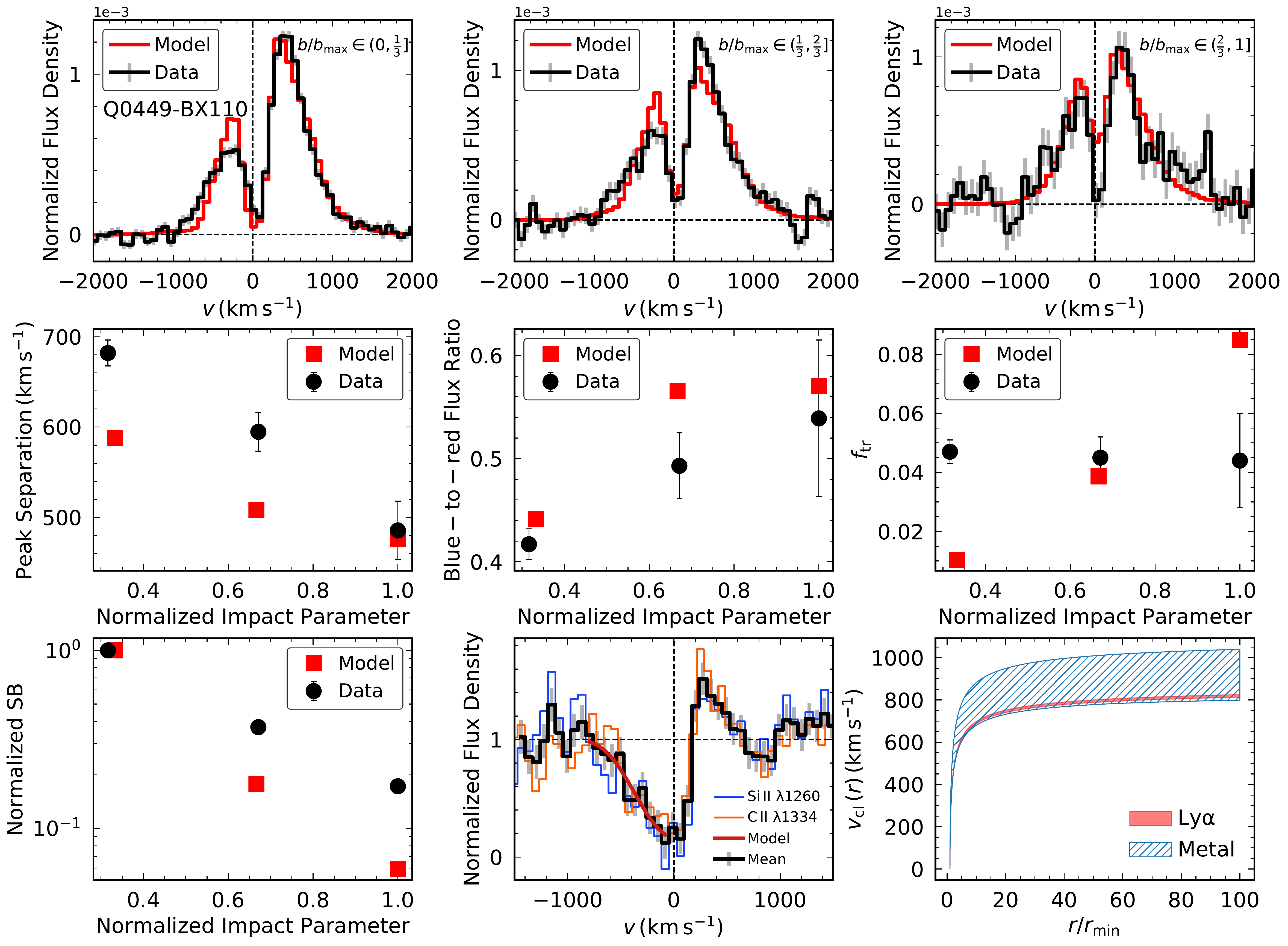}}
\caption{Same as Figure \ref{fig:fits_example}, but for Q0207-BX87 and Q0449-BX110.}
\label{fig:fits2}
\end{figure*}

\begin{figure*}[htbp]
\centerline{\epsfig{angle=00,width=0.8\hsize,file=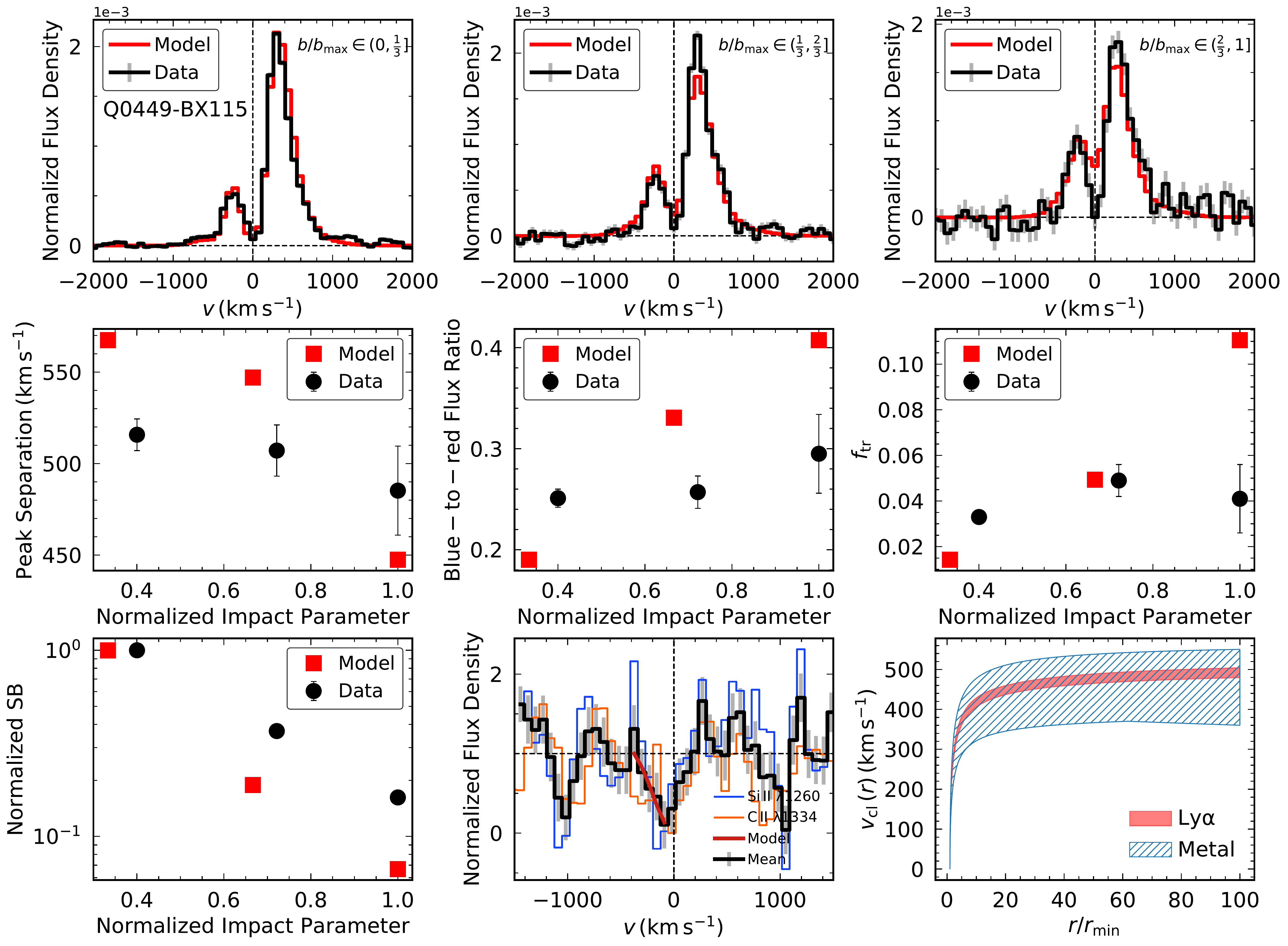}}
\vspace{0.3cm}
\centerline{\epsfig{angle=00,width=0.8\hsize,file=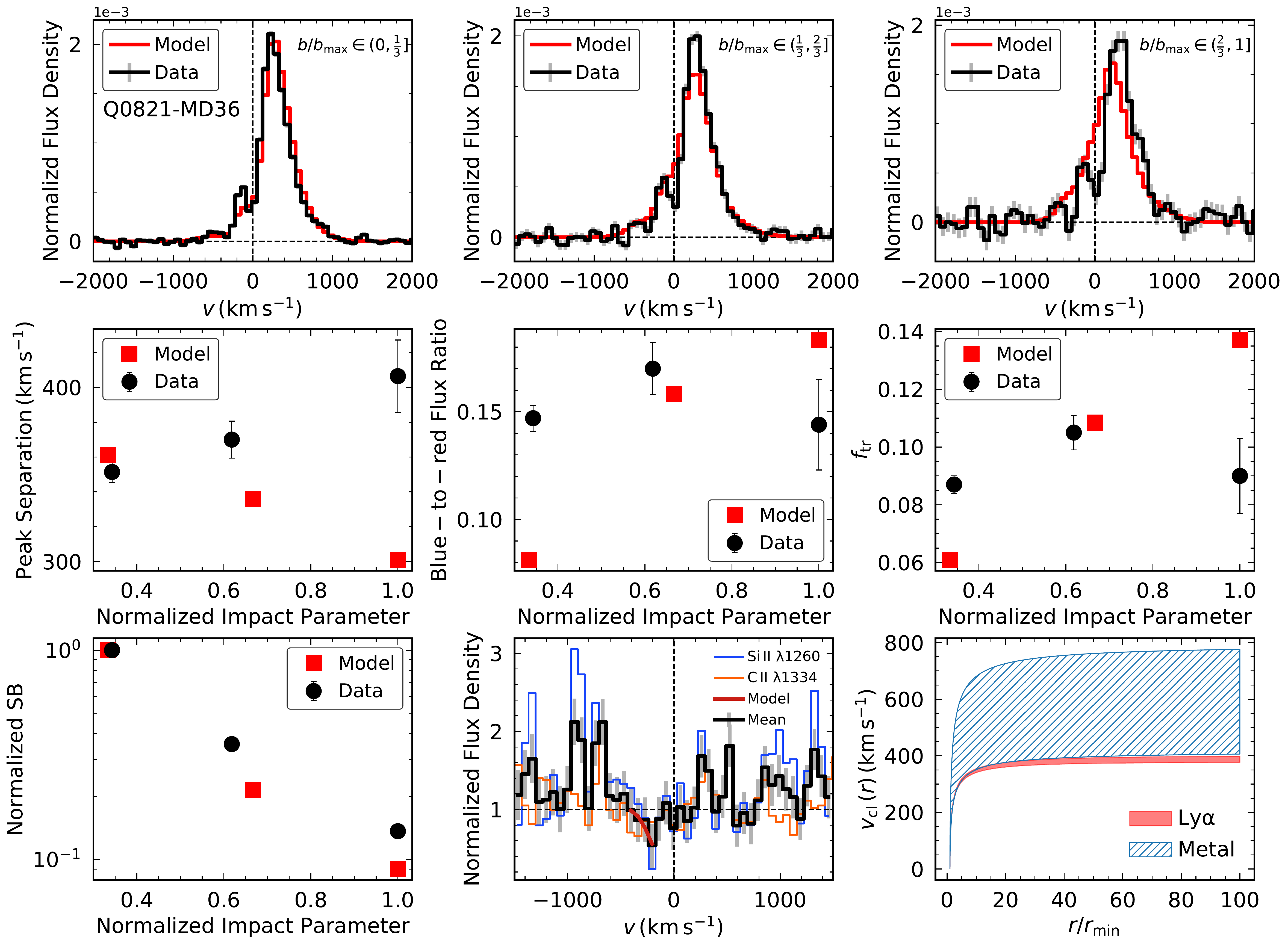}}
\caption{Same as Figure \ref{fig:fits_example}, but for Q0449-BX115 and Q0821-MD36.}
\label{fig:fits3}
\end{figure*}

\begin{figure*}[htbp]
\centerline{\epsfig{angle=00,width=0.8\hsize,file=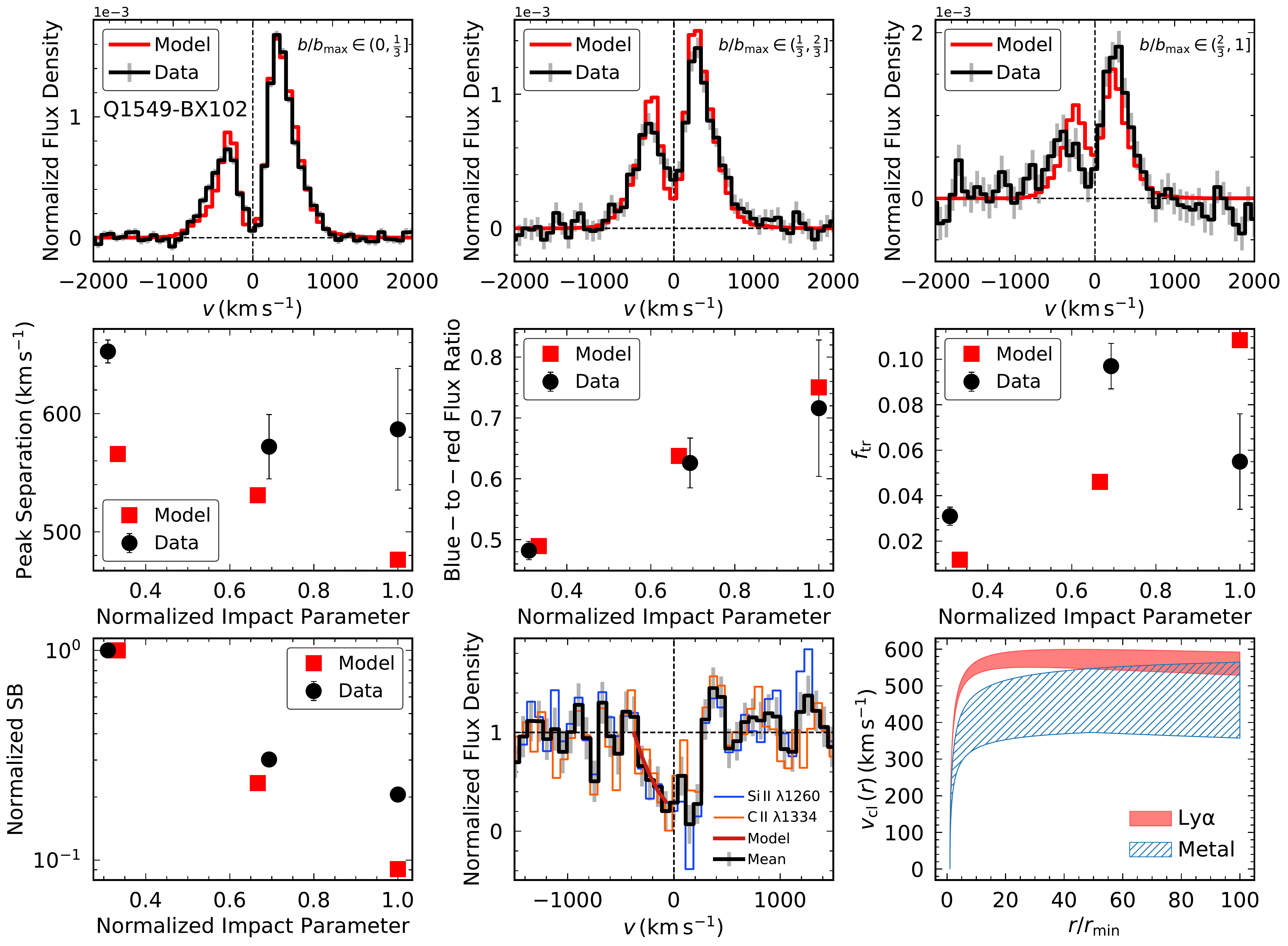}}
\vspace{0.3cm}
\centerline{\epsfig{angle=00,width=0.8\hsize,file=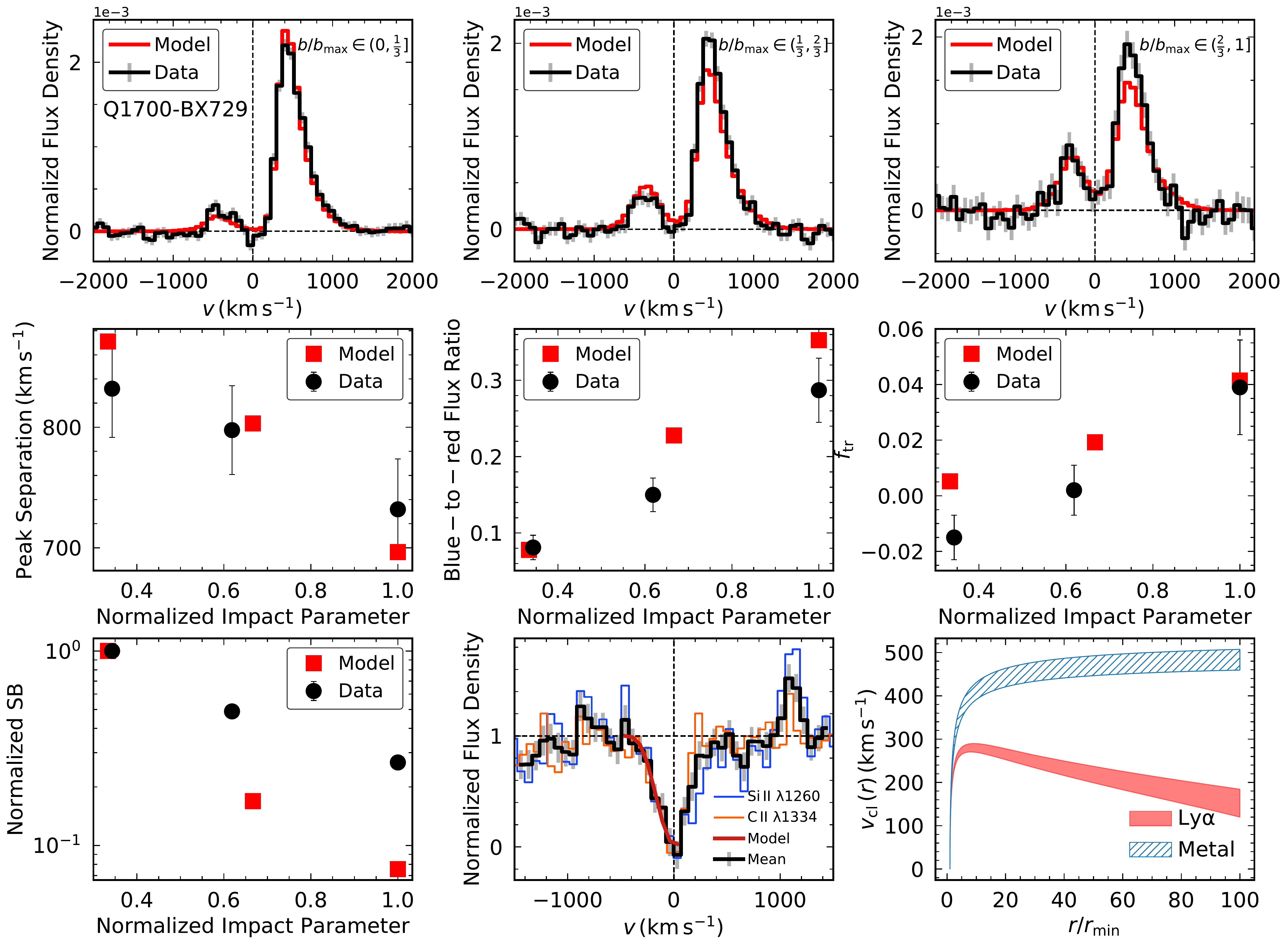}}
\caption{Same as Figure \ref{fig:fits_example}, but for Q1549-BX102 and Q1700-BX729.}
\label{fig:fits4}
\end{figure*}

\begin{figure*}[htbp]
\centerline{\epsfig{angle=00,width=0.8\hsize,file=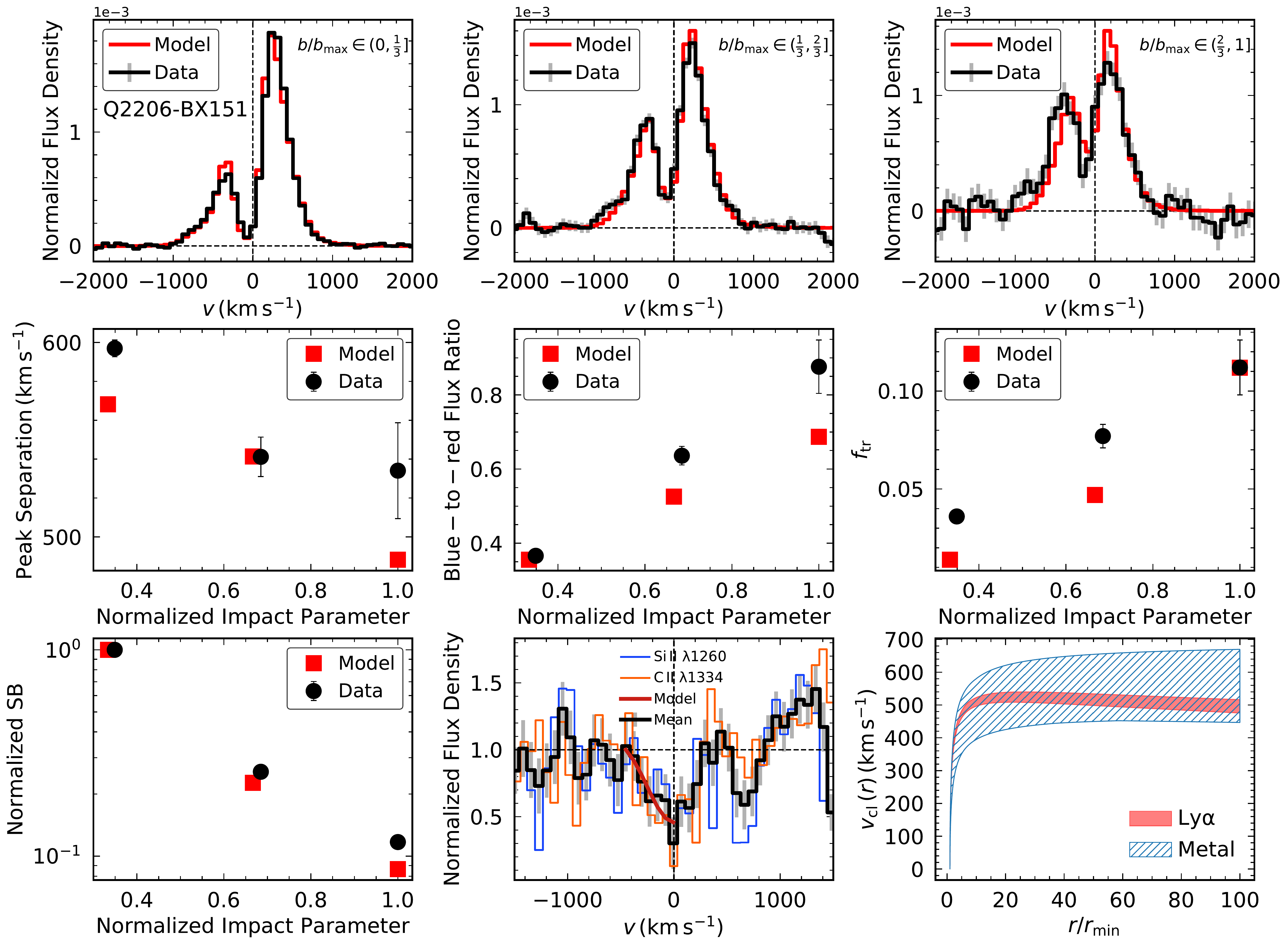}}
\vspace{0.3cm}
\centerline{\epsfig{angle=00,width=0.8\hsize,file=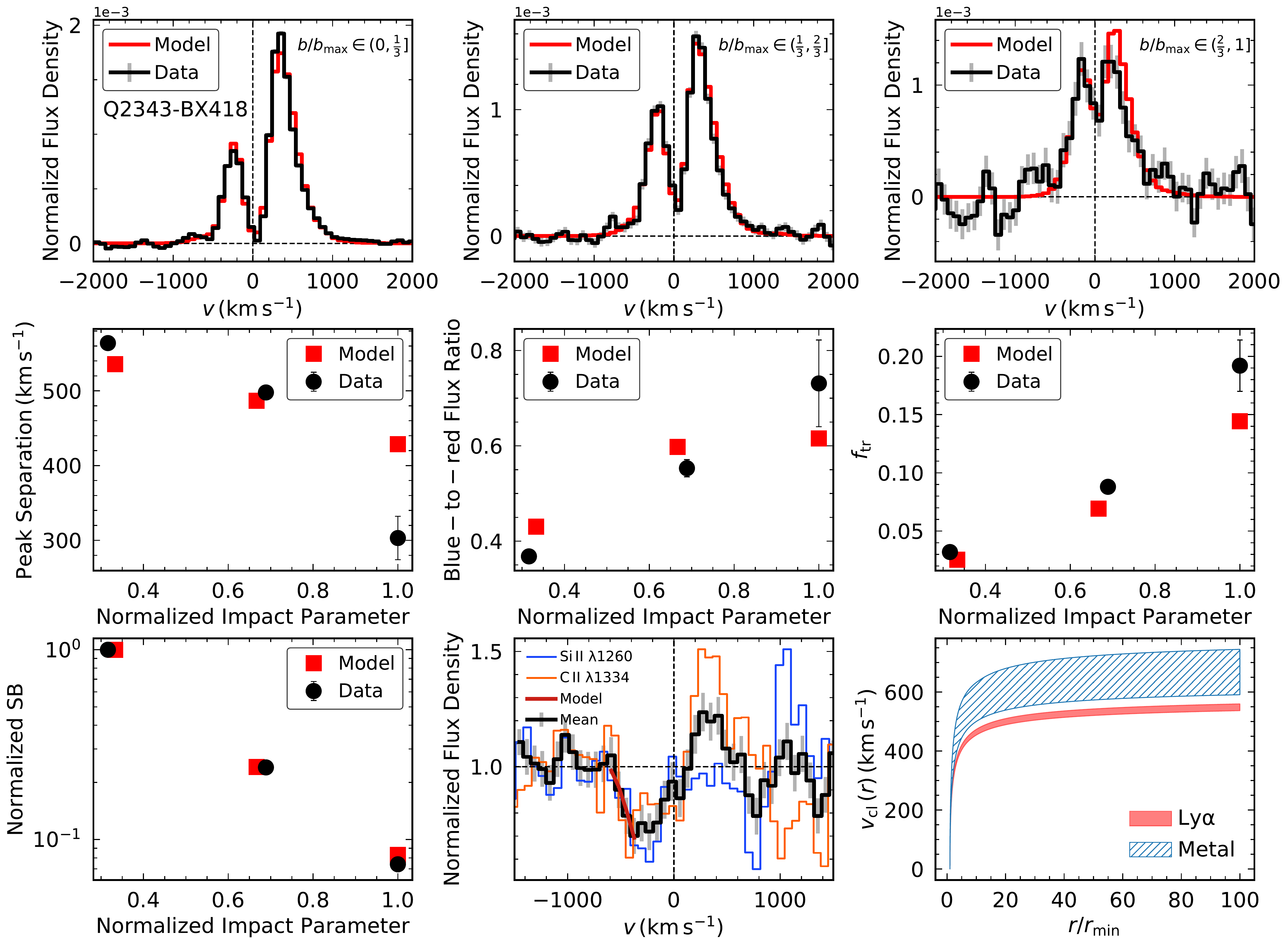}}
\caption{Same as Figure \ref{fig:fits_example}, but for Q2206-BX151 and Q2343-BX418.}
\label{fig:fits5}
\end{figure*}

\begin{figure*}[htbp]
\centerline{\epsfig{angle=00,width=0.8\hsize,file=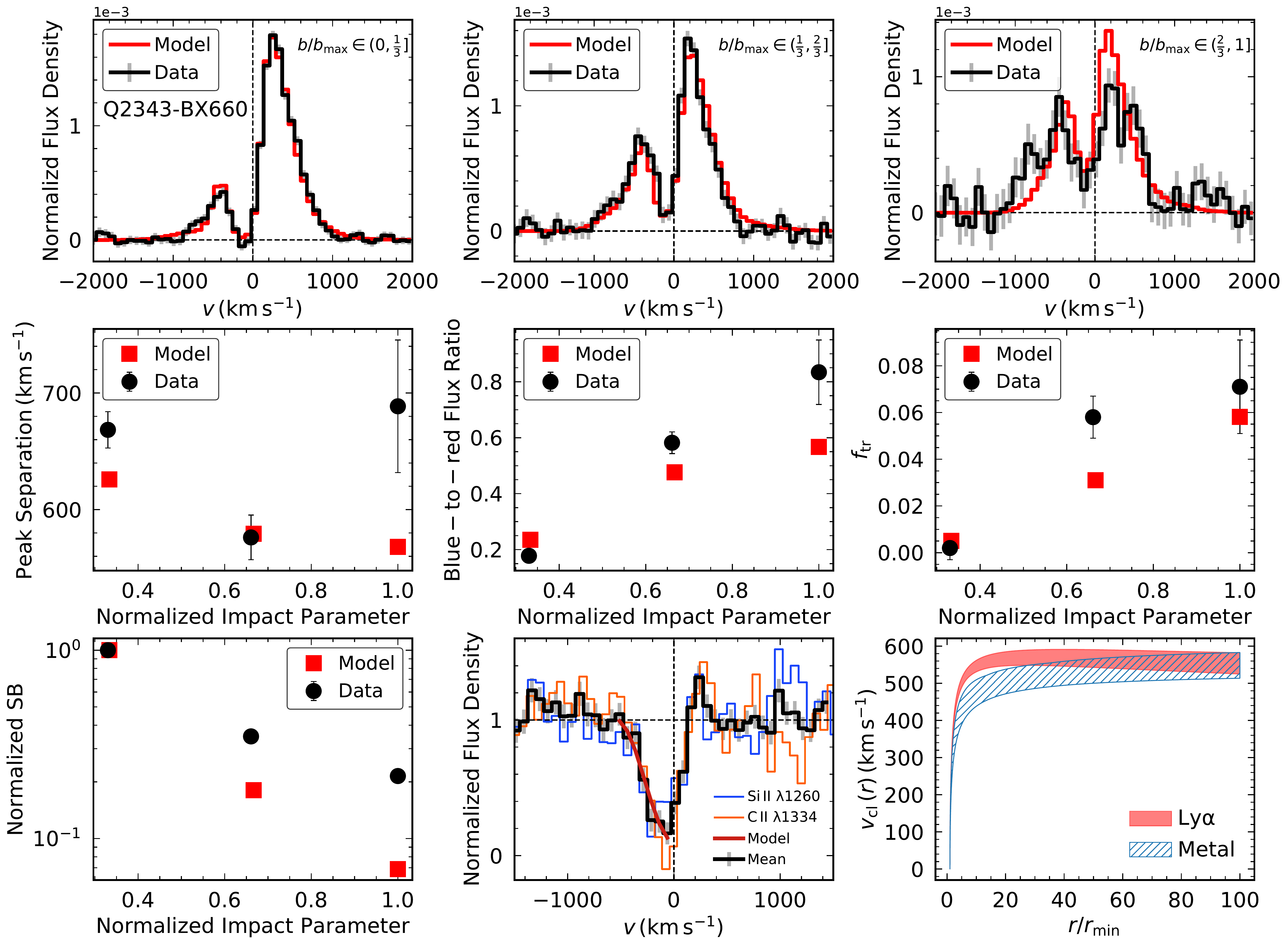}}
\caption{Same as Figure \ref{fig:fits_example}, but for Q2343-BX660.}
\label{fig:fits6}
\end{figure*}

In Figures \ref{fig:fits1} to \ref{fig:fits6} below we present the results of \lya\ and metal absorption line modeling for all objects except Q0207-BX144 (already shown in Figure \ref{fig:fits_example}). In each panel, the top row shows the best-fit RT models (red) to the spatially resolved \lya\ spectra (black); the middle row and the first panel of the bottom row show a comparison between the radial trends of peak separation, peak flux ratio, trough flux fraction, and SB predicted by the best-fit models and measured from observations; and the rest of the bottom row shows the best-fit models (red) to the average metal absorption line profile (black), as well as a comparison of clump radial outflow velocity profiles inferred from \lya\ emission and the average metal absorption line. 

\section{Posterior Distribution Example: Q0207-BX144}
\label{sec:appendix2}
\begin{figure*}[htbp]
\centerline{\epsfig{angle=00,width=\hsize,file=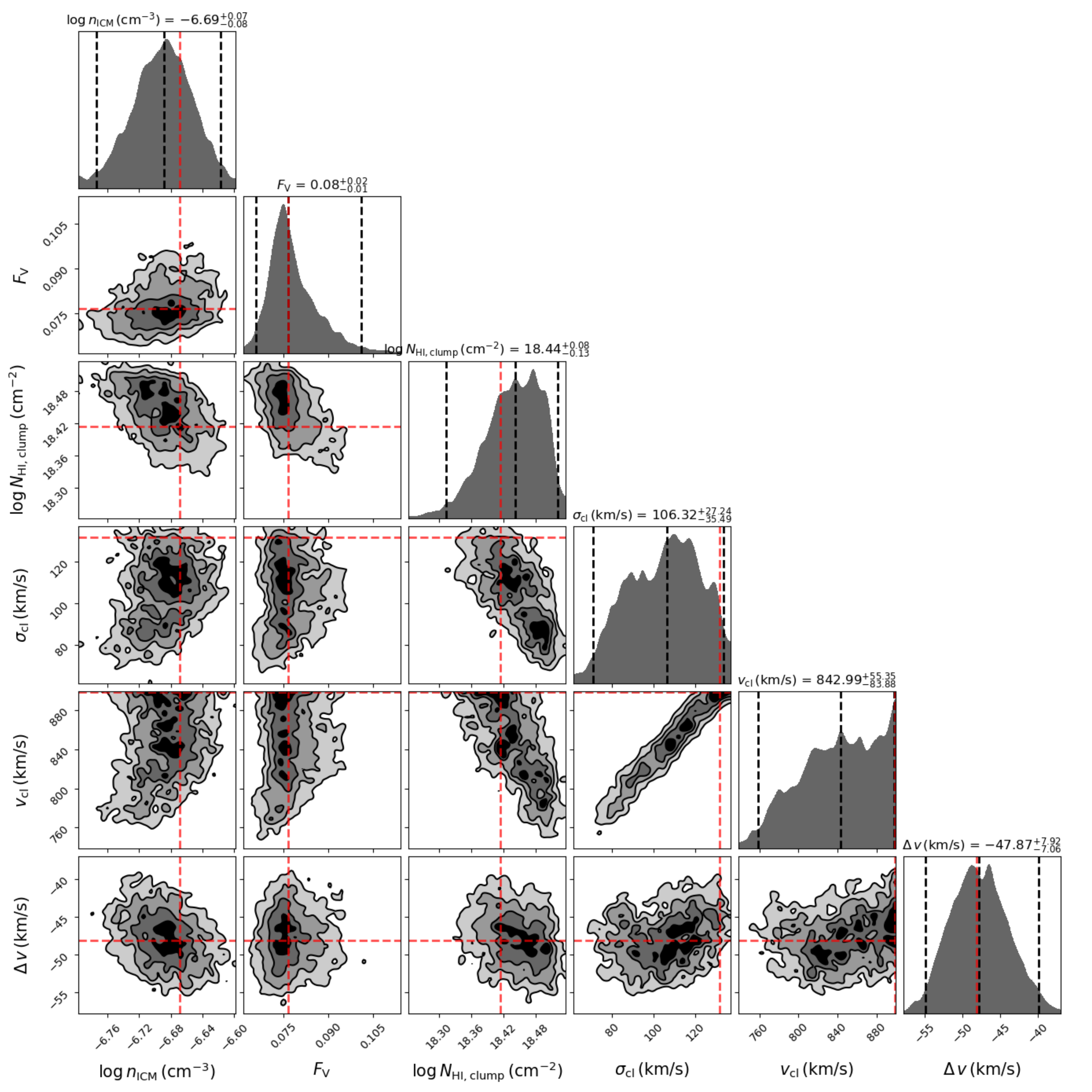}}
\caption{Posterior distribution of spatially-resolved modeling for Q0207-BX144. The [2.5\%, 50\%, 97.5\%] (i.e., 2-$\sigma$ confidence intervals) quantiles of parameters are indicated by vertical black dashed lines, and the maximum likelihood point in the parameter space is indicated by vertical red dashed lines.}
\label{fig:posterior_example}
\end{figure*}

\edit1{As an example of constraints on the model parameters, we present the posterior distribution of the spatially-resolved \lya\ modeling of Q0207-BX144 in Figure \ref{fig:posterior_example}. }

\facility{{Keck II} (KCWI)}
\software{KCWI DRP\footnote{\url{https://github.com/kcwidev/kderp}},
\texttt{astropy} \citep{astropy2013,astropy2018},
\texttt{dynesty} \citep{Skilling04, Skilling06, Speagle20},
\texttt{spectral-cube} \citep{spectralcube},
\texttt{SExtractor} \citep{sextractor},
\texttt{cmasher} \citep{cmasher},
\texttt{seaborn} \citep{seaborn}}

\bibliographystyle{aasjournal}

\begin{thebibliography}{}
\expandafter\ifx\csname natexlab\endcsname\relax\def\natexlab#1{#1}\fi
\providecommand{\url}[1]{\href{#1}{#1}}
\providecommand{\dodoi}[1]{doi:~\href{http://doi.org/#1}{\nolinkurl{#1}}}
\providecommand{\doeprint}[1]{\href{http://ascl.net/#1}{\nolinkurl{http://ascl.net/#1}}}
\providecommand{\doarXiv}[1]{\href{https://arxiv.org/abs/#1}{\nolinkurl{https://arxiv.org/abs/#1}}}

\bibitem[{{Adelberger} {et~al.}(2005){Adelberger}, {Steidel}, {Pettini},
  {Shapley}, {Reddy}, \& {Erb}}]{Adelberger2005}
{Adelberger}, K.~L., {Steidel}, C.~C., {Pettini}, M., {et~al.} 2005, \apj, 619,
  697, \dodoi{10.1086/426580}

\bibitem[{{Astropy Collaboration} {et~al.}(2013){Astropy Collaboration},
  {Robitaille}, {Tollerud}, {Greenfield}, {Droettboom}, {Bray}, {Aldcroft},
  {Davis}, {Ginsburg}, {Price-Whelan}, {Kerzendorf}, {Conley}, {Crighton},
  {Barbary}, {Muna}, {Ferguson}, {Grollier}, {Parikh}, {Nair}, {Unther},
  {Deil}, {Woillez}, {Conseil}, {Kramer}, {Turner}, {Singer}, {Fox}, {Weaver},
  {Zabalza}, {Edwards}, {Azalee Bostroem}, {Burke}, {Casey}, {Crawford},
  {Dencheva}, {Ely}, {Jenness}, {Labrie}, {Lim}, {Pierfederici}, {Pontzen},
  {Ptak}, {Refsdal}, {Servillat}, \& {Streicher}}]{astropy2013}
{Astropy Collaboration}, {Robitaille}, T.~P., {Tollerud}, E.~J., {et~al.} 2013,
  \aap, 558, A33, \dodoi{10.1051/0004-6361/201322068}

\bibitem[{{Astropy Collaboration} {et~al.}(2018){Astropy Collaboration},
  {Price-Whelan}, {Sip{\H{o}}cz}, {G{\"u}nther}, {Lim}, {Crawford}, {Conseil},
  {Shupe}, {Craig}, {Dencheva}, {Ginsburg}, {VanderPlas}, {Bradley},
  {P{\'e}rez-Su{\'a}rez}, {de Val-Borro}, {Aldcroft}, {Cruz}, {Robitaille},
  {Tollerud}, {Ardelean}, {Babej}, {Bach}, {Bachetti}, {Bakanov}, {Bamford},
  {Barentsen}, {Barmby}, {Baumbach}, {Berry}, {Biscani}, {Boquien}, {Bostroem},
  {Bouma}, {Brammer}, {Bray}, {Breytenbach}, {Buddelmeijer}, {Burke},
  {Calderone}, {Cano Rodr{\'\i}guez}, {Cara}, {Cardoso}, {Cheedella}, {Copin},
  {Corrales}, {Crichton}, {D'Avella}, {Deil}, {Depagne}, {Dietrich}, {Donath},
  {Droettboom}, {Earl}, {Erben}, {Fabbro}, {Ferreira}, {Finethy}, {Fox},
  {Garrison}, {Gibbons}, {Goldstein}, {Gommers}, {Greco}, {Greenfield},
  {Groener}, {Grollier}, {Hagen}, {Hirst}, {Homeier}, {Horton}, {Hosseinzadeh},
  {Hu}, {Hunkeler}, {Ivezi{\'c}}, {Jain}, {Jenness}, {Kanarek}, {Kendrew},
  {Kern}, {Kerzendorf}, {Khvalko}, {King}, {Kirkby}, {Kulkarni}, {Kumar},
  {Lee}, {Lenz}, {Littlefair}, {Ma}, {Macleod}, {Mastropietro}, {McCully},
  {Montagnac}, {Morris}, {Mueller}, {Mumford}, {Muna}, {Murphy}, {Nelson},
  {Nguyen}, {Ninan}, {N{\"o}the}, {Ogaz}, {Oh}, {Parejko}, {Parley}, {Pascual},
  {Patil}, {Patil}, {Plunkett}, {Prochaska}, {Rastogi}, {Reddy Janga},
  {Sabater}, {Sakurikar}, {Seifert}, {Sherbert}, {Sherwood-Taylor}, {Shih},
  {Sick}, {Silbiger}, {Singanamalla}, {Singer}, {Sladen}, {Sooley},
  {Sornarajah}, {Streicher}, {Teuben}, {Thomas}, {Tremblay}, {Turner},
  {Terr{\'o}n}, {van Kerkwijk}, {de la Vega}, {Watkins}, {Weaver}, {Whitmore},
  {Woillez}, {Zabalza}, \& {Astropy Contributors}}]{astropy2018}
{Astropy Collaboration}, {Price-Whelan}, A.~M., {Sip{\H{o}}cz}, B.~M., {et~al.}
  2018, \aj, 156, 123, \dodoi{10.3847/1538-3881/aabc4f}

\bibitem[{{Bacon} {et~al.}(2010){Bacon}, {Accardo}, {Adjali}, {Anwand},
  {Bauer}, {Biswas}, {Blaizot}, {Boudon}, {Brau-Nogue}, {Brinchmann},
  {Caillier}, {Capoani}, {Carollo}, {Contini}, {Couderc}, {Daguis{\'e}},
  {Deiries}, {Delabre}, {Dreizler}, {Dubois}, {Dupieux}, {Dupuy}, {Emsellem},
  {Fechner}, {Fleischmann}, {Fran{\c{c}}ois}, {Gallou}, {Gharsa}, {Glindemann},
  {Gojak}, {Guiderdoni}, {Hansali}, {Hahn}, {Jarno}, {Kelz}, {Koehler},
  {Kosmalski}, {Laurent}, {Le Floch}, {Lilly}, {Lizon}, {Loupias}, {Manescau},
  {Monstein}, {Nicklas}, {Olaya}, {Pares}, {Pasquini}, {P{\'e}contal-Rousset},
  {Pell{\'o}}, {Petit}, {Popow}, {Reiss}, {Remillieux}, {Renault}, {Roth},
  {Rupprecht}, {Serre}, {Schaye}, {Soucail}, {Steinmetz}, {Streicher}, {Stuik},
  {Valentin}, {Vernet}, {Weilbacher}, {Wisotzki}, \& {Yerle}}]{muse}
{Bacon}, R., {Accardo}, M., {Adjali}, L., {et~al.} 2010, in Society of
  Photo-Optical Instrumentation Engineers (SPIE) Conference Series, Vol. 7735,
  Ground-based and Airborne Instrumentation for Astronomy III, ed. I.~S.
  {McLean}, S.~K. {Ramsay}, \& H.~{Takami}, 773508, \dodoi{10.1117/12.856027}

\bibitem[{{Baldwin} {et~al.}(1981){Baldwin}, {Phillips}, \& {Terlevich}}]{bpt}
{Baldwin}, J.~A., {Phillips}, M.~M., \& {Terlevich}, R. 1981, \pasp, 93, 5,
  \dodoi{10.1086/130766}

\bibitem[{{Bertin} \& {Arnouts}(1996)}]{sextractor}
{Bertin}, E., \& {Arnouts}, S. 1996, \aaps, 117, 393,
  \dodoi{10.1051/aas:1996164}

\bibitem[{{Byrohl} {et~al.}(2021){Byrohl}, {Nelson}, {Behrens}, {Kostyuk},
  {Glatzle}, {Pillepich}, {Hernquist}, {Marinacci}, \&
  {Vogelsberger}}]{Byrohl2021}
{Byrohl}, C., {Nelson}, D., {Behrens}, C., {et~al.} 2021, \mnras,
  \dodoi{10.1093/mnras/stab1958}

\bibitem[{{Cantalupo} {et~al.}(2014){Cantalupo}, {Arrigoni-Battaia},
  {Prochaska}, {Hennawi}, \& {Madau}}]{Cantalupo2014}
{Cantalupo}, S., {Arrigoni-Battaia}, F., {Prochaska}, J.~X., {Hennawi}, J.~F.,
  \& {Madau}, P. 2014, \nat, 506, 63, \dodoi{10.1038/nature12898}

\bibitem[{{Cantalupo} {et~al.}(2012){Cantalupo}, {Lilly}, \&
  {Haehnelt}}]{Cantalupo2012}
{Cantalupo}, S., {Lilly}, S.~J., \& {Haehnelt}, M.~G. 2012, \mnras, 425, 1992,
  \dodoi{10.1111/j.1365-2966.2012.21529.x}

\bibitem[{{Cappellari} \& {Copin}(2003)}]{Cappellari2003}
{Cappellari}, M., \& {Copin}, Y. 2003, MNRAS, 342, 345,
  \dodoi{10.1046/j.1365-8711.2003.06541.x}

\bibitem[{{Charlot} \& {Fall}(1993)}]{Charlot1993}
{Charlot}, S., \& {Fall}, S.~M. 1993, \apj, 415, 580, \dodoi{10.1086/173187}

\bibitem[{{Chen} {et~al.}(2020){Chen}, {Steidel}, {Hummels}, {Rudie}, {Dong},
  {Trainor}, {Bogosavljevi{\'c}}, {Erb}, {Pettini}, {Reddy}, {Shapley},
  {Strom}, {Theios}, {Faucher-Gigu{\`e}re}, {Hopkins}, \&
  {Kere{\v{s}}}}]{Chen2020}
{Chen}, Y., {Steidel}, C.~C., {Hummels}, C.~B., {et~al.} 2020, \mnras, 499,
  1721, \dodoi{10.1093/mnras/staa2808}

\bibitem[{{Chen} {et~al.}(2021){Chen}, {Steidel}, {Erb}, {Law}, {Trainor},
  {Reddy}, {Shapley}, {Pahl}, {Strom}, {Li}, \& {Rudie}}]{Chen2021}
{Chen}, Y., {Steidel}, C.~C., {Erb}, D.~K., {et~al.} 2021, arXiv e-prints,
  arXiv:2104.10173.
\newblock \doarXiv{2104.10173}

\bibitem[{{Chen} {et~al.}(2010){Chen}, {Tremonti}, {Heckman}, {Kauffmann},
  {Weiner}, {Brinchmann}, \& {Wang}}]{Chen10}
{Chen}, Y.-M., {Tremonti}, C.~A., {Heckman}, T.~M., {et~al.} 2010, \aj, 140,
  445, \dodoi{10.1088/0004-6256/140/2/445}

\bibitem[{{Chisholm} {et~al.}(2015){Chisholm}, {Tremonti}, {Leitherer}, {Chen},
  {Wofford}, \& {Lundgren}}]{Chisholm15}
{Chisholm}, J., {Tremonti}, C.~A., {Leitherer}, C., {et~al.} 2015, \apj, 811,
  149, \dodoi{10.1088/0004-637X/811/2/149}

\bibitem[{{Chisholm} {et~al.}(2016){Chisholm}, {Tremonti Christy}, {Leitherer},
  \& {Chen}}]{Chisholm16}
{Chisholm}, J., {Tremonti Christy}, A., {Leitherer}, C., \& {Chen}, Y. 2016,
  \mnras, 463, 541, \dodoi{10.1093/mnras/stw1951}

\bibitem[{{Chisholm} {et~al.}(2018){Chisholm}, {Gazagnes}, {Schaerer},
  {Verhamme}, {Rigby}, {Bayliss}, {Sharon}, {Gladders}, \&
  {Dahle}}]{Chisholm2018}
{Chisholm}, J., {Gazagnes}, S., {Schaerer}, D., {et~al.} 2018, \aap, 616, A30,
  \dodoi{10.1051/0004-6361/201832758}

\bibitem[{{Chung} {et~al.}(2019){Chung}, {Dijkstra}, {Ciardi}, {Kakiichi}, \&
  {Naab}}]{Chung2019}
{Chung}, A.~S., {Dijkstra}, M., {Ciardi}, B., {Kakiichi}, K., \& {Naab}, T.
  2019, \mnras, 484, 2420, \dodoi{10.1093/mnras/stz149}

\bibitem[{Claeyssens {et~al.}(2019)Claeyssens, Richard, Blaizot, Garel,
  Leclercq, Patrício, Verhamme, Wisotzki, Bacon, Carton, Clément, Herenz,
  Marino, Muzahid, Saust, \& Schaye}]{Claeyssens2019}
Claeyssens, A., Richard, J., Blaizot, J., {et~al.} 2019, Monthly Notices of the
  Royal Astronomical Society, 489, 5022 , \dodoi{10.1093/mnras/stz2492}

\bibitem[{{Conroy} {et~al.}(2008){Conroy}, {Shapley}, {Tinker}, {Santos}, \&
  {Lemson}}]{Conroy2008}
{Conroy}, C., {Shapley}, A.~E., {Tinker}, J.~L., {Santos}, M.~R., \& {Lemson},
  G. 2008, \apj, 679, 1192, \dodoi{10.1086/587834}

\bibitem[{Daddi {et~al.}(2021)Daddi, Valentino, Rich, Neill, Gronke,
  O’Sullivan, Elbaz, Bournaud, Finoguenov, Marchal, Delvecchio, Jin, Liu,
  Strazzullo, Calabro, Coogan, D’Eugenio, Gobat, Kalita, Laursen, Martin,
  Puglisi, Schinnerer, \& Wang}]{Daddi2021}
Daddi, E., Valentino, F., Rich, R.~M., {et~al.} 2021, Astronomy \&
  Astrophysics, 649, A78, \dodoi{10.1051/0004-6361/202038700}

\bibitem[{{Dekel} \& {Birnboim}(2006)}]{Dekel2006}
{Dekel}, A., \& {Birnboim}, Y. 2006, \mnras, 368, 2,
  \dodoi{10.1111/j.1365-2966.2006.10145.x}

\bibitem[{{Dijkstra}(2014{\natexlab{a}})}]{Dijkstra2014}
{Dijkstra}, M. 2014{\natexlab{a}}, \pasa, 31, e040,
  \dodoi{10.1017/pasa.2014.33}

\bibitem[{{Dijkstra}(2014{\natexlab{b}})}]{Dijkstra14}
---. 2014{\natexlab{b}}, \pasa, 31, e040, \dodoi{10.1017/pasa.2014.33}

\bibitem[{{Dijkstra} {et~al.}(2016){Dijkstra}, {Gronke}, \&
  {Venkatesan}}]{Dijkstra2016}
{Dijkstra}, M., {Gronke}, M., \& {Venkatesan}, A. 2016, \apj, 828, 71,
  \dodoi{10.3847/0004-637X/828/2/71}

\bibitem[{{Dijkstra} \& {Kramer}(2012)}]{Dijkstra12}
{Dijkstra}, M., \& {Kramer}, R. 2012, \mnras, 424, 1672,
  \dodoi{10.1111/j.1365-2966.2012.21131.x}

\bibitem[{{Dijkstra} \& {Loeb}(2009)}]{Dijkstra2009}
{Dijkstra}, M., \& {Loeb}, A. 2009, \mnras, 400, 1109,
  \dodoi{10.1111/j.1365-2966.2009.15533.x}

\bibitem[{{Duval} {et~al.}(2014){Duval}, {Schaerer}, {{\"O}stlin}, \&
  {Laursen}}]{Duval14}
{Duval}, F., {Schaerer}, D., {{\"O}stlin}, G., \& {Laursen}, P. 2014, \aap,
  562, A52, \dodoi{10.1051/0004-6361/201220455}

\bibitem[{{Erb} {et~al.}(2016){Erb}, {Pettini}, {Steidel}, {Strom}, {Rudie},
  {Trainor}, {Shapley}, \& {Reddy}}]{Erb16}
{Erb}, D.~K., {Pettini}, M., {Steidel}, C.~C., {et~al.} 2016, \apj, 830, 52,
  \dodoi{10.3847/0004-637X/830/1/52}

\bibitem[{{Erb} {et~al.}(2018){Erb}, {Steidel}, \& {Chen}}]{Erb18}
{Erb}, D.~K., {Steidel}, C.~C., \& {Chen}, Y. 2018, \apjl, 862, L10,
  \dodoi{10.3847/2041-8213/aacff6}

\bibitem[{{Erb} {et~al.}(2006){Erb}, {Steidel}, {Shapley}, {Pettini}, {Reddy},
  \& {Adelberger}}]{Erb2006}
{Erb}, D.~K., {Steidel}, C.~C., {Shapley}, A.~E., {et~al.} 2006, \apj, 647,
  128, \dodoi{10.1086/505341}

\bibitem[{{Faucher-Gigu{\`e}re} {et~al.}(2010){Faucher-Gigu{\`e}re},
  {Kere{\v{s}}}, {Dijkstra}, {Hernquist}, \&
  {Zaldarriaga}}]{Faucher-Giguere2010}
{Faucher-Gigu{\`e}re}, C.-A., {Kere{\v{s}}}, D., {Dijkstra}, M., {Hernquist},
  L., \& {Zaldarriaga}, M. 2010, \apj, 725, 633,
  \dodoi{10.1088/0004-637X/725/1/633}

\bibitem[{{Fielding} \& {Bryan}(2022)}]{Fielding22}
{Fielding}, D.~B., \& {Bryan}, G.~L. 2022, \apj, 924, 82,
  \dodoi{10.3847/1538-4357/ac2f41}

\bibitem[{{Fynbo} {et~al.}(1999){Fynbo}, {M{\o}ller}, \& {Warren}}]{Fynbo1999}
{Fynbo}, J.~U., {M{\o}ller}, P., \& {Warren}, S.~J. 1999, \mnras, 305, 849,
  \dodoi{10.1046/j.1365-8711.1999.02520.x}

\bibitem[{Ginsburg {et~al.}(2019)Ginsburg, Koch, Robitaille, Beaumont,
  adamginsburg, Sipőcz, ZuHone, Patra, Jones, Lim, Stern, Rosolowsky, Earl,
  de~Val-Borro, jrobbfed, shuokong, Kepley, Sokolov, Badger, Maret, Garrido,
  Booker, \& Tollerud}]{spectralcube}
Ginsburg, A., Koch, E., Robitaille, T., {et~al.} 2019,
  radio-astro-tools/spectral-cube: Release v0.4.5, v0.4.5,  Zenodo,
  \dodoi{10.5281/zenodo.3558614}

\bibitem[{{Girelli} {et~al.}(2020){Girelli}, {Pozzetti}, {Bolzonella},
  {Giocoli}, {Marulli}, \& {Baldi}}]{Girelli2020}
{Girelli}, G., {Pozzetti}, L., {Bolzonella}, M., {et~al.} 2020, \aap, 634,
  A135, \dodoi{10.1051/0004-6361/201936329}

\bibitem[{{Gordon} {et~al.}(2003){Gordon}, {Clayton}, {Misselt}, {Landolt}, \&
  {Wolff}}]{Gordon2003}
{Gordon}, K.~D., {Clayton}, G.~C., {Misselt}, K.~A., {Landolt}, A.~U., \&
  {Wolff}, M.~J. 2003, \apj, 594, 279, \dodoi{10.1086/376774}

\bibitem[{{Gronke}(2017)}]{Gronke2017}
{Gronke}, M. 2017, \aap, 608, A139, \dodoi{10.1051/0004-6361/201731791}

\bibitem[{{Gronke} \& {Dijkstra}(2016)}]{Gronke2016a}
{Gronke}, M., \& {Dijkstra}, M. 2016, \apj, 826, 14,
  \dodoi{10.3847/0004-637X/826/1/14}

\bibitem[{{Gronke} {et~al.}(2016){Gronke}, {Dijkstra}, {McCourt}, \&
  {Oh}}]{Gronke2016b}
{Gronke}, M., {Dijkstra}, M., {McCourt}, M., \& {Oh}, S.~P. 2016, \apjl, 833,
  L26, \dodoi{10.3847/2041-8213/833/2/L26}

\bibitem[{{Gronke} {et~al.}(2017){Gronke}, {Dijkstra}, {McCourt}, \&
  {Oh}}]{Gronke17}
---. 2017, \aap, 607, A71, \dodoi{10.1051/0004-6361/201731013}

\bibitem[{{Gronke} \& {Oh}(2018)}]{Gronke18}
{Gronke}, M., \& {Oh}, S.~P. 2018, \mnras, 480, L111,
  \dodoi{10.1093/mnrasl/sly131}

\bibitem[{{Gronke} \& {Oh}(2020)}]{Gronke20}
---. 2020, \mnras, 492, 1970, \dodoi{10.1093/mnras/stz3332}

\bibitem[{{Haiman} {et~al.}(2000){Haiman}, {Spaans}, \&
  {Quataert}}]{Haiman2000}
{Haiman}, Z., {Spaans}, M., \& {Quataert}, E. 2000, \apjl, 537, L5,
  \dodoi{10.1086/312754}

\bibitem[{{Hansen} \& {Oh}(2006)}]{Hansen06}
{Hansen}, M., \& {Oh}, S.~P. 2006, \mnras, 367, 979,
  \dodoi{10.1111/j.1365-2966.2005.09870.x}

\bibitem[{{Hashimoto} {et~al.}(2015){Hashimoto}, {Verhamme}, {Ouchi},
  {Shimasaku}, {Schaerer}, {Nakajima}, {Shibuya}, {Rauch}, {Ono}, \&
  {Goto}}]{Hashimoto2015}
{Hashimoto}, T., {Verhamme}, A., {Ouchi}, M., {et~al.} 2015, \apj, 812, 157,
  \dodoi{10.1088/0004-637X/812/2/157}

\bibitem[{{Hayes}(2015)}]{Hayes2015}
{Hayes}, M. 2015, \pasa, 32, e027, \dodoi{10.1017/pasa.2015.25}

\bibitem[{{Hayes} {et~al.}(2021){Hayes}, {Runnholm}, {Gronke}, \&
  {Scarlata}}]{Hayes2021}
{Hayes}, M.~J., {Runnholm}, A., {Gronke}, M., \& {Scarlata}, C. 2021, \apj,
  908, 36, \dodoi{10.3847/1538-4357/abd246}

\bibitem[{{Heckman} {et~al.}(2015){Heckman}, {Alexandroff}, {Borthakur},
  {Overzier}, \& {Leitherer}}]{Heckman2015}
{Heckman}, T.~M., {Alexandroff}, R.~M., {Borthakur}, S., {Overzier}, R., \&
  {Leitherer}, C. 2015, \apj, 809, 147, \dodoi{10.1088/0004-637X/809/2/147}

\bibitem[{{Heckman} {et~al.}(2011){Heckman}, {Borthakur}, {Overzier},
  {Kauffmann}, {Basu-Zych}, {Leitherer}, {Sembach}, {Martin}, {Rich},
  {Schiminovich}, \& {Seibert}}]{Heckman2011}
{Heckman}, T.~M., {Borthakur}, S., {Overzier}, R., {et~al.} 2011, \apj, 730, 5,
  \dodoi{10.1088/0004-637X/730/1/5}

\bibitem[{{Henry} {et~al.}(2015){Henry}, {Scarlata}, {Martin}, \&
  {Erb}}]{Henry2015}
{Henry}, A., {Scarlata}, C., {Martin}, C.~L., \& {Erb}, D. 2015, \apj, 809, 19,
  \dodoi{10.1088/0004-637X/809/1/19}

\bibitem[{{Inoue} {et~al.}(2014){Inoue}, {Shimizu}, {Iwata}, \&
  {Tanaka}}]{Inoue2014}
{Inoue}, A.~K., {Shimizu}, I., {Iwata}, I., \& {Tanaka}, M. 2014, \mnras, 442,
  1805, \dodoi{10.1093/mnras/stu936}

\bibitem[{{Izotov} {et~al.}(2021){Izotov}, {Worseck}, {Schaerer}, {Guseva},
  {Chisholm}, {Thuan}, {Fricke}, \& {Verhamme}}]{Izotov2021}
{Izotov}, Y.~I., {Worseck}, G., {Schaerer}, D., {et~al.} 2021, \mnras, 503,
  1734, \dodoi{10.1093/mnras/stab612}

\bibitem[{Izotov {et~al.}(2018)Izotov, Worseck, Schaerer, Guseva, Thuan,
  Fricke, \& Orlitova}]{Izotov2018}
Izotov, Y.~I., Worseck, G., Schaerer, D., {et~al.} 2018, Monthly Notices of the
  Royal Astronomical Society, 478, 4851 , \dodoi{10.1093/mnras/sty1378}

\bibitem[{{Kere{\v{s}}} {et~al.}(2005){Kere{\v{s}}}, {Katz}, {Weinberg}, \&
  {Dav{\'e}}}]{Keres2005}
{Kere{\v{s}}}, D., {Katz}, N., {Weinberg}, D.~H., \& {Dav{\'e}}, R. 2005,
  \mnras, 363, 2, \dodoi{10.1111/j.1365-2966.2005.09451.x}

\bibitem[{{Kollmeier} {et~al.}(2010){Kollmeier}, {Zheng}, {Dav{\'e}}, {Gould},
  {Katz}, {Miralda-Escud{\'e}}, \& {Weinberg}}]{Kollmeier2010}
{Kollmeier}, J.~A., {Zheng}, Z., {Dav{\'e}}, R., {et~al.} 2010, \apj, 708,
  1048, \dodoi{10.1088/0004-637X/708/2/1048}

\bibitem[{{Kornei} {et~al.}(2010){Kornei}, {Shapley}, {Erb}, {Steidel},
  {Reddy}, {Pettini}, \& {Bogosavljevi{\'c}}}]{Kornei2010}
{Kornei}, K.~A., {Shapley}, A.~E., {Erb}, D.~K., {et~al.} 2010, \apj, 711, 693,
  \dodoi{10.1088/0004-637X/711/2/693}

\bibitem[{{Kuhlen} \& {Faucher-Gigu{\`e}re}(2012)}]{Kuhlen12}
{Kuhlen}, M., \& {Faucher-Gigu{\`e}re}, C.-A. 2012, \mnras, 423, 862,
  \dodoi{10.1111/j.1365-2966.2012.20924.x}

\bibitem[{{Kulas} {et~al.}(2012){Kulas}, {Shapley}, {Kollmeier}, {Zheng},
  {Steidel}, \& {Hainline}}]{Kulas2012}
{Kulas}, K.~R., {Shapley}, A.~E., {Kollmeier}, J.~A., {et~al.} 2012, \apj, 745,
  33, \dodoi{10.1088/0004-637X/745/1/33}

\bibitem[{{Kusakabe} {et~al.}(2019){Kusakabe}, {Shimasaku}, {Momose}, {Ouchi},
  {Nakajima}, {Hashimoto}, {Harikane}, {Silverman}, \& {Capak}}]{Kusakabe2019}
{Kusakabe}, H., {Shimasaku}, K., {Momose}, R., {et~al.} 2019, \pasj, 71, 55,
  \dodoi{10.1093/pasj/psz029}

\bibitem[{{Lake} {et~al.}(2015){Lake}, {Zheng}, {Cen}, {Sadoun}, {Momose}, \&
  {Ouchi}}]{Lake2015}
{Lake}, E., {Zheng}, Z., {Cen}, R., {et~al.} 2015, \apj, 806, 46,
  \dodoi{10.1088/0004-637X/806/1/46}

\bibitem[{{Laursen} {et~al.}(2013){Laursen}, {Duval}, \&
  {{\"O}stlin}}]{Laursen13}
{Laursen}, P., {Duval}, F., \& {{\"O}stlin}, G. 2013, \apj, 766, 124,
  \dodoi{10.1088/0004-637X/766/2/124}

\bibitem[{{Laursen} {et~al.}(2011){Laursen}, {Sommer-Larsen}, \&
  {Razoumov}}]{Laursen2011}
{Laursen}, P., {Sommer-Larsen}, J., \& {Razoumov}, A.~O. 2011, \apj, 728, 52,
  \dodoi{10.1088/0004-637X/728/1/52}

\bibitem[{{Leclercq} {et~al.}(2017){Leclercq}, {Bacon}, {Wisotzki}, {Mitchell},
  {Garel}, {Verhamme}, {Blaizot}, {Hashimoto}, {Herenz}, {Conseil},
  {Cantalupo}, {Inami}, {Contini}, {Richard}, {Maseda}, {Schaye}, {Marino},
  {Akhlaghi}, {Brinchmann}, \& {Carollo}}]{Leclercq2017}
{Leclercq}, F., {Bacon}, R., {Wisotzki}, L., {et~al.} 2017, \aap, 608, A8,
  \dodoi{10.1051/0004-6361/201731480}

\bibitem[{{Leclercq} {et~al.}(2020){Leclercq}, {Bacon}, {Verhamme}, {Garel},
  {Blaizot}, {Brinchmann}, {Cantalupo}, {Claeyssens}, {Conseil}, {Contini},
  {Hashimoto}, {Herenz}, {Kusakabe}, {Marino}, {Maseda}, {Matthee}, {Mitchell},
  {Pezzulli}, {Richard}, {Schmidt}, \& {Wisotzki}}]{Leclercq2020}
{Leclercq}, F., {Bacon}, R., {Verhamme}, A., {et~al.} 2020, \aap, 635, A82,
  \dodoi{10.1051/0004-6361/201937339}

\bibitem[{{Leitherer} {et~al.}(2013){Leitherer}, {Chandar}, {Tremonti},
  {Wofford}, \& {Schaerer}}]{Leitherer13}
{Leitherer}, C., {Chandar}, R., {Tremonti}, C.~A., {Wofford}, A., \&
  {Schaerer}, D. 2013, \apj, 772, 120, \dodoi{10.1088/0004-637X/772/2/120}

\bibitem[{{Li} \& {Gronke}(2022)}]{Li22b}
{Li}, Z., \& {Gronke}, M. 2022, \mnras, 513, 5034,
  \dodoi{10.1093/mnras/stac1207}

\bibitem[{{Li} {et~al.}(2021){Li}, {Steidel}, {Gronke}, \& {Chen}}]{Li21}
{Li}, Z., {Steidel}, C.~C., {Gronke}, M., \& {Chen}, Y. 2021, \mnras, 502,
  2389, \dodoi{10.1093/mnras/staa3951}

\bibitem[{{Li} {et~al.}(2022){Li}, {Steidel}, {Gronke}, {Chen}, \&
  {Matsuda}}]{Li22a}
{Li}, Z., {Steidel}, C.~C., {Gronke}, M., {Chen}, Y., \& {Matsuda}, Y. 2022,
  \mnras, 513, 3414, \dodoi{10.1093/mnras/stac958}

\bibitem[{{Martin} {et~al.}(2010){Martin}, {Moore}, {Morrissey}, {Matuszewski},
  {Rahman}, {Adkins}, \& {Epps}}]{kcwi2010}
{Martin}, C., {Moore}, A., {Morrissey}, P., {et~al.} 2010, in Society of
  Photo-Optical Instrumentation Engineers (SPIE) Conference Series, Vol. 7735,
  Ground-based and Airborne Instrumentation for Astronomy III, ed. I.~S.
  {McLean}, S.~K. {Ramsay}, \& H.~{Takami}, 77350M, \dodoi{10.1117/12.858227}

\bibitem[{{Martin}(2005)}]{Martin05}
{Martin}, C.~L. 2005, \apj, 621, 227, \dodoi{10.1086/427277}

\bibitem[{{Martin} {et~al.}(2012){Martin}, {Shapley}, {Coil}, {Kornei},
  {Bundy}, {Weiner}, {Noeske}, \& {Schiminovich}}]{Martin12}
{Martin}, C.~L., {Shapley}, A.~E., {Coil}, A.~L., {et~al.} 2012, \apj, 760,
  127, \dodoi{10.1088/0004-637X/760/2/127}

\bibitem[{{Mas-Ribas} \& {Dijkstra}(2016)}]{Mas-Ribas2016}
{Mas-Ribas}, L., \& {Dijkstra}, M. 2016, \apj, 822, 84,
  \dodoi{10.3847/0004-637X/822/2/84}

\bibitem[{{Mas-Ribas} {et~al.}(2017){Mas-Ribas}, {Dijkstra}, {Hennawi},
  {Trenti}, {Momose}, \& {Ouchi}}]{Mas-Ribas2017}
{Mas-Ribas}, L., {Dijkstra}, M., {Hennawi}, J.~F., {et~al.} 2017, \apj, 841,
  19, \dodoi{10.3847/1538-4357/aa704e}

\bibitem[{{Matthee} {et~al.}(2021){Matthee}, {Sobral}, {Hayes}, {Pezzulli},
  {Gronke}, {Schaerer}, {Naidu}, {R{\"o}ttgering}, {Calhau}, {Paulino-Afonso},
  {Santos}, \& {Amor{\'\i}n}}]{Matthee2021}
{Matthee}, J., {Sobral}, D., {Hayes}, M., {et~al.} 2021, \mnras, 505, 1382,
  \dodoi{10.1093/mnras/stab1304}

\bibitem[{{Mitchell} {et~al.}(2021){Mitchell}, {Blaizot}, {Cadiou}, {Dubois},
  {Garel}, \& {Rosdahl}}]{Mitchell2021}
{Mitchell}, P.~D., {Blaizot}, J., {Cadiou}, C., {et~al.} 2021, \mnras, 501,
  5757, \dodoi{10.1093/mnras/stab035}

\bibitem[{Momose {et~al.}(2014)Momose, Ouchi, Nakajima, Ono, Shibuya,
  Shimasaku, Yuma, Mori, \& Umemura}]{Momose2014}
Momose, R., Ouchi, M., Nakajima, K., {et~al.} 2014, Monthly Notices of the
  Royal Astronomical Society, 442, 110 , \dodoi{10.1093/mnras/stu825}

\bibitem[{Momose {et~al.}(2016)Momose, Ouchi, Nakajima, Ono, Shibuya,
  Shimasaku, Yuma, Mori, \& Umemura}]{Momose2016}
---. 2016, Monthly Notices of the Royal Astronomical Society, 457, 2318 ,
  \dodoi{10.1093/mnras/stw021}

\bibitem[{{Morrissey} {et~al.}(2012){Morrissey}, {Matuszewski}, {Martin},
  {Moore}, {Adkins}, {Epps}, {Bartos}, {Cabak}, {Cowley}, {Davis}, {Delacroix},
  {Fucik}, {Hilliard}, {James}, {Kaye}, {Lingner}, {Neill}, {Pistor},
  {Phillips}, {Rockosi}, \& {Weber}}]{kcwi2012}
{Morrissey}, P., {Matuszewski}, M., {Martin}, C., {et~al.} 2012, in Society of
  Photo-Optical Instrumentation Engineers (SPIE) Conference Series, Vol. 8446,
  Ground-based and Airborne Instrumentation for Astronomy IV, ed. I.~S.
  {McLean}, S.~K. {Ramsay}, \& H.~{Takami}, 844613, \dodoi{10.1117/12.924729}

\bibitem[{{Murray} {et~al.}(2005){Murray}, {Quataert}, \&
  {Thompson}}]{Murray05}
{Murray}, N., {Quataert}, E., \& {Thompson}, T.~A. 2005, \apj, 618, 569,
  \dodoi{10.1086/426067}

\bibitem[{{Naidu} {et~al.}(2022){Naidu}, {Matthee}, {Oesch}, {Conroy},
  {Sobral}, {Pezzulli}, {Hayes}, {Erb}, {Amor{\'\i}n}, {Gronke}, {Schaerer},
  {Tacchella}, {Kerutt}, {Paulino-Afonso}, {Calhau}, {Llerena}, \&
  {R{\"o}ttgering}}]{Naidu2022}
{Naidu}, R.~P., {Matthee}, J., {Oesch}, P.~A., {et~al.} 2022, \mnras, 510,
  4582, \dodoi{10.1093/mnras/stab3601}

\bibitem[{{Nakajima} {et~al.}(2016){Nakajima}, {Ellis}, {Iwata}, {Inoue},
  {Kusakabe}, {Ouchi}, \& {Robertson}}]{Nakajima2016}
{Nakajima}, K., {Ellis}, R.~S., {Iwata}, I., {et~al.} 2016, \apjl, 831, L9,
  \dodoi{10.3847/2041-8205/831/1/L9}

\bibitem[{{Neufeld}(1991)}]{Neufeld91}
{Neufeld}, D.~A. 1991, \apjl, 370, L85, \dodoi{10.1086/185983}

\bibitem[{{Orlitov{\'a}} {et~al.}(2018){Orlitov{\'a}}, {Verhamme}, {Henry},
  {Scarlata}, {Jaskot}, {Oey}, \& {Schaerer}}]{Orlitova2018}
{Orlitov{\'a}}, I., {Verhamme}, A., {Henry}, A., {et~al.} 2018, \aap, 616, A60,
  \dodoi{10.1051/0004-6361/201732478}

\bibitem[{{Ouchi} {et~al.}(2020){Ouchi}, {Ono}, \& {Shibuya}}]{Ouchi2020}
{Ouchi}, M., {Ono}, Y., \& {Shibuya}, T. 2020, \araa, 58, 617,
  \dodoi{10.1146/annurev-astro-032620-021859}

\bibitem[{{Patr{\'\i}cio} {et~al.}(2016){Patr{\'\i}cio}, {Richard}, {Verhamme},
  {Wisotzki}, {Brinchmann}, {Turner}, {Christensen}, {Weilbacher}, {Blaizot},
  {Bacon}, {Contini}, {Lagattuta}, {Cantalupo}, {Cl{\'e}ment}, \&
  {Soucail}}]{Patricio2016}
{Patr{\'\i}cio}, V., {Richard}, J., {Verhamme}, A., {et~al.} 2016, \mnras, 456,
  4191, \dodoi{10.1093/mnras/stv2859}

\bibitem[{{Planck Collaboration} {et~al.}(2020){Planck Collaboration},
  {Aghanim}, {Akrami}, {Ashdown}, {Aumont}, {Baccigalupi}, {Ballardini},
  {Banday}, {Barreiro}, {Bartolo}, {Basak}, {Battye}, {Benabed}, {Bernard},
  {Bersanelli}, {Bielewicz}, {Bock}, {Bond}, {Borrill}, {Bouchet}, {Boulanger},
  {Bucher}, {Burigana}, {Butler}, {Calabrese}, {Cardoso}, {Carron},
  {Challinor}, {Chiang}, {Chluba}, {Colombo}, {Combet}, {Contreras}, {Crill},
  {Cuttaia}, {de Bernardis}, {de Zotti}, {Delabrouille}, {Delouis}, {Di
  Valentino}, {Diego}, {Dor{\'e}}, {Douspis}, {Ducout}, {Dupac}, {Dusini},
  {Efstathiou}, {Elsner}, {En{\ss}lin}, {Eriksen}, {Fantaye}, {Farhang},
  {Fergusson}, {Fernandez-Cobos}, {Finelli}, {Forastieri}, {Frailis},
  {Fraisse}, {Franceschi}, {Frolov}, {Galeotta}, {Galli}, {Ganga},
  {G{\'e}nova-Santos}, {Gerbino}, {Ghosh}, {Gonz{\'a}lez-Nuevo}, {G{\'o}rski},
  {Gratton}, {Gruppuso}, {Gudmundsson}, {Hamann}, {Handley}, {Hansen},
  {Herranz}, {Hildebrandt}, {Hivon}, {Huang}, {Jaffe}, {Jones}, {Karakci},
  {Keih{\"a}nen}, {Keskitalo}, {Kiiveri}, {Kim}, {Kisner}, {Knox},
  {Krachmalnicoff}, {Kunz}, {Kurki-Suonio}, {Lagache}, {Lamarre}, {Lasenby},
  {Lattanzi}, {Lawrence}, {Le Jeune}, {Lemos}, {Lesgourgues}, {Levrier},
  {Lewis}, {Liguori}, {Lilje}, {Lilley}, {Lindholm}, {L{\'o}pez-Caniego},
  {Lubin}, {Ma}, {Mac{\'\i}as-P{\'e}rez}, {Maggio}, {Maino}, {Mandolesi},
  {Mangilli}, {Marcos-Caballero}, {Maris}, {Martin}, {Martinelli},
  {Mart{\'\i}nez-Gonz{\'a}lez}, {Matarrese}, {Mauri}, {McEwen}, {Meinhold},
  {Melchiorri}, {Mennella}, {Migliaccio}, {Millea}, {Mitra},
  {Miville-Desch{\^e}nes}, {Molinari}, {Montier}, {Morgante}, {Moss}, {Natoli},
  {N{\o}rgaard-Nielsen}, {Pagano}, {Paoletti}, {Partridge}, {Patanchon},
  {Peiris}, {Perrotta}, {Pettorino}, {Piacentini}, {Polastri}, {Polenta},
  {Puget}, {Rachen}, {Reinecke}, {Remazeilles}, {Renzi}, {Rocha}, {Rosset},
  {Roudier}, {Rubi{\~n}o-Mart{\'\i}n}, {Ruiz-Granados}, {Salvati}, {Sandri},
  {Savelainen}, {Scott}, {Shellard}, {Sirignano}, {Sirri}, {Spencer},
  {Sunyaev}, {Suur-Uski}, {Tauber}, {Tavagnacco}, {Tenti}, {Toffolatti},
  {Tomasi}, {Trombetti}, {Valenziano}, {Valiviita}, {Van Tent}, {Vibert},
  {Vielva}, {Villa}, {Vittorio}, {Wandelt}, {Wehus}, {White}, {White},
  {Zacchei}, \& {Zonca}}]{planck2018cosmo}
{Planck Collaboration}, {Aghanim}, N., {Akrami}, Y., {et~al.} 2020, \aap, 641,
  A6, \dodoi{10.1051/0004-6361/201833910}

\bibitem[{{Reddy} {et~al.}(2012){Reddy}, {Pettini}, {Steidel}, {Shapley},
  {Erb}, \& {Law}}]{Reddy2012}
{Reddy}, N.~A., {Pettini}, M., {Steidel}, C.~C., {et~al.} 2012, \apj, 754, 25,
  \dodoi{10.1088/0004-637X/754/1/25}

\bibitem[{{Reddy} \& {Steidel}(2009)}]{Reddy2009}
{Reddy}, N.~A., \& {Steidel}, C.~C. 2009, \apj, 692, 778,
  \dodoi{10.1088/0004-637X/692/1/778}

\bibitem[{{Reddy} {et~al.}(2016){Reddy}, {Steidel}, {Pettini},
  {Bogosavljevi{\'c}}, \& {Shapley}}]{Reddy2016}
{Reddy}, N.~A., {Steidel}, C.~C., {Pettini}, M., {Bogosavljevi{\'c}}, M., \&
  {Shapley}, A.~E. 2016, \apj, 828, 108, \dodoi{10.3847/0004-637X/828/2/108}

\bibitem[{{Rivera-Thorsen} {et~al.}(2019){Rivera-Thorsen}, {Dahle}, {Chisholm},
  {Florian}, {Gronke}, {Rigby}, {Gladders}, {Mahler}, {Sharon}, \&
  {Bayliss}}]{Rivera-Thorsen2019}
{Rivera-Thorsen}, T.~E., {Dahle}, H., {Chisholm}, J., {et~al.} 2019, Science,
  366, 738, \dodoi{10.1126/science.aaw0978}

\bibitem[{{Robertson} {et~al.}(2015){Robertson}, {Ellis}, {Furlanetto}, \&
  {Dunlop}}]{Robertson2015}
{Robertson}, B.~E., {Ellis}, R.~S., {Furlanetto}, S.~R., \& {Dunlop}, J.~S.
  2015, \apjl, 802, L19, \dodoi{10.1088/2041-8205/802/2/L19}

\bibitem[{{Rubin} {et~al.}(2014){Rubin}, {Prochaska}, {Koo}, {Phillips},
  {Martin}, \& {Winstrom}}]{Rubin14}
{Rubin}, K. H.~R., {Prochaska}, J.~X., {Koo}, D.~C., {et~al.} 2014, \apj, 794,
  156, \dodoi{10.1088/0004-637X/794/2/156}

\bibitem[{{Rudie} {et~al.}(2019){Rudie}, {Steidel}, {Pettini}, {Trainor},
  {Strom}, {Hummels}, {Reddy}, \& {Shapley}}]{Rudie19}
{Rudie}, G.~C., {Steidel}, C.~C., {Pettini}, M., {et~al.} 2019, \apj, 885, 61,
  \dodoi{10.3847/1538-4357/ab4255}

\bibitem[{{Rudie} {et~al.}(2013){Rudie}, {Steidel}, {Shapley}, \&
  {Pettini}}]{Rudie2013}
{Rudie}, G.~C., {Steidel}, C.~C., {Shapley}, A.~E., \& {Pettini}, M. 2013,
  \apj, 769, 146, \dodoi{10.1088/0004-637X/769/2/146}

\bibitem[{{Rudie} {et~al.}(2012){Rudie}, {Steidel}, {Trainor}, {Rakic},
  {Bogosavljevi{\'c}}, {Pettini}, {Reddy}, {Shapley}, {Erb}, \&
  {Law}}]{Rudie2012}
{Rudie}, G.~C., {Steidel}, C.~C., {Trainor}, R.~F., {et~al.} 2012, \apj, 750,
  67, \dodoi{10.1088/0004-637X/750/1/67}

\bibitem[{{Rupke} {et~al.}(2005){Rupke}, {Veilleux}, \& {Sanders}}]{Rupke05}
{Rupke}, D.~S., {Veilleux}, S., \& {Sanders}, D.~B. 2005, \apjs, 160, 115,
  \dodoi{10.1086/432889}

\bibitem[{{Sanders} {et~al.}(2016){Sanders}, {Shapley}, {Kriek}, {Reddy},
  {Freeman}, {Coil}, {Siana}, {Mobasher}, {Shivaei}, {Price}, \& {de
  Groot}}]{Sanders2016}
{Sanders}, R.~L., {Shapley}, A.~E., {Kriek}, M., {et~al.} 2016, \apj, 816, 23,
  \dodoi{10.3847/0004-637X/816/1/23}

\bibitem[{{Shapley} {et~al.}(2003){Shapley}, {Steidel}, {Pettini}, \&
  {Adelberger}}]{Shapley2003}
{Shapley}, A.~E., {Steidel}, C.~C., {Pettini}, M., \& {Adelberger}, K.~L. 2003,
  \apj, 588, 65, \dodoi{10.1086/373922}

\bibitem[{{Shibuya} {et~al.}(2014){Shibuya}, {Ouchi}, {Nakajima}, {Hashimoto},
  {Ono}, {Rauch}, {Gauthier}, {Shimasaku}, {Goto}, {Mori}, \&
  {Umemura.}}]{Shibuya2014}
{Shibuya}, T., {Ouchi}, M., {Nakajima}, K., {et~al.} 2014, \apj, 788, 74,
  \dodoi{10.1088/0004-637X/788/1/74}

\bibitem[{{Skilling}(2004)}]{Skilling04}
{Skilling}, J. 2004, in American Institute of Physics Conference Series, Vol.
  735, American Institute of Physics Conference Series, ed. R.~{Fischer},
  R.~{Preuss}, \& U.~V. {Toussaint}, 395--405, \dodoi{10.1063/1.1835238}

\bibitem[{{Skilling}(2006)}]{Skilling06}
{Skilling}, J. 2006, Bayesian Anal., 1, 833, \dodoi{10.1214/06-BA127}

\bibitem[{{Solimano} {et~al.}(2022){Solimano}, {Gonz{\'a}lez-L{\'o}pez},
  {Aravena}, {Johnston}, {Moya-Sierralta}, {Barrientos}, {Bayliss}, {Gladders},
  {Infante}, {Ledoux}, {L{\'o}pez}, {Poudel}, {Rigby}, {Sharon}, \&
  {Tejos}}]{Solimano2022}
{Solimano}, M., {Gonz{\'a}lez-L{\'o}pez}, J., {Aravena}, M., {et~al.} 2022,
  arXiv e-prints, arXiv:2206.02949.
\newblock \doarXiv{2206.02949}

\bibitem[{{Speagle}(2020)}]{Speagle20}
{Speagle}, J.~S. 2020, \mnras, 493, 3132, \dodoi{10.1093/mnras/staa278}

\bibitem[{{Stanway} \& {Eldridge}(2018)}]{BPASSv2.2}
{Stanway}, E.~R., \& {Eldridge}, J.~J. 2018, \mnras, 479, 75,
  \dodoi{10.1093/mnras/sty1353}

\bibitem[{{Stark} {et~al.}(2017){Stark}, {Ellis}, {Charlot}, {Chevallard},
  {Tang}, {Belli}, {Zitrin}, {Mainali}, {Gutkin}, {Vidal-Garc{\'\i}a},
  {Bouwens}, \& {Oesch}}]{Stark2017}
{Stark}, D.~P., {Ellis}, R.~S., {Charlot}, S., {et~al.} 2017, \mnras, 464, 469,
  \dodoi{10.1093/mnras/stw2233}

\bibitem[{{Steidel} {et~al.}(2000){Steidel}, {Adelberger}, {Shapley},
  {Pettini}, {Dickinson}, \& {Giavalisco}}]{Steidel2000}
{Steidel}, C.~C., {Adelberger}, K.~L., {Shapley}, A.~E., {et~al.} 2000, \apj,
  532, 170, \dodoi{10.1086/308568}

\bibitem[{{Steidel} {et~al.}(2011){Steidel}, {Bogosavljevi{\'c}}, {Shapley},
  {Kollmeier}, {Reddy}, {Erb}, \& {Pettini}}]{Steidel11}
{Steidel}, C.~C., {Bogosavljevi{\'c}}, M., {Shapley}, A.~E., {et~al.} 2011,
  \apj, 736, 160, \dodoi{10.1088/0004-637X/736/2/160}

\bibitem[{{Steidel} {et~al.}(2018){Steidel}, {Bogosavljevi{\'c}}, {Shapley},
  {Reddy}, {Rudie}, {Pettini}, {Trainor}, \& {Strom}}]{Steidel2018}
---. 2018, \apj, 869, 123, \dodoi{10.3847/1538-4357/aaed28}

\bibitem[{{Steidel} {et~al.}(2010){Steidel}, {Erb}, {Shapley}, {Pettini},
  {Reddy}, {Bogosavljevi{\'c}}, {Rudie}, \& {Rakic}}]{Steidel10}
{Steidel}, C.~C., {Erb}, D.~K., {Shapley}, A.~E., {et~al.} 2010, \apj, 717,
  289, \dodoi{10.1088/0004-637X/717/1/289}

\bibitem[{{Steidel} {et~al.}(2014){Steidel}, {Rudie}, {Strom}, {Pettini},
  {Reddy}, {Shapley}, {Trainor}, {Erb}, {Turner}, {Konidaris}, {Kulas}, {Mace},
  {Matthews}, \& {McLean}}]{Steidel2014}
{Steidel}, C.~C., {Rudie}, G.~C., {Strom}, A.~L., {et~al.} 2014, \apj, 795,
  165, \dodoi{10.1088/0004-637X/795/2/165}

\bibitem[{{Strom} {et~al.}(2017){Strom}, {Steidel}, {Rudie}, {Trainor},
  {Pettini}, \& {Reddy}}]{Strom2017}
{Strom}, A.~L., {Steidel}, C.~C., {Rudie}, G.~C., {et~al.} 2017, \apj, 836,
  164, \dodoi{10.3847/1538-4357/836/2/164}

\bibitem[{{Tang} {et~al.}(2019){Tang}, {Stark}, {Chevallard}, \&
  {Charlot}}]{Tang2019}
{Tang}, M., {Stark}, D.~P., {Chevallard}, J., \& {Charlot}, S. 2019, \mnras,
  489, 2572, \dodoi{10.1093/mnras/stz2236}

\bibitem[{{Theios} {et~al.}(2019){Theios}, {Steidel}, {Strom}, {Rudie},
  {Trainor}, \& {Reddy}}]{Theios2019}
{Theios}, R.~L., {Steidel}, C.~C., {Strom}, A.~L., {et~al.} 2019, \apj, 871,
  128, \dodoi{10.3847/1538-4357/aaf386}

\bibitem[{{Trainor} \& {Steidel}(2012)}]{Trainor2012}
{Trainor}, R.~F., \& {Steidel}, C.~C. 2012, \apj, 752, 39,
  \dodoi{10.1088/0004-637X/752/1/39}

\bibitem[{{Trainor} {et~al.}(2015){Trainor}, {Steidel}, {Strom}, \&
  {Rudie}}]{Trainor15}
{Trainor}, R.~F., {Steidel}, C.~C., {Strom}, A.~L., \& {Rudie}, G.~C. 2015,
  \apj, 809, 89, \dodoi{10.1088/0004-637X/809/1/89}

\bibitem[{{Trainor} {et~al.}(2016){Trainor}, {Strom}, {Steidel}, \&
  {Rudie}}]{Trainor2016}
{Trainor}, R.~F., {Strom}, A.~L., {Steidel}, C.~C., \& {Rudie}, G.~C. 2016,
  \apj, 832, 171, \dodoi{10.3847/0004-637X/832/2/171}

\bibitem[{{Trainor} {et~al.}(2019){Trainor}, {Strom}, {Steidel}, {Rudie},
  {Chen}, \& {Theios}}]{Trainor2019}
{Trainor}, R.~F., {Strom}, A.~L., {Steidel}, C.~C., {et~al.} 2019, \apj, 887,
  85, \dodoi{10.3847/1538-4357/ab4993}

\bibitem[{{Tumlinson} {et~al.}(2017){Tumlinson}, {Peeples}, \&
  {Werk}}]{Tumlinson2017}
{Tumlinson}, J., {Peeples}, M.~S., \& {Werk}, J.~K. 2017, \araa, 55, 389,
  \dodoi{10.1146/annurev-astro-091916-055240}

\bibitem[{Umehata {et~al.}(2019)Umehata, Fumagalli, Smail, Matsuda, Swinbank,
  Cantalupo, Sykes, Ivison, Steidel, Shapley, Vernet, Yamada, Tamura, Kubo,
  Nakanishi, Kajisawa, Hatsukade, \& Kohno}]{Umehata2019}
Umehata, H., Fumagalli, M., Smail, I., {et~al.} 2019, Science, 366, 97,
  \dodoi{10.1126/science.aaw5949}

\bibitem[{van~der Velden(2020)}]{cmasher}
van~der Velden, E. 2020, Journal of Open Source Software, 5, 2004,
  \dodoi{10.21105/joss.02004}

\bibitem[{{Veilleux} {et~al.}(2020){Veilleux}, {Maiolino}, {Bolatto}, \&
  {Aalto}}]{Veilleux2020}
{Veilleux}, S., {Maiolino}, R., {Bolatto}, A.~D., \& {Aalto}, S. 2020, \aapr,
  28, 2, \dodoi{10.1007/s00159-019-0121-9}

\bibitem[{{Verhamme} {et~al.}(2015){Verhamme}, {Orlitov{\'a}}, {Schaerer}, \&
  {Hayes}}]{Verhamme2015}
{Verhamme}, A., {Orlitov{\'a}}, I., {Schaerer}, D., \& {Hayes}, M. 2015, \aap,
  578, A7, \dodoi{10.1051/0004-6361/201423978}

\bibitem[{{Verhamme} {et~al.}(2017){Verhamme}, {Orlitov{\'a}}, {Schaerer},
  {Izotov}, {Worseck}, {Thuan}, \& {Guseva}}]{Verhamme2017}
{Verhamme}, A., {Orlitov{\'a}}, I., {Schaerer}, D., {et~al.} 2017, \aap, 597,
  A13, \dodoi{10.1051/0004-6361/201629264}

\bibitem[{Verhamme {et~al.}(2008)Verhamme, Schaerer, Atek, \&
  Tapken}]{Verhamme2008}
Verhamme, A., Schaerer, D., Atek, H., \& Tapken, C. 2008, Astronomy and
  Astrophysics, 491, 89 , \dodoi{10.1051/0004-6361:200809648}

\bibitem[{Waskom {et~al.}(2020)Waskom, Gelbart, Botvinnik, Ostblom, Hobson,
  Lukauskas, Gemperline, Augspurger, Halchenko, Warmenhoven, Cole, de~Ruiter,
  Vanderplas, Hoyer, Pye, Miles, Swain, Meyer, Martin, Bachant, Quintero,
  Kunter, Villalba, Brian, Fitzgerald, Evans, Williams, O'Kane, Yarkoni, \&
  Brunner}]{seaborn}
Waskom, M., Gelbart, M., Botvinnik, O., {et~al.} 2020, mwaskom/seaborn: v0.11.1
  (December 2020), v0.11.1,  Zenodo, \dodoi{10.5281/zenodo.4379347}

\bibitem[{{Weiner} {et~al.}(2009){Weiner}, {Coil}, {Prochaska}, {Newman},
  {Cooper}, {Bundy}, {Conselice}, {Dutton}, {Faber}, {Koo}, {Lotz}, {Rieke}, \&
  {Rubin}}]{Weiner09}
{Weiner}, B.~J., {Coil}, A.~L., {Prochaska}, J.~X., {et~al.} 2009, \apj, 692,
  187, \dodoi{10.1088/0004-637X/692/1/187}

\bibitem[{{Whitaker} {et~al.}(2014){Whitaker}, {Franx}, {Leja}, {van Dokkum},
  {Henry}, {Skelton}, {Fumagalli}, {Momcheva}, {Brammer}, {Labb{\'e}},
  {Nelson}, \& {Rigby}}]{Whitaker2014}
{Whitaker}, K.~E., {Franx}, M., {Leja}, J., {et~al.} 2014, \apj, 795, 104,
  \dodoi{10.1088/0004-637X/795/2/104}

\bibitem[{{Wisotzki} {et~al.}(2016){Wisotzki}, {Bacon}, {Blaizot},
  {Brinchmann}, {Herenz}, {Schaye}, {Bouch{\'e}}, {Cantalupo}, {Contini},
  {Carollo}, {Caruana}, {Courbot}, {Emsellem}, {Kamann}, {Kerutt}, {Leclercq},
  {Lilly}, {Patr{\'{\i}}cio}, {Sandin}, {Steinmetz}, {Straka}, {Urrutia},
  {Verhamme}, {Weilbacher}, \& {Wendt}}]{Wisotzki2016}
{Wisotzki}, L., {Bacon}, R., {Blaizot}, J., {et~al.} 2016, \aap, 587, A98,
  \dodoi{10.1051/0004-6361/201527384}

\bibitem[{{Wisotzki} {et~al.}(2018){Wisotzki}, {Bacon}, {Brinchmann},
  {Cantalupo}, {Richter}, {Schaye}, {Schmidt}, {Urrutia}, {Weilbacher},
  {Akhlaghi}, {Bouch{\'e}}, {Contini}, {Guiderdoni}, {Herenz}, {Inami},
  {Kerutt}, {Leclercq}, {Marino}, {Maseda}, {Monreal-Ibero}, {Nanayakkara},
  {Richard}, {Saust}, {Steinmetz}, \& {Wendt}}]{Wisotzki2018}
{Wisotzki}, L., {Bacon}, R., {Brinchmann}, J., {et~al.} 2018, \nat, 562, 229,
  \dodoi{10.1038/s41586-018-0564-6}

\bibitem[{{Xue} {et~al.}(2017){Xue}, {Lee}, {Dey}, {Reddy}, {Hong}, {Prescott},
  {Inami}, {Jannuzi}, \& {Gonzalez}}]{Xue2017}
{Xue}, R., {Lee}, K.-S., {Dey}, A., {et~al.} 2017, \apj, 837, 172,
  \dodoi{10.3847/1538-4357/837/2/172}

\bibitem[{{Yang} {et~al.}(2017){Yang}, {Malhotra}, {Gronke}, {Rhoads},
  {Leitherer}, {Wofford}, {Jiang}, {Dijkstra}, {Tilvi}, \& {Wang}}]{Yang2017}
{Yang}, H., {Malhotra}, S., {Gronke}, M., {et~al.} 2017, \apj, 844, 171,
  \dodoi{10.3847/1538-4357/aa7d4d}

\bibitem[{{Zheng} {et~al.}(2011){Zheng}, {Cen}, {Weinberg}, {Trac}, \&
  {Miralda-Escud{\'e}}}]{Zheng2011}
{Zheng}, Z., {Cen}, R., {Weinberg}, D., {Trac}, H., \& {Miralda-Escud{\'e}}, J.
  2011, \apj, 739, 62, \dodoi{10.1088/0004-637X/739/2/62}

\end{thebibliography}

\end{CJK*}
\end{document}